\documentclass{aastex631}

\usepackage{silence}
\usepackage{float}
\usepackage{amsmath}
\usepackage{tasks}
\usepackage{booktabs}
\usepackage{rotating}
\usepackage{makecell}
\usepackage{xspace}
\usepackage{enumitem}
\usepackage{textcomp}
\setlist[itemize]{noitemsep}

\shorttitle{A Targeted Search for Variable Gravitationally Lensed Quasars}
\shortauthors{Sheu et al.}

\graphicspath{{./}{figures/}}

\newcommand{\ipg}{\emph{g}\xspace}
\newcommand{\ipr}{\emph{r}\xspace}
\newcommand{\ipi}{\emph{i}\xspace}
\newcommand{\ipz}{\emph{z}\xspace}

\newcommand{\txrd}{\textcolor{black}}
\newcommand{\txrr}{\textcolor{black}}
\newcommand{\txr}{\textcolor{black}}
\newcommand{\txo}{\textcolor{black}}
\newcommand{\txg}{\textcolor{black}}
\newcommand{\txb}{\textcolor{black}}
\newcommand{\txp}{\textcolor{black}}

\begin{document}

\title{A Targeted Search for Variable Gravitationally Lensed Quasars}

\correspondingauthor{William Sheu}
\email{wsheu@astro.ucla.edu}

\author[0000-0003-1889-0227]{William Sheu}
\affiliation{Department of Physics \& Astronomy, University of California, Los Angeles \\ 430 Portola Plaza, Los Angeles, CA 90095, USA}
\affiliation{Physics Division, Lawrence Berkeley National Laboratory \\ 1 Cyclotron Road, Berkeley, CA 94720, USA}

\author[0000-0001-8156-0330]{Xiaosheng Huang}
\affiliation{Department of Physics \& Astronomy, University of San Francisco \\ 2130 Fulton Street, San Francisco, CA 94117-1080, USA}
\affiliation{Physics Division, Lawrence Berkeley National Laboratory \\ 1 Cyclotron Road, Berkeley, CA 94720, USA}

\author[0000-0001-7101-9831]{Aleksandar Cikota}
\affiliation{Gemini Observatory / NSF's NOIRLab \\ Casilla 603, La Serena, Chile}

\author[0000-0001-7266-930X]{Nao Suzuki}
\affiliation{Physics Division, Lawrence Berkeley National Laboratory \\ 1 Cyclotron Road, Berkeley, CA 94720, USA}
\affiliation{Kavli Institute for the Physics and Mathematics of the Universe, University of Tokyo \\ Kashiwa 277-8583, Japan}

\author[0000-0002-6011-0530]{Antonella Palmese}
\affiliation{Department of Physics, Carnegie Mellon University \\ 5000 Forbes Ave
Pittsburgh, PA 15213-3890, USA}

\author[0000-0002-5042-5088]{David J. Schlegel}
\affiliation{Physics Division, Lawrence Berkeley National Laboratory \\ 1 Cyclotron Road, Berkeley, CA 94720, USA}

\author[0000-0002-0385-0014]{Christopher Storfer}
\affiliation{Physics Division, Lawrence Berkeley National Laboratory \\ 1 Cyclotron Road, Berkeley, CA 94720, USA}
\affiliation{Institute for Astronomy, University of Hawaii \\ 2680 Woodlawn Drive, Honolulu, HI 96822-1897, USA}

\begin{abstract}

We present a pipeline to identify photometric variability within strong \txrr{gravitationally} lensing candidates, in the DESI Legacy Imaging Surveys.  In our first paper \citep{sheu2023}, we \txg{laid out} our pipeline and presented seven new \txrr{gravitationally} lensed supernovae candidates in a retrospective search.  In this \txrr{companion} paper, we apply \txrr{a modified version of that} pipeline to search for \txrr{gravitationally} lensed quasars.  \txrd{From a sample of 5807 strong lenses, we have identified 13 new gravitationally lensed quasar candidates (three of them quadruply-lensed).}  We note that our methodology differs from most lensed quasar search algorithms that solely rely on the morphology, location, and color of the \txb{candidate systems}.  By also taking into account the \txb{temporal photometric} variability of the posited lensed images in our search via \txrd{difference imaging}, we have discovered new lensed quasar \txg{candidates}.  \txrr{While variability searches using \txrd{difference imaging} algorithms have been done in the past, they are typically preformed over vast swathes of sky, whereas we specifically target strong gravitationally lensed candidates.}  We also have applied our pipeline to \txrd{655} known \txrr{gravitationally} lensed quasar candidates from past lensed quasar searches, of which we identify 13 that display significant variability (one of them quadruply-lensed).  This pipeline demonstrates a promising search strategy to \txrr{discover gravitationally} lensed quasars in other existing and upcoming surveys. 
\end{abstract}


\section{Introduction} \label{sec:introduction}
Currently there is a significant discrepancy between early and late universe observations of the Hubble constant ($H_0$), the expansion rate of the universe at the present day.  Assuming a flat $\Lambda$CMB model, the measurement from the Planck cosmic microwave background (CMB) observations yields $H_0 = 67.4 \pm 0.5$ km/s/Mpc \citep{planck2018}.  On the other hand, local measurements report significantly higher values \citep[e.g., $H_0 = 73.30 \pm 1.04$ km/s/Mpc;][]{reiss2021}.  These predictions disagree by approximately $5\sigma$.  While recent Tip of the Red Giant Branch (TRGB) measurements give $H_0 = 69.8 \pm 1.7$ km/s/Mpc \citep{freedman2021}, there remains potentially \txb{strong} ambiguity with respect to the true value of $H_0$.

\txrr{Gravitationally} lensed transients may hold the key to resolving this tension.  In a strong \txrr{gravitational} lensing system, the light from a background galaxy is bent around a foreground galaxy, producing multiple images of the background source around the foreground lens.  If the background galaxy hosts a transient event, the time delays between the images can be measured and used to constrain $H_0$, when accurately modeled \citep[e.g.][]{wong2019, kelly2023}.  \txrr{Gravitationally} lensed supernovae are particularly valuable in this context due to their well-defined light curves, and in the case of Type Ia supernovae, their standardizable peak magnitudes.  However, these events are extremely rare, with only eight confirmed cases reported \citep{quimby2014, kelly2015, goobar, rodney2021, astronotekelly, astronotegoobar, kellysn2, snh0pe}.  In our first paper \citep[][hereafter Paper~I]{sheu2023}, we have identified seven new candidates for \txg{retrospective} lensed supernovae using targeted image subtraction and transient detection methods applied to individual exposures of the Dark Energy Spectroscopic Instrument \citep[DESI;][]{desi2016} Legacy Imaging Surveys \citep{dr9}.

Compared to supernovae, quasars have stochastic light curves than lensed supernovae.  However, \txrr{gravitationally} lensed quasars are much more common.
 Utilizing six quasars, \citet{wong2019} was able to constrain $H_0$ with a $2.4 \%$ precision with mass profile assumptions.  By expanding the current catalog of \txrr{gravitationally} lensed quasars, it will be possible to constrain the Hubble constant to sub-percent level precision \citep[e.g.,][]{liao2019}, \txb{with accurate lens modeling} \citep[e.g.,][]{shajib2023}. 

In addition to constraining $H_0$, \txrr{gravitationally} lensed quasars can provide valuable insights into the properties of galaxies and quasars in the early universe, shedding light on the formation and evolution of galaxies and black holes.  The \txrr{gravitational} lensing phenomenon itself allows for the detection of high redshift quasars, due to lens magnification. Additionally, we can achieve \txg{higher cadence} and extended measurements of a quasar's \txg{combined} light curve, as there are typically two or four lensed images \citep[e.g.,][]{dahle2015}.  By leveraging spectroscopic variability, reverberation mapping of \txrr{gravitationally} lensed quasars \txg{also} enables us to constrain the masses of black holes at high redshifts \citep[e.g.,][]{kelly2013}.  

\txrr{In this paper, we show the applicability of finding new lensed quasars through a targeted \txrd{difference imaging} search algorithm, identifying variability within candidate systems.  Specifically, we do two searches: one that identifies lensed quasar candidates in a set of strong lens candidates via variability, and another that identifies lensed quasar candidates with strong variability in a set of previously identified lensed quasar candidates.  Oftentimes in this paper we use the word ``targeted" in the sense that we focus observations on specific objects, in contrast to other search tactics that cover vast swaths of the sky to search for (lensed) quasars.  Therefore if deployed in future surveys, the search utilized in this paper should be supplied with a set of lens candidates within the survey footprint, and is intended to identify variable, gravitationally lensed quasar candidates from this set.  While the concept of a variability search for gravitationally lensed quasars is not new \citep{kochanek2006} and have been deployed on previous searches with much success \citep[e.g.,][]{lacki2009, kostrzewa2018, dux2023}, our work is the first to deploy such a targeted \txrd{difference imaging} search algorithm in the DESI Legacy Imaging Surveys.}

Presently there are \txb{$\sim 200$} confirmed \txrr{gravitationally} lensed quasar systems \citep[e.g.,][]{walsh1979, weymann1979, weymann1980, inada2012, more2016, agnello2018, lemon2019, lemon2018, jaelani2021, lemon2023}.  In this paper, we present \txrd{13} new \txrr{gravitationally} lensed quasar candidates (\txrd{three} being quadruply-lensed) to demonstrate our pipeline's capability to detect lensed transients, augmenting the catalog of \txrr{gravitationally} lensed quasars that can be found with current data.  Additionally, we \txrr{confirm significant} variability for 13 \txrr{gravitationally} lensed quasar candidates identified by \citet[][hereafter D22]{dawes2022} and \citet[][hereafter H23]{he2023}, demonstrating our pipeline's ability to provide further confirmation for \txrr{gravitationally} lensed quasar candidates using existing photometric data.  \txrr{Followup spectroscopic observations of these candidates may gleam light onto whether these systems are truly lensed quasars, and thus would give an estimate of our result's accuracy.}




\section{Observation} \label{sec:observation}
\begin{figure}
\begin{center}
\includegraphics[width=175mm]{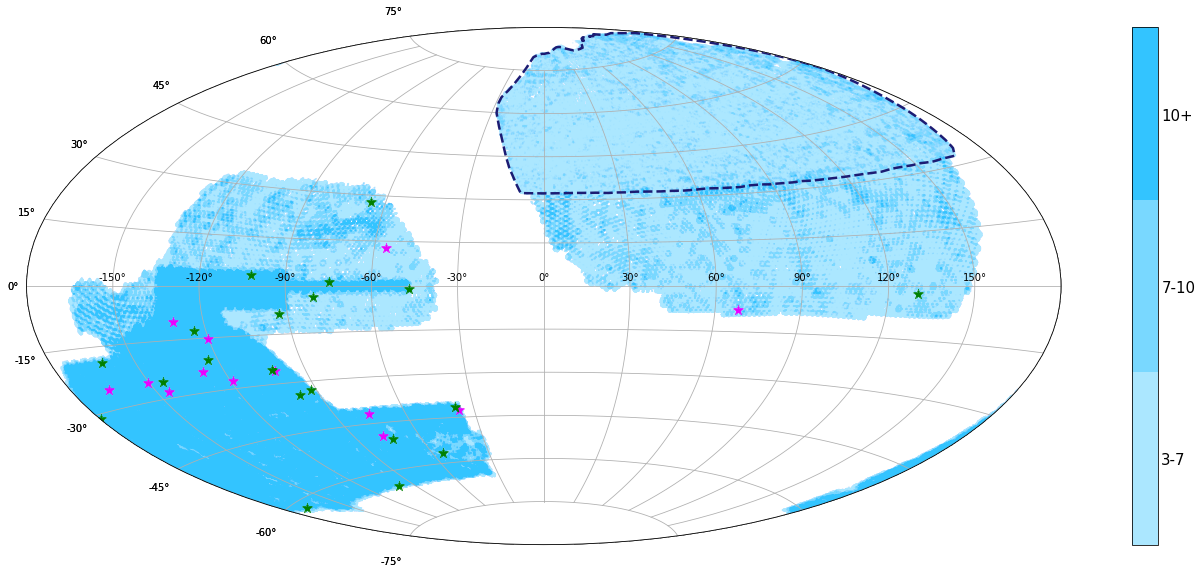}
\caption{\txrd{The Legacy Imaging Surveys DR9 footprint, color-coded by the \ipz~band observation depth.  The footprint of the Mayall $z$~band Legacy Survey and Beijing--Arizona Sky Survey is outlined in blue dashed line, whose observations were excluded from our search.  The region with 10+ exposures (bottom left) highlights the DES footprint.  The green stars denote the variable lensed quasar candidates identified by our first search (targeted search on strong lens candidates; see \S\ref{res_targeted_search}).  The purple stars denote the variable lensed quasar candidates identified by our second search (targeted search on lensed quasar candidates; see \S\ref{conflq}). }}\label{footprint}
\end{center}
\end{figure}

\begin{figure}
\begin{center}
\includegraphics[width=110mm]{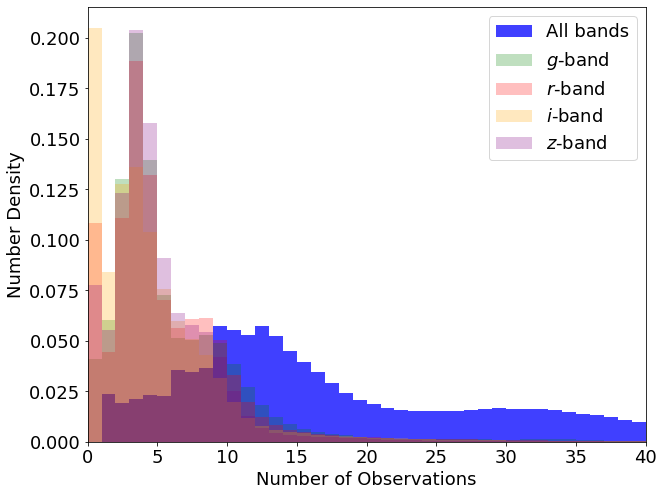}
\caption{\txrr{A histogram of the number of observations for $10^8$ random points sampled from the DR10 footprint.  As the number of Y-band observations are substantially less than other bands, they are not included in this plot.  }}\label{observations_hist}
\end{center}
\end{figure}

We use the same observations \txp{as} in Paper~I.  We utilize the DESI Legacy Imaging Surveys data releases (DR) 9 and 10 datasets\footnote{\url{https://www.legacysurvey.org}}, specifically using observations by the Dark Energy Camera \citep[DECam;][]{flaugher2015} on the 4 meter Blanco telescope.  These observations cover $\sim 9,000$ deg$^2$ of the extragalactic sky below Dec $\sim 32$\textdegree, across DECam \ipg, \ipr, \ipi, \ipz, and Y bands. 
 \txrd{See Figure~\ref{footprint} for a depiction of the Legacy Imaging Surveys DECam footprint.}  \txrr{On average, a random point within the Legacy Imaging Surveys footprint receives $15^{+18}_{-8}$ observations over about five years\footnote{derived from \url{https://www.legacysurvey.org/dr10/files/\#random-catalogs-randoms}} (see Figure~\ref{observations_hist}; it should be noted that oftentimes a single night captures multiple bands of a particular pointing, and so the average total number of nights observed is lower than this).}  

\txg{In this paper, \txrr{firstly} we conduct a targeted search \txrr{on strong lens candidates, to find new lensed quasar candidates}.  As such,} we compile a database of 5807 strong lenses and candidates, with the majority \txrr{(about 93\%)} from \citet{huang2020}, \citet{huang2021}, and \citeauthor{storfer2022} (\citeyear{storfer2022}, hereafter SLS~I, SLS~II, and SLS~III, respectively; shorthand for ``strong lenses search'')\txrr{.  The remaining 7\% are strong lens candidates selected} from \citet{moustakas2012}, \citet{carrasco2017}, \citet{diehl2017}, \citet{jacobs2017}, \citet{pourrahmani2018}, \citet{sonnenfeld2018}, \citet{wong2018}, \citet{jacobs2019}, and \citet{sonnenfeld2020}.  \txrr{This 5807 systems sample is the same set of strong lens targets as in Paper~I. }

\txrr{Secondly, }we target the 655 grade A and B lensed quasar candidates identified by D22\footnote{\url{https://sites.google.com/usfca.edu/neuralens/publications/lensed-qso-candidates-dawes-2022}} and H23\footnote{\url{https://github.com/EigenHermit/lensed_qso_cand_catalogue_He-22/blob/main/Candidates/Catalogues/H22-details.csv}}.  In these two papers, the candidates were discovered and graded by the morphology and location of the lensed images \txrr{in the DESI Legacy Imaging Surveys}.  In this paper, we further examine these candidates using individual exposures in individual bands observed at different times and identify those with significant variability for all posited lensed images.

\section{Methodology} \label{sec:methodology}

We take a targeted approach to search for lensed transients\txg{, as laid out} in \S3 of Paper~I.  \txg{To describe succinctly}, we apply two image subtraction algorithms \citep{bramich2008, hu2021} between each exposure and the median coadd across all exposures for each band, for every strong lensing system or candidate.  \txrr{After generating two sets of difference images (for each \txrd{difference imaging} algorithm) of every exposure, we use a Python implementation (SEP; \citealp{sep}) of the source extraction algorithm from \citet{bertin1996} to detect any potential sources in all difference images, with thresholds ranging from 1.0 to 2.5$\sigma$ in 0.25 increments as determined by SEP (with detections $>$ 2.5$\sigma$ treated as the same detection level as 2.5$\sigma$).  Henceforth, we denote a detection in a single difference algorithm as a ``sub-detection" (a given transient event in a single exposure can generate two sub-detections, by being detected in difference images produced by both image subtraction algorithms).  All sub-detections (from both subtraction algorithms, across all bands and across all exposures) are then grouped together spatially and temporally.  These groupings contain all sub-detections that are spatially and temporally close to other sub-detections in the group.  If a given group has less than three sub-detections, the pipeline disregards them.  If a group has three or more difference image sub-detections, it is labelled as a possible transient detection.  For more specificity and clarification on how we use \txrd{difference imaging} algorithms in our pipeline, we again refer the reader to \S3 of Paper~I.}

Since we search for lensed quasars in this \txrr{companion} paper \txrr{(in contrast to lensed supernovae in Paper~I)}, we make the following changes to the selection criteria \txrr{and apply them to our search on the strong lens candidates sample}: 
\begin{enumerate}
  \item \emph{Location} - The posited lensed quasar should be located at the core of the posited \txrr{lensed source galaxy} image (if visible)\txrr{, for all lensed images}.  
  \item \emph{Color} - The posited lensed quasar should appear blue or white in the coadded RGB image \txrr{(i.e., there should be significantly higher flux in the \ipg~band than in the redder bands)}.  \txo{The lensed quasar image is usually brighter than their host galaxy, though this is not always the case.}  \txr{The coadded RGB images are made with the Legacy Surveys\txrr{/SDSS} RGB image generation code\footnote{\url{https://github.com/legacysurvey/imagine/blob/main/map/views.py\#L5364}}, \txrr{which weighs and assigns} the \ipg, \ipr, \ipi, and \ipz bands \txrr{to an associated RGB channel}.}  While there are cases of red quasars \citep[e.g.,][]{fawcett2020}, \txb{these are rare}.  All posited lensed quasar images should have similar colors.  
  \item \emph{Variability} - We expect the lensed quasar to exhibit a significant level of variability between exposures.  Thus, all posited lensed quasar images should display over and under subtraction between exposures. 
 \txrr{The variability should also be detected by our pipeline, and visually verified to be possible fluctuations caused by a lensed quasar (as opposed to artifacts or cosmic rays).}
\end{enumerate}

\txrr{Note that these criteria were visually assessed by one person, and therefore is subject to human error.}
Of these selected lensed quasar candidates, we then compute the point spread function (PSF) photometry of each posited images and present their light curves.  A PSF must be resolved using SEP at the lensed image location; if not, we omit that exposure data from a given lensed image's light curve.  We filter out photometric data with a signal-to-noise ratio $< 5$.  From this, we identify 20 systems (\txrd{13} new lensed quasar candidates, and \txrd{seven} previously identified lensed quasars).  This methodology differs from other lensed quasar searches in the Legacy Imaging Surveys (e.g., D22 and H23), which solely rely on morphology (PSF-like), location, and color of identified objects.  While taking these factor into account, our pipeline also relies on the \txb{photometric} variability of the posited quasar images.  As we will show below \txo{(see \S\ref{newlq})}, we therefore are able to identify \txrr{lensed quasar} candidates previously overlooked by past searches. 

\txo{In addition to the targeted search, we also apply our pipeline to 655 grade A and B lensed quasar candidates identified by D22 and H23, and} \txr{select candidates that show high variability for} each posited lensed image.  \txo{The purpose of this second search is to demonstrate the potential of utilizing targeted variability for further confirmation of lensed quasar candidates found using traditional lensed quasar search methods}\txb{; thus this is not meant to be an exhaustive search.}  \txr{This is done, for each observed band, by setting a threshold on their standard deviation from their mean magnitude\txg{:}}

\begin{equation}\label{eq:sd_equation}
    \overline{\sigma} = \frac{1}{N} \sum_b \sqrt{N_b \sum_{m_b} (m_b - \mu_b)^2}, 
\end{equation}
\txr{where} $\overline{\sigma}$ is the average magnitude standard deviation across all bands (weighted by the number of \txr{observation passes} of each band), $m_b$ is the measured PSF magnitudes for band $b$, $\mu_b$ is the mean PSF magnitude for band $b$, $N_b$ is the total number of PSF photometry observations for band $b$, and $N$ is the total number of epochs across all bands, or $\sum_b N_b$.  We calculate $\overline{\sigma}$ for each image on every system, and only retain systems where all posited \txr{lensed} images have a higher $\overline{\sigma}$ than twice the average across our 20 lensed quasars and lensed quasar candidates from the first pipeline run \txr{in \S\ref{newlq}}.  \txo{We use this fiducial $\overline{\sigma}$ threshold to limit our selected candidates to ones that exhibit high variability, \txb{so} as \txg{to} demonstrate our pipeline's ability to provide variability confirmation \txg{for} already-identified lensed quasar candidates.}

\section{Results} \label{sec:results}
\txrr{In Table~\ref{table:filter}, we show by how much the selection criteria filters our initial sample for the strong lens candidates and lensed quasar candidates samples. As outlined in \S\ref{sec:methodology}, for the strong lens candidates sample, all criteria were judged via visual inspection; for the lensed quasar candidates sample, we only apply an automated $\overline{\sigma}$ threshold set by the average of the 20 systems in the first search (as the initial sample is already filtered by a location and color criteria in D22 and H23).  By comparing the ``Variability Criterion" between the two samples in Table~\ref{table:filter}, we see that there is a large difference in the percentage filtered out by the criterion.  This is caused by three main reasons.  Firstly, we set the fiducial $\overline{\sigma}$ threshold for the second sample to be the average of the 20 candidates found by the first, meaning that by definition, if applied to the first sample, only about half of the 20 candidates fulfill this threshold.  Secondly, many of the systems in the lensed quasar candidates sample are cases where the quasar images are very close together, but are distinguished as two separate PSF-like objects by the Tractor \citep[a forward modeling source extraction algorithm;][]{lang2016}.  However, the \txrd{difference imaging} algorithms employed in this work struggle to successfully resolve these situations, and so these systems are filtered out as well.  Lastly, an inherent variability bias \citep[where quadruply-lensed systems/quads are expected to have higher variability than doubly-lensed systems/doubles;][]{lemon2020} works against the lensed quasar candidates sample, as a large majority of the D22 and H23 sample are doubly lensed quasar candidates.}

\begin{deluxetable}{lccc}[H]
\tablecaption{Selection Criteria Filters\label{table:filter}}
\tablehead{
     &
    Initial Size &
    Location and Color Criteria &
    Variability Criterion}
\startdata
Strong Lens Candidates Sample & 5807 & 123 (2.12\%) & 20 (16.26\%)\\
Lensed Quasar Candidates Sample & 655 & - & 13 (1.98\%)\\
\enddata
\tablecomments{\txrr{Number of candidate systems that remain after imposing specific criteria on the initial sample.  The percentages shown are relative to the number of samples in the previous step.  }}
\end{deluxetable}

We divide our results into two sections: identified lensed quasar candidates \txo{from a targeted search on strong-lens candidates} (\S\ref{res_targeted_search}) and previously identified lensed quasars candidates \txo{for which} we provide additional variability confirmation (\S\ref{conflq}). See Table~\ref{table} for an overview of all 33 systems.  \txrd{Of these, 27 systems lie within the DES footprint (see Figure~\ref{footprint}).} 
\begin{center}
\begin{deluxetable}{ccccccc}
\tablecaption{Lensed Quasar Candidates\label{table}}
\tablehead{
    System Name &
    \multicolumn{1}{p{1.3cm}}{\centering RA\\{[}deg{]}} &
    \multicolumn{1}{p{1.3cm}}{\centering Dec\\{[}deg{]}} &
    \multicolumn{1}{p{5.5cm}}{\centering Prior Discovery\\Status}&
    \multicolumn{1}{p{1.9cm}}{\centering Number of\\Observations}&
    \multicolumn{1}{p{1.3cm}}{\centering DECam\\\ipr~band\\{[}mag{]}}&
    \multicolumn{1}{p{1.3cm}}{\centering $\langle\overline{\sigma}\rangle$\\{[}mag{]}}
    }
\decimalcolnumbers 
\startdata
\hyperref[p0]{WG021416.37-210535.3}     &    33.5681 &  $-$21.0931 &     Confirmed LQ \citep{agnello2018_single} & 18/17/15/17 & 19.23 & 0.19 \\
\hyperref[p1]{DESJ0340$-$2545}    &   55.0363 &  $-$25.7609 &    Confirmed LQ \citep{lemon2020} &             41/39/37 & 20.16 & 0.18 \\
\hyperref[p2]{SDSS J2222+2745}          &   335.5357 &  $+$27.7599 &     Confirmed LQ \citep{dahle2013}  &               15/15/14 & 20.95 & 0.38 \\
\hyperref[p3]{KIDSJ0008$-$3237}    &    2.0670 &  $-$32.6207 &    LQ Candidate \citep{khramtsov2019} &           34/29/26 & 19.41 & 0.21 \\
\hyperref[p4]{J0011-0845}               &    2.8344 &   $-$8.7643 &    Confirmed LQ \citep{lemon2018} &               5/14 & 20.29 & 0.10 \\
\hyperref[p5]{J0030-3358}    &    7.6741 &  $-$33.9765 &     Confirmed LQ \citep{lemon2023} &            29/22 & 21.01 & 0.14 \\
\hyperref[p6]{HSCJ091843.38-022007.3}   &  139.6802 &  $-$2.3354 &  Confirmed LQ \citep{sonnenfeld2020}  &              26/21 & 19.98 & 0.14 \\
\hline
\hyperref[n1]{DESI-343.6329-50.4884}    &  343.6329 &  $-$50.4884 &     SL Candidate (SLS III) &            46/46/36 & 20.23 & 0.26 \\
\hyperref[n0]{DESI-350.3458-03.5082}    &  350.3458 &  $-$3.5082 &     SL Candidate (SLS III) & 15/25/8 & 21.50 & 0.39 \\
\hyperref[n3]{DESI-344.8782+01.2913}    &  344.8782 &   $+$1.2913 &     SL Candidate (SLS III) &         35/22 & 20.77 & 0.19 \\
\hyperref[n16]{DESI-009.5736-64.5942}   &    9.5736 &  $-$64.5942 &     SL Candidate (SLS III) &          37/38 & 22.56 & 0.17 \\
\hyperref[n6]{DESI-011.9124+03.3756}    &   11.9124 &   $+$3.3756 &     SL Candidate (SLS III) &           14/24 & 21.14 & 0.24 \\
\hyperref[n9]{DESI-033.9735-12.6841}    &   33.9735 &  $-$12.6841 &     SL Candidate (SLS I) &            26/27 & 21.58 & 0.23 \\
\hyperref[n5]{DESI-071.8595-18.9863}    &   71.8595 &  $-$18.9863 &     SL Candidate (SLS III) &           25/31 & 21.68 & 0.20 \\
\hyperref[n10]{DESI-080.2447-61.8266}   &   80.2447 &  $-$61.8266 &     SL Candidate (SLS III) &           17/18 & 20.13 & 0.13 \\
\hyperref[n15]{DESI-089.5700-30.9485}   &   89.5700 &  $-$30.9485 &     SL Candidate (SLS II) &            34/34 & 22.06 & 0.14 \\
\hyperref[n12]{DESI-308.0432-41.5946}   &  308.0432 &  $-$41.5946 &     SL Candidate (SLS III) &          35/33 & 21.03 & 0.14 \\
\hyperref[n11]{DESI-316.8445-00.9920}   &  316.8445 &   $-$0.9920 &     SL Candidate (SLS II) &            50/47 & 21.07 & 0.14 \\
\hyperref[n13]{DESI-324.5771-56.6459}   &  324.5771 &  $-$56.6459 &     SL Candidate (SLS III) &            44/26 & 21.22 & 0.24 \\
\hyperref[n14]{DESI-012.3074-25.6429}   &   12.3074 &  $-$25.6429 &     SL Candidate (SLS III) &             27 & 19.61 & 0.24 \\
\hline
\hyperref[c0]{J0343-2828}    &   55.7976 &   $-$28.4777 &     Disconfirmed LQ \citep{lemon2023} &             34/34/34 & 20.96 & 0.24 \\
\hyperref[c6]{DESI-011.5839-26.1241}    &   11.5839 &  $-$26.1241 &     LQ Candidate (D22, H23) &           32/41 & 20.74 & 0.27 \\
\hyperref[c1]{DESI-029.1039-27.8562}    &   29.1039 &  $-$27.8562 &     LQ Candidate (D22, H23) &             30/31 & 19.75 & 0.35 \\
\hyperref[c2]{DESI-030.0872-15.1609}    &   30.0872 &  $-$15.1609 &     LQ Candidate (D22) &            36/35 & 19.12 & 0.32 \\
\hyperref[c3]{DESI-038.0655-24.4942}    &   38.0655 &  $-$24.4942 &     LQ Candidate (D22, H23) &          31/21 & 18.39 & 0.25 \\
\hyperref[c10]{DESI-040.6886-10.0492}   &   40.6886 &  $-$10.0492 &     LQ Candidate (D22, H23) &           34/33 & 19.39 & 0.22 \\
\hyperref[c4]{DESI-060.4504-25.2439}    &   60.4504 &  $-$25.2439 &     LQ Candidate (D22) &            29/24 & 18.74 & 0.28 \\
\hyperref[c7]{DESI-076.5562-25.5135}    &   76.5562 &  $-$25.5135 &     LQ Candidate (D22, H23) &            33/34 & 19.87 & 0.17 \\
\hyperref[c5]{DESI-202.0009-07.8030}    &  202.0009 &    $-$7.8030 &     LQ Candidate (D22, H23) &        34/24 & 19.75 & 0.23 \\
\hyperref[c8]{DESI-306.7073-42.4719}    &  306.7073 &  $-$42.4719 &     LQ Candidate (D22) &             44/46 & 19.48 & 0.33 \\
\hyperref[c11]{DESI-325.8843$+$12.5745} &  325.8843 &  $+$12.5745 &     LQ Candidate (D22, H23) &           33/36 & 18.52 & 0.36 \\
\hyperref[c12]{DESI-345.6309-41.9157}   &  345.6309 &  $-$41.9157 &     LQ Candidate (D22, H23) &          36/23 & 19.34 & 0.33 \\
\hyperref[c9]{DESI-346.9550-49.2117}    &  346.9550 &  $-$49.2117 &     LQ Candidate (D22) &          44/43 & 20.38 & 0.40 \\
\enddata
\tablecomments{Summary table for our lensed quasar systems and candidates.  \txr{Column 1: system name. Column 2 and 3: the RA and Dec of each system, respectively.}  Column~4: the discovery status of each system \txrr{and the initial discovery paper}, with SL and LQ referring to ``strong lens'' and ``lensed quasar'' respectively.  \txrr{Column~5: the number of resolved PSF photometry observations across all bands, given for each resolved image.}  \txrd{Column~6: time-averaged DECam \ipr~band magnitude of the brightest lensed image.}  Column~7: average magnitude standard deviation \txb{(Equation~\ref{eq:sd_equation}), averaged across all resolved images and bands, as indicated by the angular brackets}.  The horizontal lines separate the table into three sections, corresponding to the differing discovery statuses and sections in this paper (\S\ref{prevlq}, \S\ref{newlq}, and \S\ref{conflq}, respectively).  \txrd{The sections are then sorted by the system's number of observed lensed images (with exception to DESI-344.8782+01.2913, which is thought to be quadruply lensed), then by their RA.}} 
\end{deluxetable}
\end{center}

\txrr{Of the objects identified by the Tractor, the photometric redshifts are estimated in \citet{zhou2020}}.  \txr{\txo{If the Tractor is unable to deblend the lens galaxy from the lensed image(s) for a given system, then no lens redshift \txb{is} provided.}  As many of the photometric redshifts for the lensed quasar images \txo{are not \txg{very} reliable} (even for confirmed lensed quasar systems), \txo{these are not presented in this paper}.}  While we use two different \txrd{difference imaging} algorithms in our detection pipeline, all difference images shown in this paper are \txr{from the \txg{\txrd{difference imaging} algorithm} of} \citet{hu2021} alone (see \S3.2 of Paper~I for our reasoning).  \txrr{All light curve data presented in this paper are available here\footnote{\url{https://portal.nersc.gov/cfs/deepsrch/lquasars/}}.} 

\txo{In this discovery paper, we observe broadly coherent light curve behavior in one or more bands, for the vast majority of the systems.}  The \txo{lack of coherence} in a minority of our candidate's light curves is likely due to \txo{one or more of these factors:} irregularities in the exposures (oversaturation, cosmic rays, etc.), poor signal to noise\txr{, and/or in the case where the lensed image is \txo{very close} to the lens or other lensed images, the lens or additional lensed galaxy light \txg{causing} inconsistencies in calculating photometry}.  \txo{Followup monitoring of} these systems \txo{is necessary for full confirmation}. 

\newpage

\subsection{Lensed Quasar Candidates from Targeted Search}\label{res_targeted_search}
\begin{figure}
\begin{center}
\includegraphics[width=165.5mm]{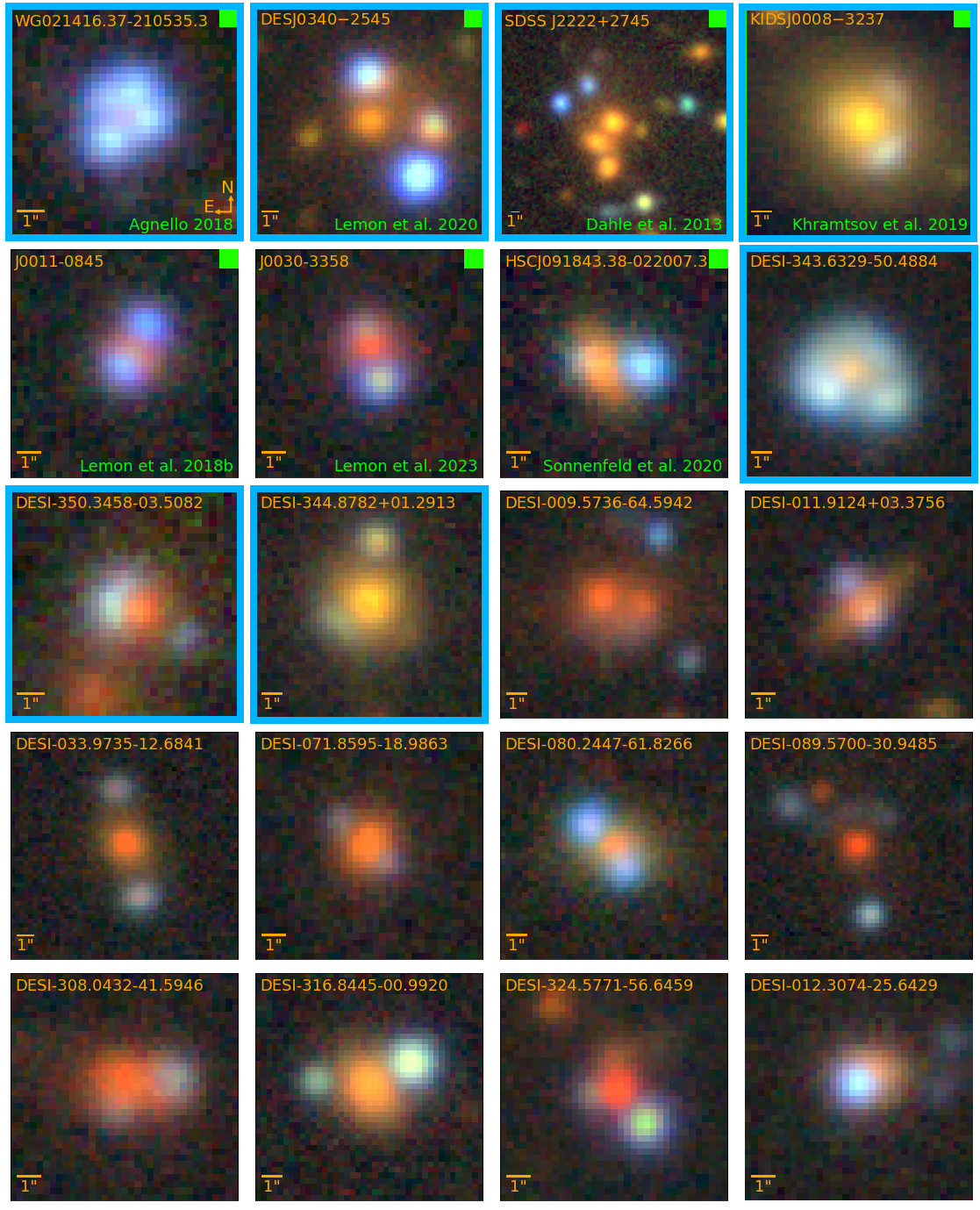}
\caption{\txrr{Our 20 identified lensed quasar candidates.  The first seven are previously identified lensed quasars (with a green square at the top right corner and green text in the bottom right citing their discovery paper\txrd{; all but KIDSJ0008$-$3237 have been spectroscopically confirmed}), and the following 16 are lensed quasar candidates newly identified by our pipeline.}  \txr{Quadruply-lensed systems are with \txb{a} blue border (in the case of SDSS~J2222+2745, sextuply lensed)}.  \txrr{The RGB image generation code used is discussed in \S\ref{sec:methodology}.}}\label{ws_montage}
\end{center}
\end{figure}

\txo{After applying our transient detection pipeline \txr{to the} 5807 strong lensing candidates, we find \txrd{seven previously identified lensed quasar candidates (six of which were already spectroscopically confirmed) as well as 13 new lensed quasar candidates (with ten doubles and three quads).}  Figure~\ref{ws_montage} presents all 20 of these systems.  \txrr{The RGB image generation code used is discussed in \S\ref{sec:methodology}}.}

\subsubsection{Previously Identified Lensed Quasars}\label{prevlq}
\txrd{Seven} systems, identified by our pipeline \txo{in our targeted search} as lensed quasar candidates, were found \txb{(through a followup literature search)} to have been previously discovered lensed quasar \txrd{candidates.  All but one of these seven systems (KIDSJ0008$-$3237) have also been spectroscopically confirmed}.  These systems are shown in the \txrd{first seven panels of} Figure~\ref{ws_montage}.  In this section, we present the DECam PSF photometry light curves for one lensed quasar (WG021416.37-210535.3), with the others presented in Appendix~\ref{append_prev}.

\begin{figure}[H]
\begin{center}
\includegraphics[width=161mm]{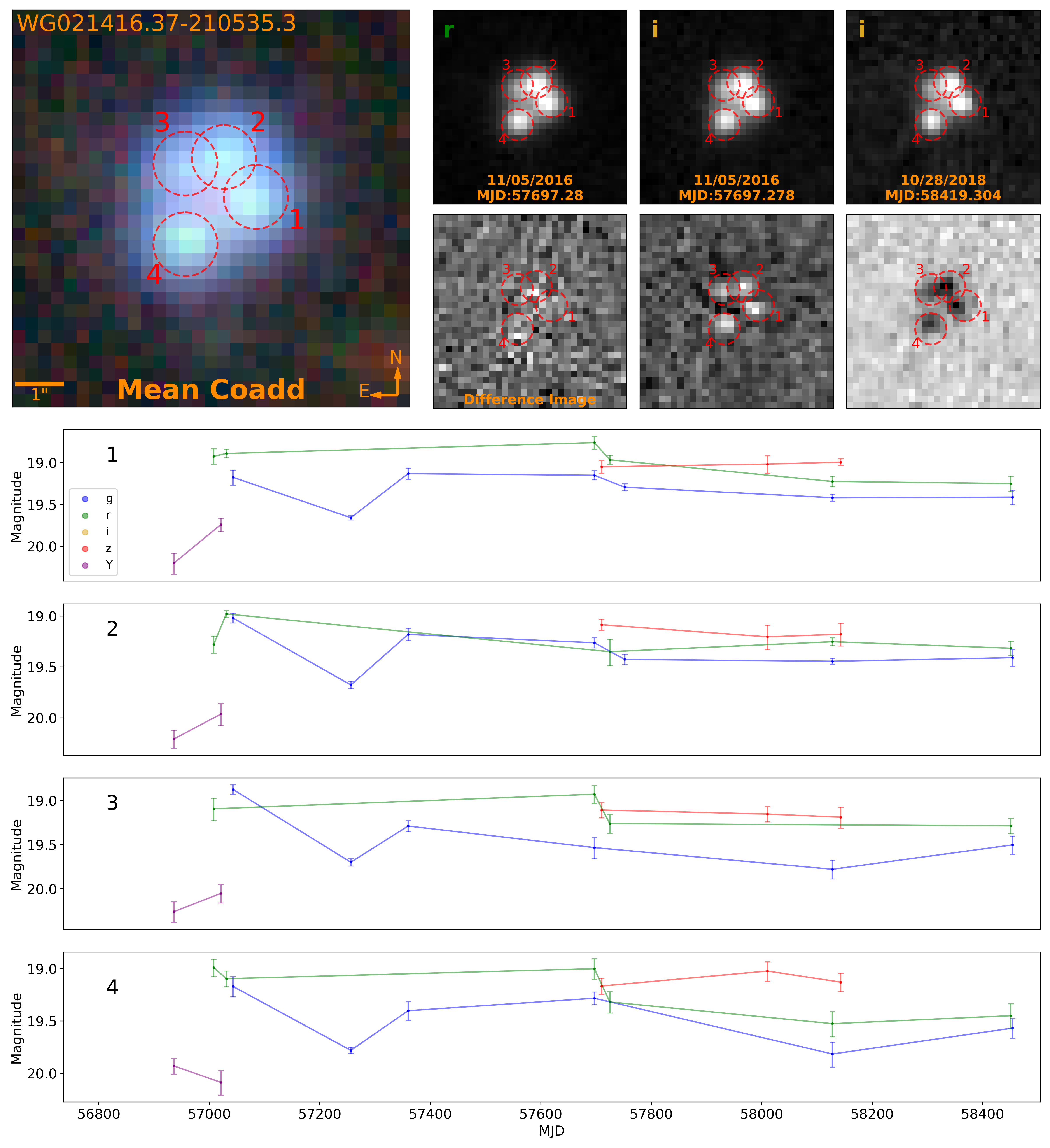}
\caption{\txr{Lensed quasar} WG021416.37-210535.3. For this and subsequent figures\txr{, we use the following} formatting: the top left image is the coadded RGB image (utilizing \ipg, \ipr, \ipi, and \ipz~bands), with the posited resolved lensed quasar images labelled.  \txrr{The RGB image generation code used is discussed in \S\ref{sec:methodology}}.  If available, the redshift of the lens galaxy ($z_L$) is given below the system name; photometric measurements are given uncertainties whereas spectroscopic measurements are shown without uncertainties.  The plots to the right of the RGB image show select single-epoch exposures (top) from various bands that exhibit variability and their corresponding difference images (bottom).  The bottom rows are the PSF photometry light curves of the posited images.  
}\label{454_quasars}
\end{center}
\end{figure}

\paragraph{WG021416.37-210535.3}\label{p0}
This system is a quadruply-lensed quasar (Figure~\ref{454_quasars}), discovered and spectroscopically confirmed by \citet{agnello2018_single}.  \txo{This system was independently identified in SLS~III as a strong lens candidate, and therefore it was included in our search.  We subsequently found that this is a known system.}  \txr{Though this system was previously discovered, \txo{here} we show new \txo{photometry} for this system.}  

\subsubsection{New Lensed Quasar Candidates}\label{newlq}
In this \txo{sub}section, we present the DECam PSF photometry light curves for one new lensed quasar candidate (DESI-350.3458-03.5082), with the other candidates presented in Appendix~\ref{append_new}.

\begin{figure}
\begin{center}
\includegraphics[width=120mm]{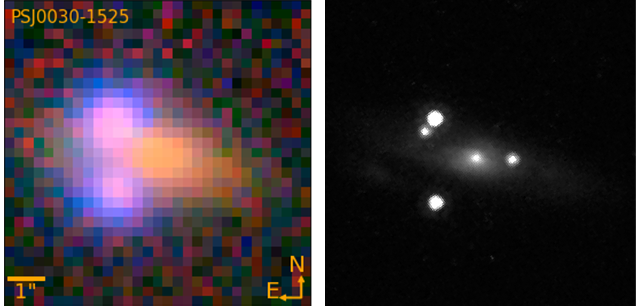}
\caption{Side-by-side imaging of PSJ0030-1525 from the Legacy Imaging Surveys and the \textit{Hubble Space Telescope} (\textit{HST}).  The Legacy Surveys RGB image (utilizing the \ipg, \ipr, \ipi, and \ipz band exposures\txrr{; see \S\ref{sec:methodology}}) is shown on the left, with the \textit{HST} F814W imaging (Prop ID: 15652; PI: T. Treu) on the right.  While \txr{there are only two resolved} lensed quasar images from the Legacy Imaging Surveys RGB image, \textit{HST} imaging reveals two additional images. 
}\label{example_hst}
\end{center}
\end{figure}

We also note that while we present \txrd{three} new quad and \txrd{ten} new double candidates, it is anticipated that one or \txr{more} of these doubles candidates are actually quadruply-imaged, but only two images can be source extracted from the DECam data in the Legacy Imaging Surveys.  See Figure~\ref{example_hst} for an example such a quadruply-lensed quasar system.

\begin{figure}[H]
\begin{center}
\includegraphics[width=161mm]{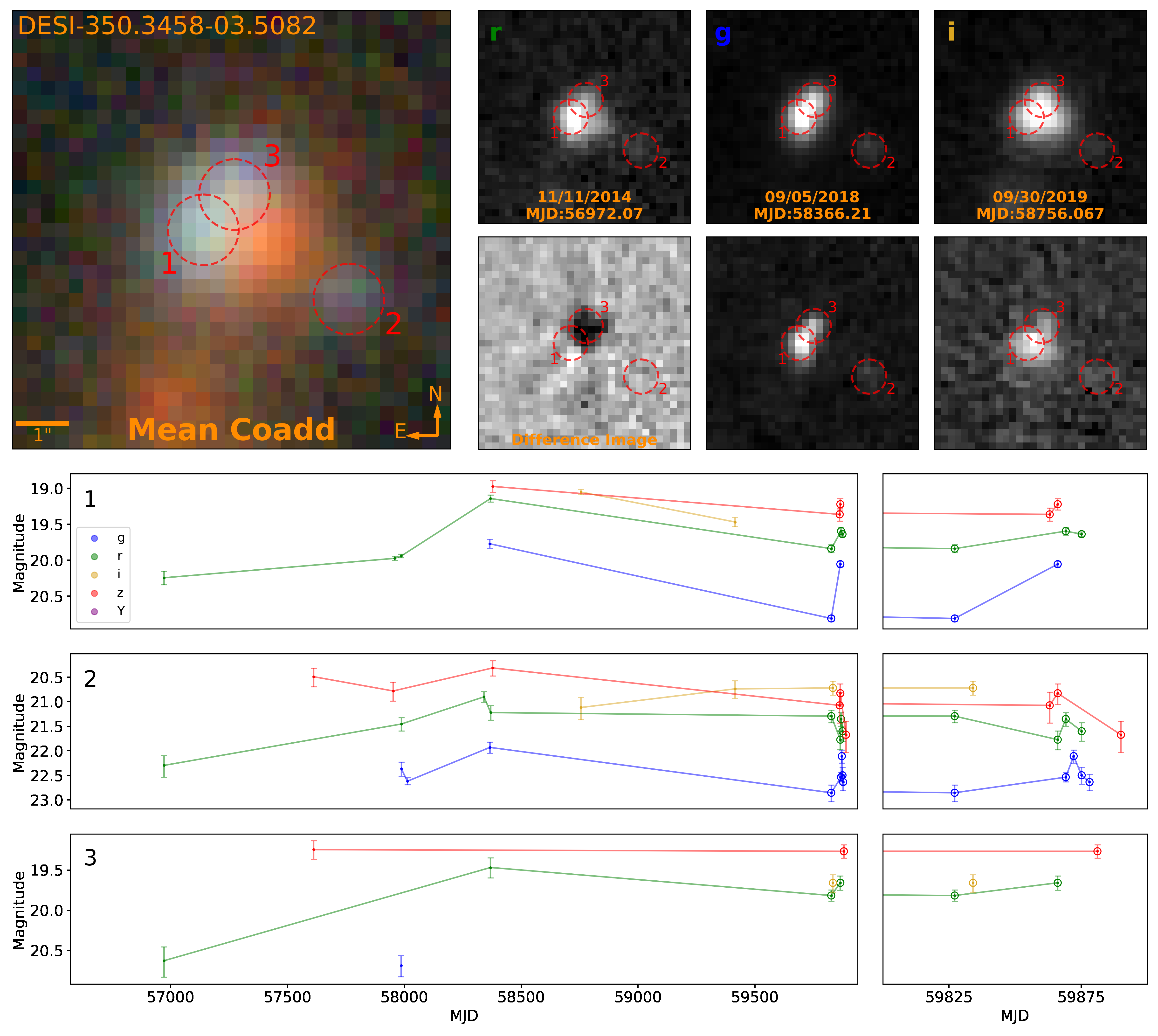}
\caption{\txr{New lensed quasar candidate} DESI-350.3458-03.5082 (see caption of Figure~\ref{454_quasars} for the full description of each subplot).  This figure also displays \txo{ zoomed-in panels of the light curves} \txrr{to the right of the light curves}, in order to highlight \txr{recent follow-up observation} data from 2022B-297190 (in open data points).}\label{435_quasars}
\end{center}
\end{figure}

\newpage
\paragraph{DESI-350.3458-03.5082}\label{n0}
This system is a quadruply-lensed quasar candidate (Figure~\ref{435_quasars}), with three lensed images comprising the Eastward arc and the counterimage located West of the lens galaxy.  DESI-350.3458-03.5082 was initially identified as a grade B strong lensing candidate in SLS~III.  In addition to the variability detected in the \txo{Legacy Imaging Surveys} presented in this paper, \txr{Pan-STARRS data provide additional evidence for variability} (C. Storfer, private communication).  \txr{Without having done variability analysis, this system would have been mistaken as simply a \txo{galaxy-galaxy lensing} system that displays a typical arc-counterarc configuration.}   For this system, recent \txr{follow-up} DECam \txr{observations} in the \ipg, \ipr, \ipi, and \ipz bands \txr{were analyzed} (\txr{NOAO Prop ID: 2022B-297190; co-PIs: A. Palmese and L. Wang),} \txo{where the variability further} supports the identity of this system being a strongly lensed quasar \txo{system}.

\newpage
\subsection{Variability Confirmation of Previously Discovered Lensed Quasar Candidates}\label{conflq}

\begin{figure}[H]
\begin{center}
\includegraphics[width=165.5mm]{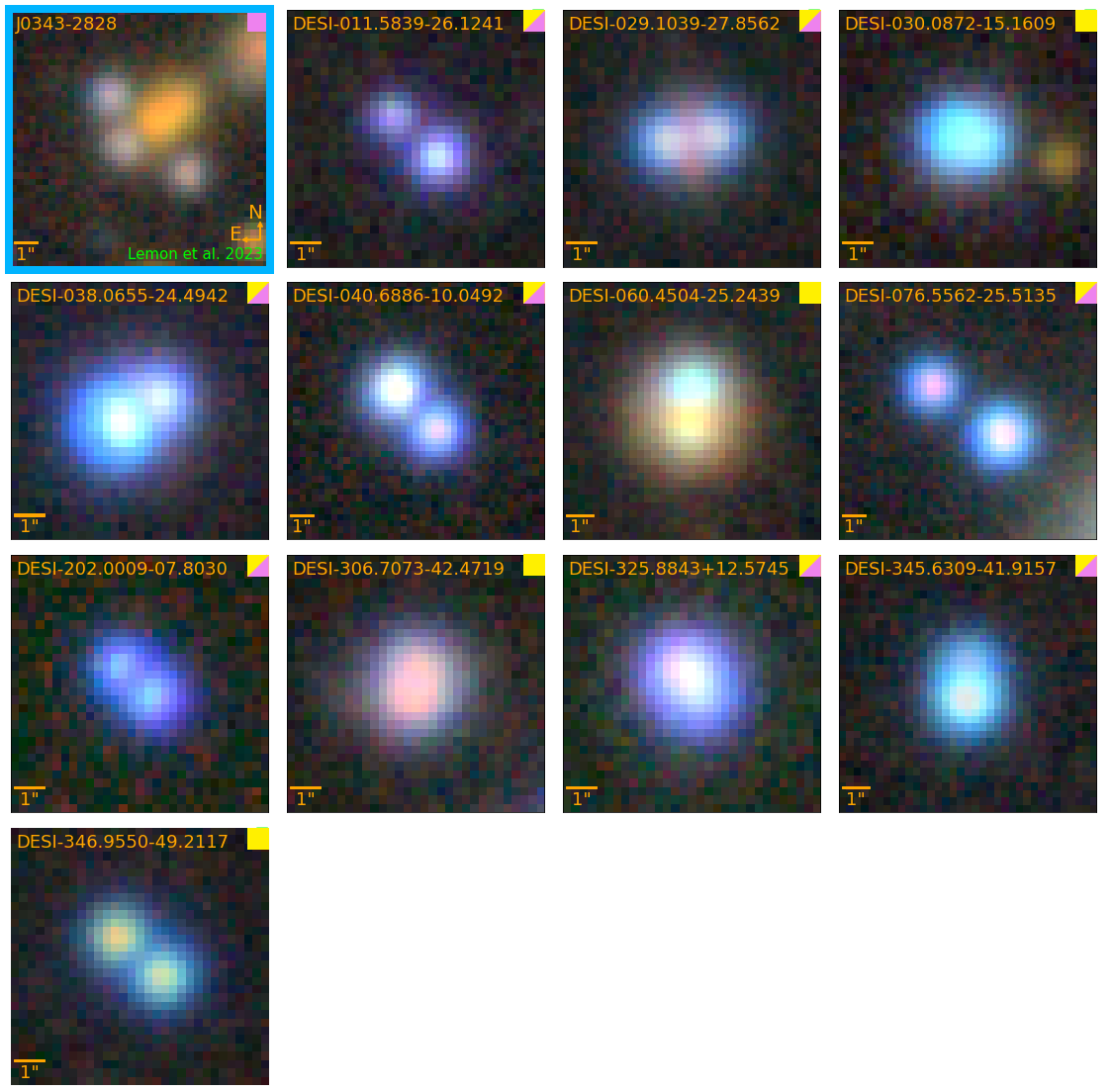}
\caption{Our 13 previously identified lensed quasar candidates (one \txr{quad, with a blue border}), \txr{for which} our pipeline detects significant variability within all posited lensed quasar images.  Systems with a \txrd{yellow and/or magenta} square on the upper right of the image are systems identified as lensed quasar candidates by D22 \txr{and/or H23, respectively}.  \txrd{In the case of J0343-2828 (top left), the system was identified and disconfirmed as a lensed quasar candidate in \citet{lemon2023}. }
}\label{cd_montage}
\end{center}
\end{figure}

The following 13 systems in Figure~\ref{cd_montage} are lensed quasar candidates discovered by D22 and/or H23, where our pipeline has identified significant variability within the DESI Legacy Imaging Surveys.  In this section, we present the DECam PSF photometry light curves for one candidate (J0343-2828), with the other candidates presented in Appendix~\ref{append_conf}.  In D22 and H23, the 655 grade A and B lensed quasar candidates were selected using location and color based algorithms to identify \txo{close} pairs/groups of \txo{PSF-like objects}.  Our paper provides variability confirmation for 13 of these systems, one of which is a posited quadruply-lensed quasar.  Our findings \txb{do} not imply that all other candidates from D22 and H23 not identified by our pipeline are not lensed quasars.  Variability detection depends on many factors, such as the activity of the quasar, the \txb{timing}/cadence of exposures, and the time delays.  \txr{Especially, the Legacy Imaging Surveys is not designed to detect transients; as shown in \S\ref{sec:observation}, the average cadence \txrr{is approximately 100} days.} 
 As our intention is to demonstrate the viability of our methodology (detecting variability in survey data), we select the candidates where our pipeline most confidently identifies variability. 

\begin{figure}[H]
\begin{center}
\includegraphics[width=161mm]{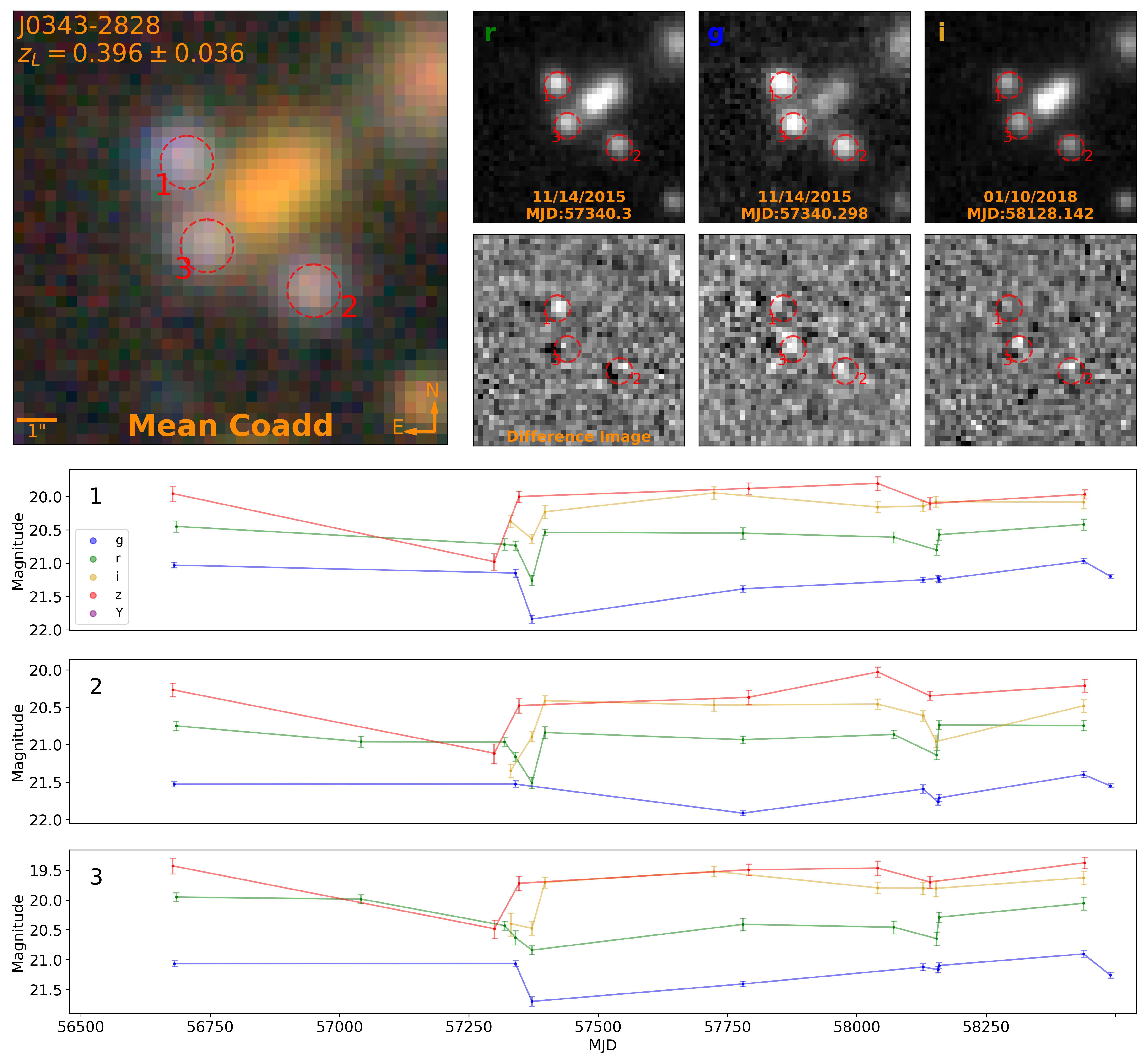}
\caption{\txr{Lensed quasar candidate} J0343-2828 (see caption of Figure~\ref{454_quasars} for the full description of each subplot).   }\label{476_quasars}
\end{center}
\end{figure}

\begin{figure}[H]
\begin{center}
\includegraphics[width=161mm]{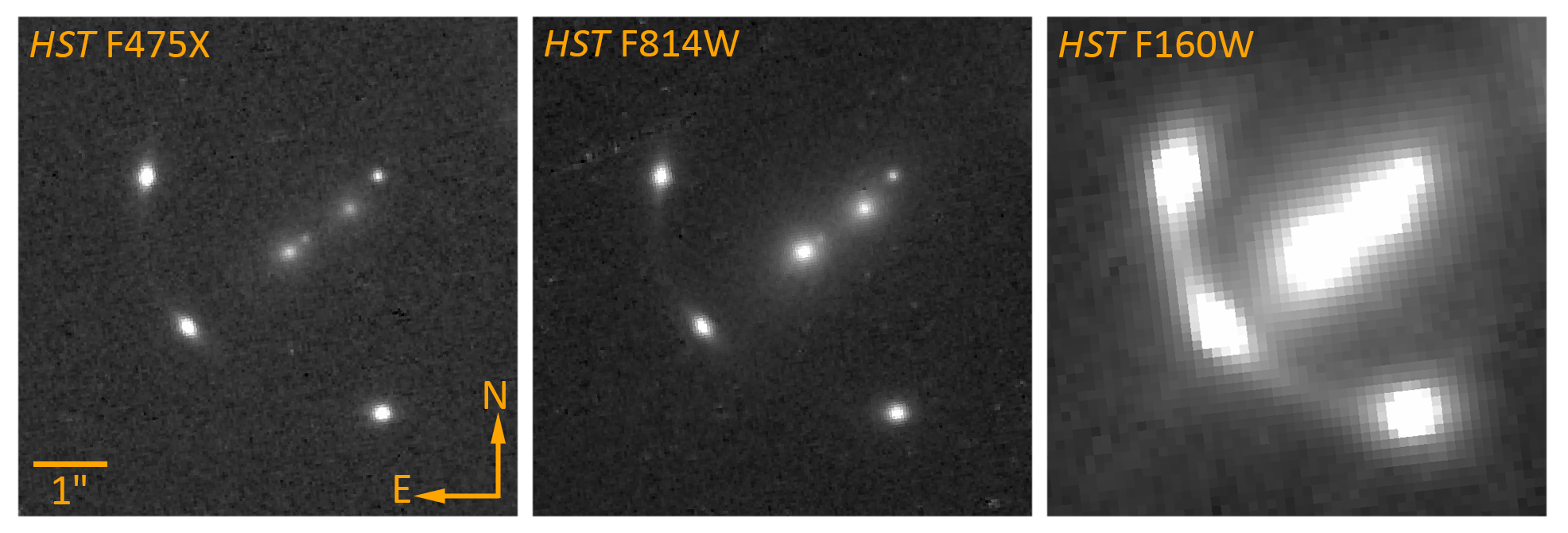}
\caption{\textit{HST} imaging of J0343-2828, in the F475X, F814W, and F160W bands (Prop ID: 15652; PI: T. Treu).   }\label{476_quasars_hst}
\end{center}
\end{figure}

\newpage
\paragraph{J0343-2828}\label{c0}
This system is a quadruply-lensed quasar candidate (Figure~\ref{476_quasars}).  J0343-2828 was initially identified as a grade A strong lensing candidate in SLS~II (named as DESI-055.7976-28.4777), then as a lensed quasar candidate in H23.  Three posited lensed images are visible, \txrd{and a faint fourth image on the opposite side of the two lensed galaxies.  Additional \textit{HST} (Prop ID: 15652; PI: T. Treu) imaging reveals a even-fainter fifth image in-between the two lenses (see Figure~\ref{476_quasars_hst})}.  \txrd{We observe significant, and broadly coherent, variability in the three brightest images, as presented in Figure~\ref{476_quasars}.}

\txrd{A literature search shows that this system was also independently identified as a lensed quasar candidate using \textit{Gaia} DR2 PSF identification in \citet{lemon2023}.  However, in their work, they also presented evidence against this being a lensed quasar system: 1.) the followup \textit{HST} imaging of the system do not show diffraction patterns, and 2.) spectroscopic observations do not reveal any prominent QSO emission lines.}

\txrd{In regard to the first point, while clear diffraction spikes are not present in the F475X and F814W bands, they are visible in F160W band (as shown in Figure~\ref{476_quasars_hst}).  This alone does not mean it is a point source; it is possible that the object is barely resolved in the optical but within the diffraction limit in the IR.  However, \citet{schmidt2023} did well-model the system with a PSF light profile in the source plane across all three \textit{HST} bands, which gives additional credence to the lensed quasar hypothesis.}

\txrd{In regard to the second point, the spectrum \citep[Figure~15 of][]{lemon2023} was observed using the European Southern Observatory (ESO) Faint Object Spectrograph and Camera version 2.  Indeed, the spectrum presented from $3650 - 8000$\AA\ observed wavelength is relatively flat and displays no distinct QSO emission lines.
We note that its signal-to-noise ratio is comparatively lower than other spectra presented in the paper.  }

\txrd{By aggregating the information at hand (variability, \textit{HST} imaging/modeling, and spectroscopy), we propose that this system could be a more-exotic type of lensed quasar\footnote{Through private communications with C. Lemon and T. Schmidt}.  One such possibility is a changing-look quasar (CLQ) that can exhibit disappearing broad emission lines \citep{macLeod2019, potts2021, green2022}.  Another possibility is that of a low-ionization broad absorption line quasar (LoBALQ), where the broad emission lines may not be very obvious \citep{schulze2017, leighly2018, leighly2019, wethers2020}.  In summary, from the variability and other evidence presented, we believe it is appropriate to reopen the discussion around J0343-2828, and reclassify it as a lensed quasar candidate.
We emphasize that this reconsideration is made possible through the variability search method presented in this work. }

\section{Conclusion} \label{sec:conclusion}
Using our transient search pipeline, we perform a targeted search on 5807 \txr{strong lens candidate} systems, and \txrr{identify \txrd{seven} previously confirmed lensed quasars, \txrd{three} new quadruply lensed, and \txrd{ten} new doubly} \txb{lensed} quasar candidates.  \txo{We also} provide additional variability confirmation to 13 lensed quasar candidates \txrr{previously identified by D22 and H23}.  \txrr{We again stress that this does not in any way imply that the candidates presented by D22 and H23 but not presented in this paper are not lensed quasars, and acknowledge that the 13 shown in this work is likely highly incomplete to the lensed quasar candidates initial sample (see \S\ref{sec:results} and \S\ref{conflq})}.  \txrd{The lensed source of J0343-2828 (one of these 13 systems) was previously considered to be disconfirmed as a quasar \citep{lemon2023}.  Now with additional variability evidence, we reclassify J0343-2828 as a lensed quasar candidate.}

Our pipeline, rather than just relying on potential lensed image positions\txo{, morphology,} and colors, also assess each \txo{image's} variability, allowing us to find new candidates \txr{not discovered by other lensed quasar search pipelines}.  \txrr{With follow-up spectroscopy, it becomes possible to properly assess the correctness and completeness of our results, but we nevertheless} believe these results demonstrate the promising viability of our pipeline and its applicability to future surveys such as the Vera C. Rubin Observatory Legacy Survey of Space and Time (LSST) and the Nancy Grace Roman Space Telescope (RST).  As \S\ref{newlq} demonstrates, applying our pipeline on future searches \citep[with improved depth and cadence; e.g.,][]{suyu2020} will discover new lensed quasar \txrr{candidates, when supplied with a set of strong lens candidates in the survey footprint}.  \txb{Furthermore, \S\ref{conflq} shows that} our methodology will allow us to give additional variability \txb{confirmation} for lensed quasar candidates \txb{identified by more traditional lensed quasar search methods}.  From discovery to confirmation, we expect our pipeline to give high yield in future lensed quasar searches, and with this, contribute significantly to constraining $H_0$.  

\section*{Acknowledgments} \label{sec:acknowledgments}

This work was supported in part by the Director, Office of Science, Office of High Energy Physics of the US
Department of Energy under contract No. DE-AC025CH11231. This research used resources of the National Energy
Research Scientific Computing Center (NERSC), a U.S. Department of Energy Office of Science User Facility operated
under the same contract as above and the Computational HEP program in The Department of Energy’s Science Office
of High Energy Physics provided resources through the “Cosmology Data Repository” project (Grant \#KA2401022).  X. Huang acknowledges the University of San Francisco Faculty Development Fund.  The work of A. Cikota is supported by NOIRLab, which is managed by the Association of Universities for Research in Astronomy (AURA) under a cooperative agreement with the National Science Foundation.

This Paper is based on observations at Cerro Tololo Inter-American Observatory, National Optical Astronomy Observatory (NOAO Prop ID: 2014B-0404 with co-PIs: D. J. Schlegel and A. Dey; NOAO Prop IDs: 2022A-388025 and 2022B-297190 with co-PIs: A. Palmese and L. Wang), which is operated by the Association of Universities for Research in Astronomy (AURA) under a cooperative agreement with the National Science Foundation.

This project uses data obtained with the Dark Energy Camera, which was constructed by the Dark Energy Survey collaboration. Funding for the DES Projects has been provided by the U.S. Department of Energy, the U.S. National Science Foundation, the Ministry of Science and Education of Spain, the Science and Technology Facilities Council of the United Kingdom, the Higher Education Funding Council for England, the National Center for Supercomputing Applications at the University of Illinois at Urbana-Champaign, the Kavli Institute of Cosmological Physics at the University of Chicago, the Center for Cosmology and Astro-Particle Physics at the Ohio State University, the Mitchell Institute for Fundamental Physics and Astronomy at Texas A\&M University, Financiadora de Estudos e Projetos, Fundação Carlos Chagas Filho de Amparo à Pesquisa do Estado do Rio de Janeiro, Conselho Nacional de Desenvolvimento Científico e Tecnológico and the Ministério da Ciência, Tecnologia e Inovacão, the Deutsche Forschungsgemeinschaft, and the Collaborating Institutions in the Dark Energy Survey. The Collaborating Institutions are Argonne National Laboratory, the University of California at Santa Cruz, the University of Cambridge, Centro de Investigaciones Enérgeticas, Medioambientales y Tecnológicas-Madrid, the University of Chicago, University College London, the DES-Brazil Consortium, the University of Edinburgh, the Eidgenössische Technische Hochschule (ETH) Zürich, Fermi National Accelerator Laboratory, the University of Illinois at Urbana-Champaign, the Institut de Ciències de l’Espai (IEEC/CSIC), the Institut de Física d’Altes Energies, Lawrence Berkeley National Laboratory, the Ludwig-Maximilians Universität München and the associated Excellence Cluster Universe, the University of Michigan, the National Optical Astronomy Observatory, the University of Nottingham, the Ohio State University, the OzDES Membership Consortium the University of Pennsylvania, the University of Portsmouth, SLAC National Accelerator Laboratory, Stanford University, the University of Sussex, and Texas A\&M University.

\txrr{We thank the referee and data editor for their insightful comments on reorganizing and improving the quality of our paper.}  \txrd{We also thank C. Lemon and T. Schmidt for cross-referencing our candidates with catalogs of previously identified lensed quasar candidates, and for thought-provoking discussions regarding J0343-2828.}

\software{Astropy \citep{astropy2013, astropy2018},  
          Montage \citep{montage},  
          SEP \citep{bertin1996, sep},
          SNCosmo \citep{sncosmo}, 
          NumPy \citep{numpy}, 
          Matplotlib \citep{matplotlib}}

\newpage
\appendix \label{appendix:a}
\restartappendixnumbering
\section{Lensed Quasar Candidates from Targeted Search}

Here, we present the candidates systems identified by our first targeted search on strong lens candidates, shown in Figure~\ref{ws_montage} (with the exception of WG021416.37-210535.3 and DESI-350.3458-03.5082, which are presented in \S\ref{prevlq} and \S\ref{newlq} respectively).  

\subsection{Previously Identified Lensed Quasars} \label{append_prev}

\begin{figure}[H]
\figurenum{A1.1}
\begin{center}
\includegraphics[width=161mm]{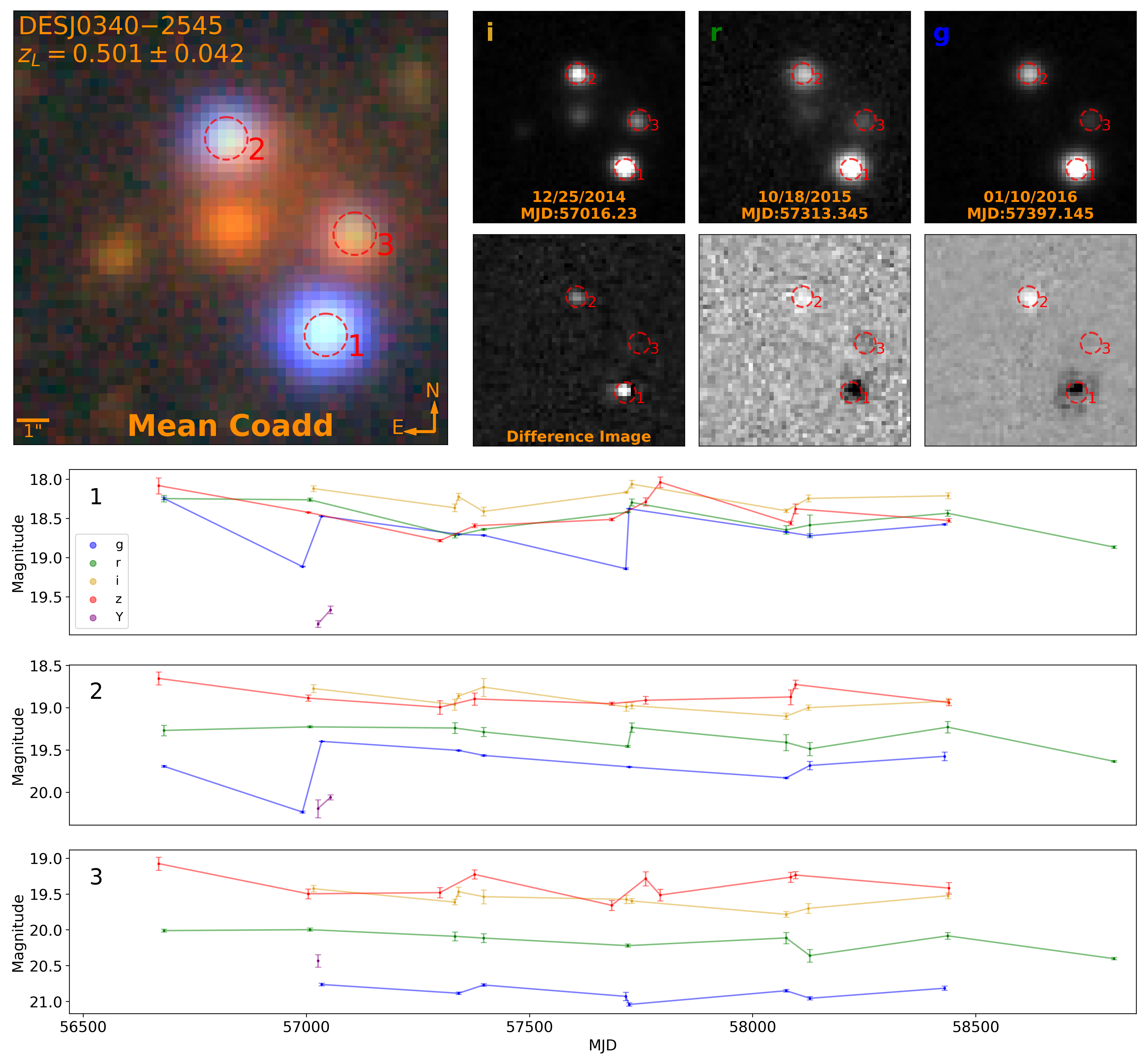}
\caption{Lensed quasar DESJ0340$-$2545 (see caption of Figure~\ref{454_quasars} for the full description of each subplot).}\label{442_quasars}
\end{center}
\end{figure}

\paragraph{DESJ0340$-$2545}\label{p1}
\txrd{This system is a quadruply-lensed quasar (Figure~\ref{442_quasars}), discovered and spectroscopically confirmed by \citet{lemon2020}.  This system was ``rediscovered'' in SLS~II as a strong lens candidate, and therefore it was included in our search.  We subsequently found that this is a known system.}  Unlike most other systems presented, this \txo{is likely} a cluster-lens lensed quasar system, with three visibly resolved images North, West, and Southwest of the main lensing galaxy respectively.  

\pagebreak

\begin{figure}[H]
\figurenum{A1.2}
\begin{center}
\includegraphics[width=161mm]{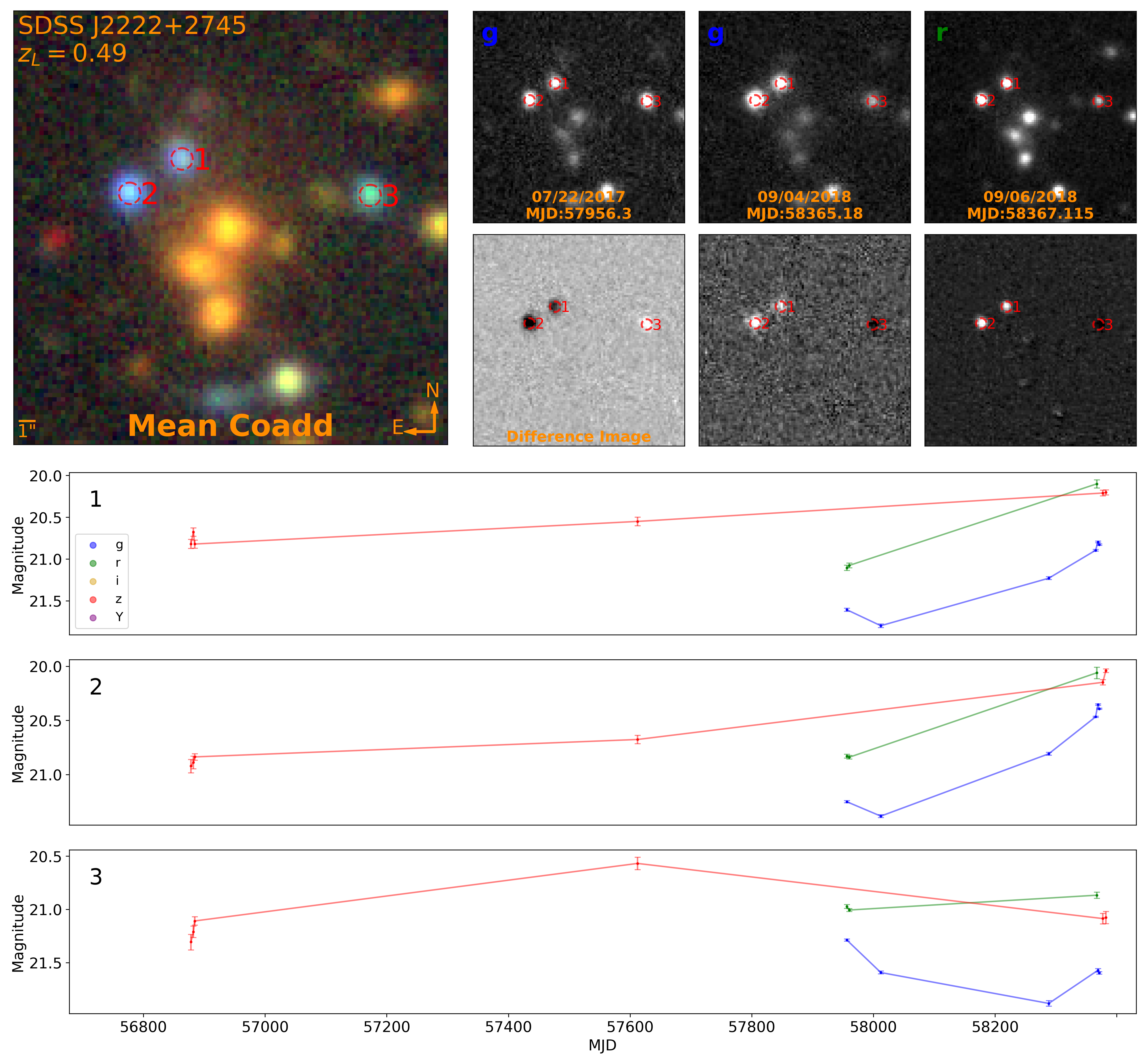}
\caption{Lensed quasar SDSS J2222+2745 (see caption of Figure~\ref{454_quasars} for the full description of each subplot).}\label{441_quasars}
\end{center}
\end{figure}

\paragraph{SDSS J2222+2745}\label{p2}
This system is a sextuply-lensed quasar (Figure~\ref{441_quasars}), discovered and spectroscopically confirmed by \citet{dahle2013}.  \txo{This system was ``rediscovered'' in SLS~II as a strong lens candidate, and therefore it was included in our search.  We subsequently found that this is a known system.}  \txr{\txo{Light curves of this system were shown in} \citet{dahle2015}.}  In Figure~\ref{441_quasars}, \txr{there are three identifiable images in the Legacy Imaging Surveys, and the remaining three are too faint and/or too close to the cluster center.}  The bright \txo{PSF-like object} South of the cluster lens has been spectroscopically confirmed to be \txo{a} white dwarf \citep{dahle2013}.

\pagebreak

\begin{figure}[H]
\figurenum{A1.3}
\begin{center}
\includegraphics[width=161mm]{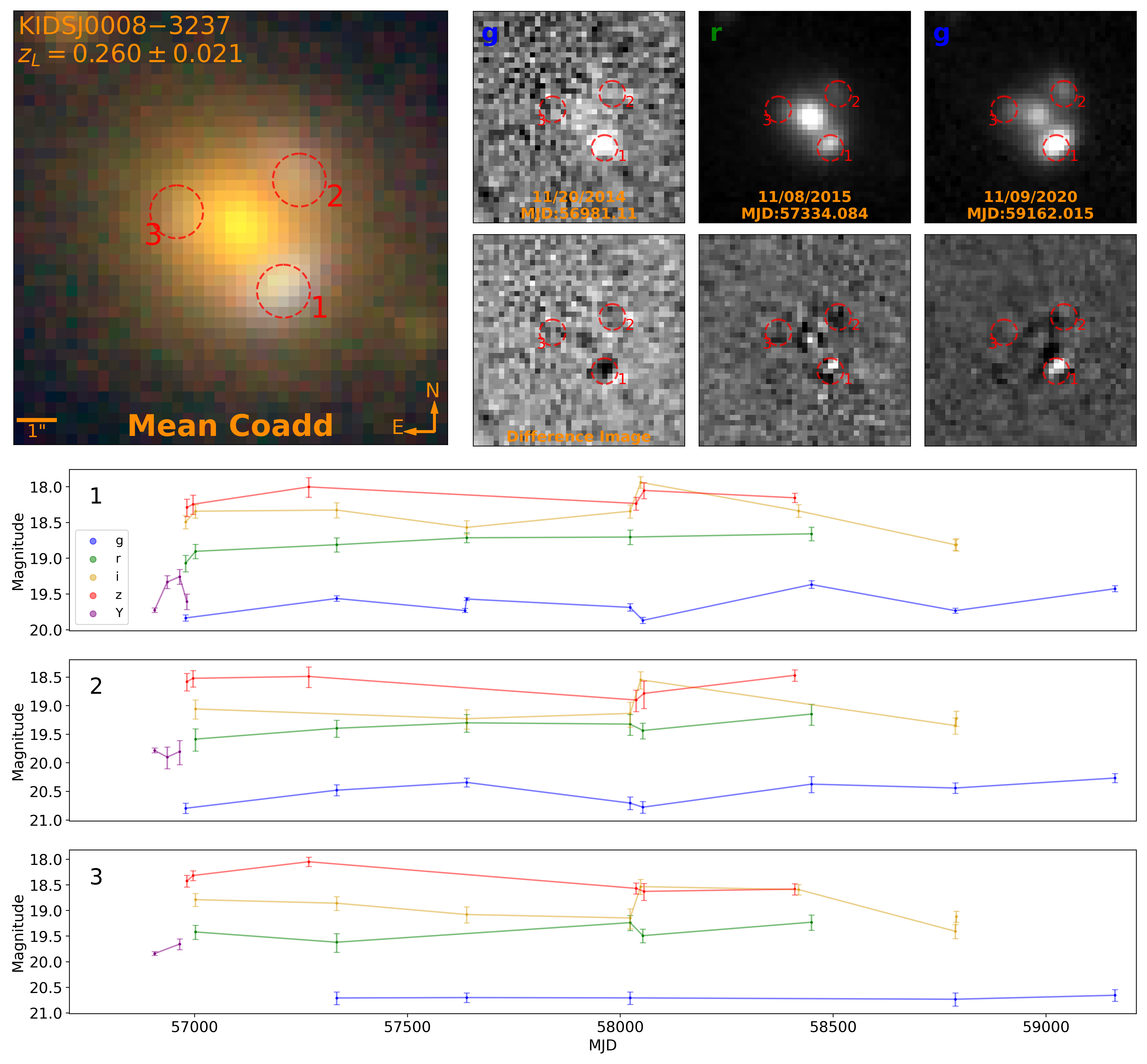}
\caption{Lensed quasar KIDSJ0008$-$3237 (see caption of Figure~\ref{454_quasars} for the full description of each subplot).}\label{450_quasars}
\end{center}
\end{figure}

 \paragraph{KIDSJ0008$-$3237}\label{p3}
\txrd{This system is a quadruply-lensed quasar (Figure~\ref{450_quasars}), initially discovered as a lensed quasar candidate by \citet{khramtsov2019}.  This system was identified in SLS~III as a strong lens candidate, and therefore it was included in our search.  We subsequently found that this is a known lensed quasar candidate.}  \txo{We posit} that the ``image'' on the Southwest of the lens is the result of two lensed quasar images \txr{merging}, with two other images seen East and Northwest of the lens.  

\pagebreak

\begin{figure}[H]
\figurenum{A1.4}
\begin{center}
\includegraphics[width=161mm]{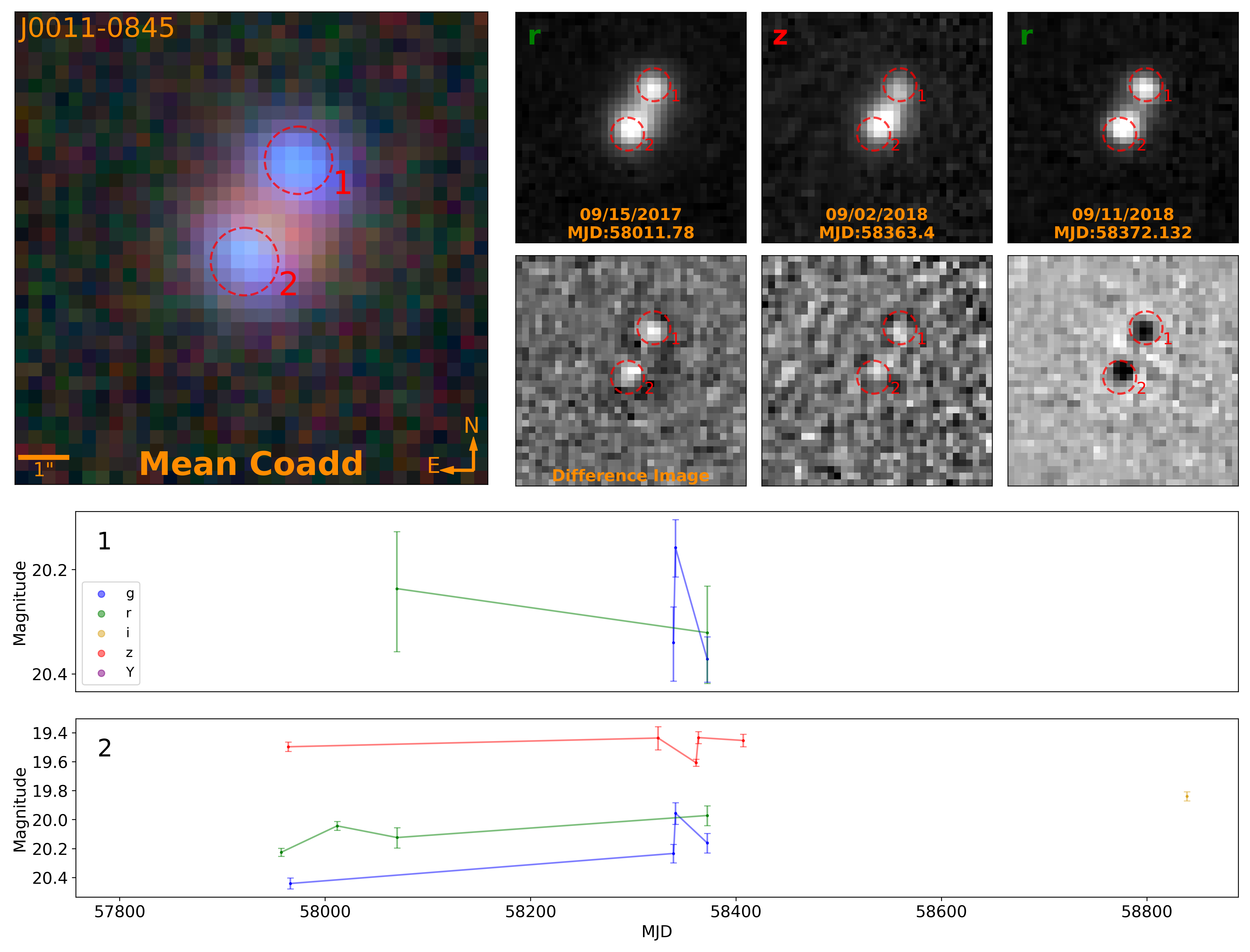}
\caption{Lensed quasar J0011-0845 (see caption of Figure~\ref{454_quasars} for the full description of each subplot).}\label{457_quasars}
\end{center}
\end{figure}

\paragraph{J0011-0845}\label{p4}
This system is a doubly-lensed quasar (Figure~\ref{457_quasars}), initially identified and spectroscopically confirmed by \citet{lemon2018}, but no light curves were provided.  This system was ``rediscovered'' in SLS~III as a strong lens candidate, and therefore it was included in our search.  We subsequently found that this is a known system.  Though this system was previously discovered, we show new light curves for this system. 
 This system exhibits a typical double lensed quasar morphology, with two bright, PSF-like images on directly opposite sides of a posited lens galaxy.

 \pagebreak

\begin{figure}[H]
\figurenum{A1.5}
\begin{center}
\includegraphics[width=161mm]{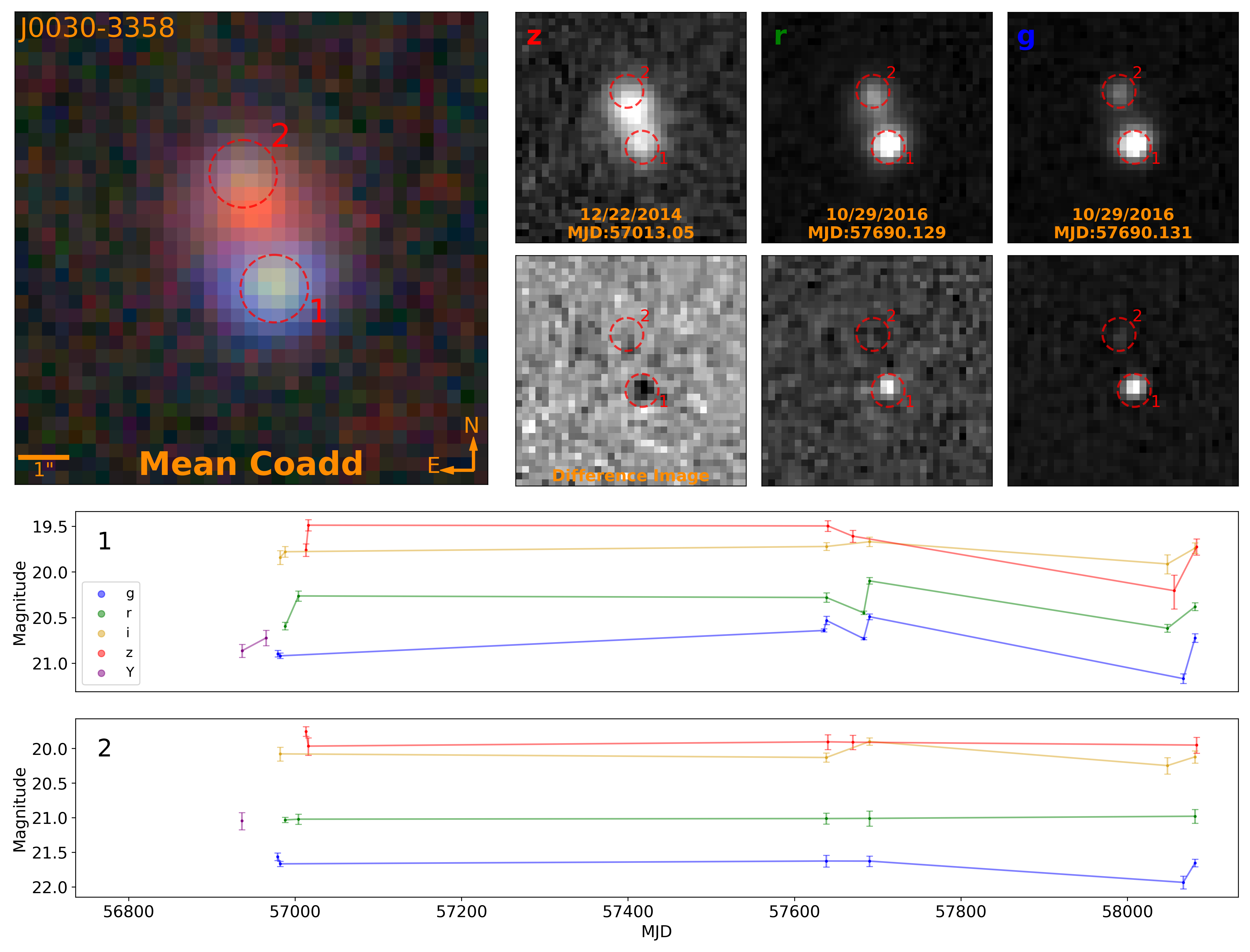}
\caption{Lensed quasar J0030-3358 (see caption of Figure~\ref{454_quasars} for the full description of each subplot).}\label{449_quasars}
\end{center}
\end{figure}

\paragraph{J0030-3358}\label{p5}
\txrd{This system is a doubly-lensed quasar (Figure~\ref{449_quasars}), initially identified as a strong lens candidate in SLS~II.  It was later spectroscopically confirmed as a lensed quasar in \citet{lemon2023}, though no light curves have been provided.}  
Though J0030-3358 exhibits a double structure, the image-to-lens angular separations of the two images clearly differ, and the Southern image is significantly brighter than the Northern image, which may indicate a longer time delay (as a strong difference in magnification implies a larger $\Delta \kappa$ and hence gravitational time delay between images).  Also similarly, both images are bright\txb{, blue}, PSF-like, and variable.  

\pagebreak

\begin{figure}[H]
\figurenum{A1.6}
\begin{center}
\includegraphics[width=161mm]{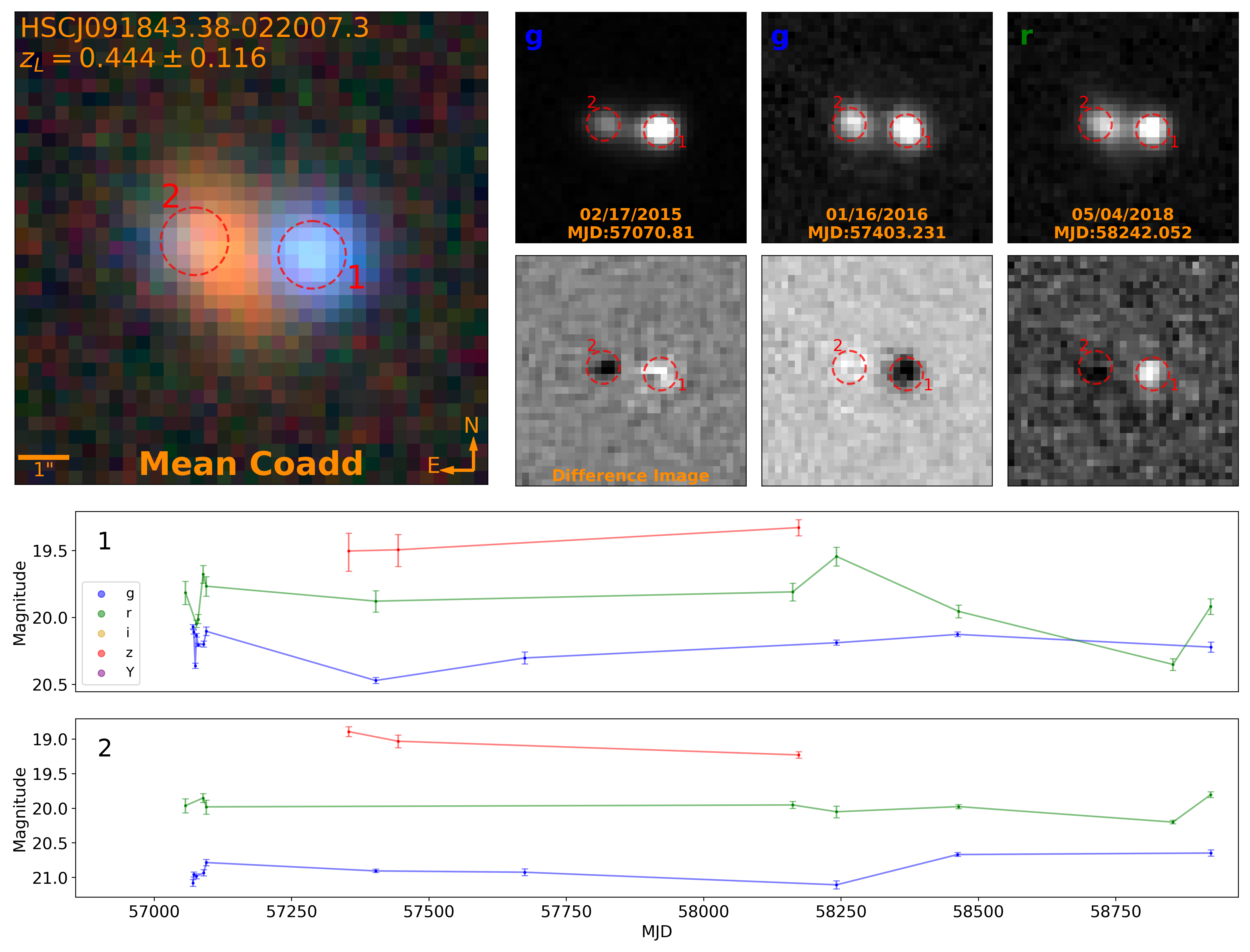}
\caption{Lensed quasar HSCJ091843.38-022007.3 (see caption of Figure~\ref{454_quasars} for the full description of each subplot).}\label{445_quasars}
\end{center}
\end{figure}

\paragraph{HSCJ091843.38-022007.3}\label{p6}
\txp{This system is a doubly-lensed quasar (Figure~\ref{445_quasars}), initially identified as a strong lens candidate in \citet{sonnenfeld2020}.  It was later spectroscopically confirmed as a lensed quasar in \citet{jaelani2021}, but no light curves have been provided.}  This \txo{system} shares a similar structure to J0030-3358 (see Figure~\ref{449_quasars}), as both exhibit a dim image near the lens galaxy, and a brighter, more distant image on the opposite side of the lens galaxy.

\pagebreak

\subsection{New Lensed Quasar Candidates} \label{append_new}

\begin{figure}[H]
\figurenum{A2.1}
\begin{center}
\includegraphics[width=161mm]{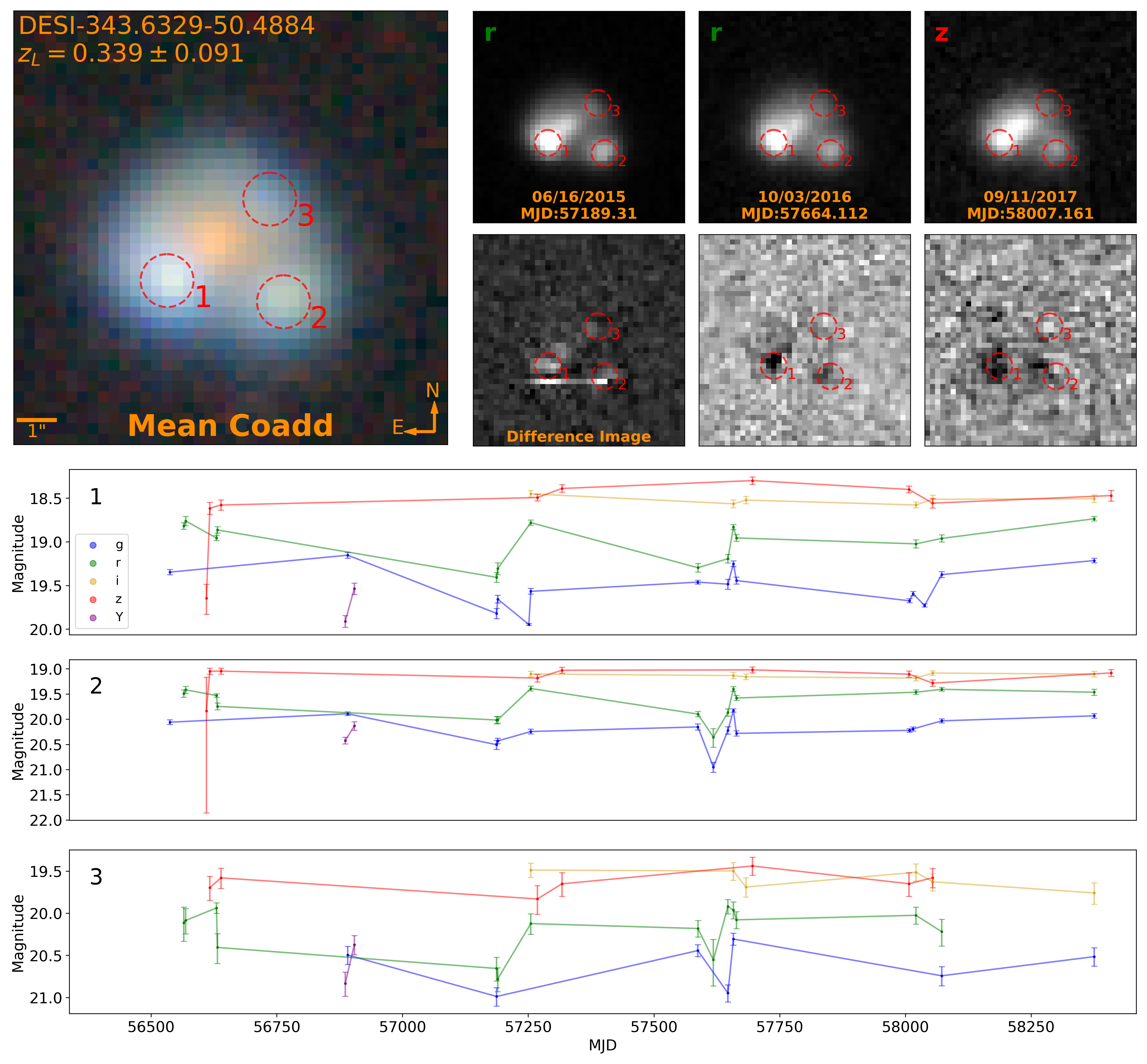}
\caption{Lensed quasar DESI-343.6329-50.4884 (see caption of Figure~\ref{454_quasars} for the full description of each subplot).}\label{456_quasars}
\end{center}
\end{figure}

\paragraph{DESI-343.6329-50.4884}\label{n1}
This system is a quadruply-lensed quasar candidate (Figure~\ref{456_quasars}), initially identified as a grade A strong lensing candidate in SLS~III.  With DR10 data, \txr{the Tractor only identifies three images, with a fourth one} likely merging with another image.  \txb{While at first glace, these systems do not have the appearance of a typical lensed quasar system, our pipeline is still able to identify PSF sources within the images; this also applies to the next system.} 

\pagebreak

\begin{figure}[H]
\figurenum{A2.2}
\begin{center}
\includegraphics[width=161mm]{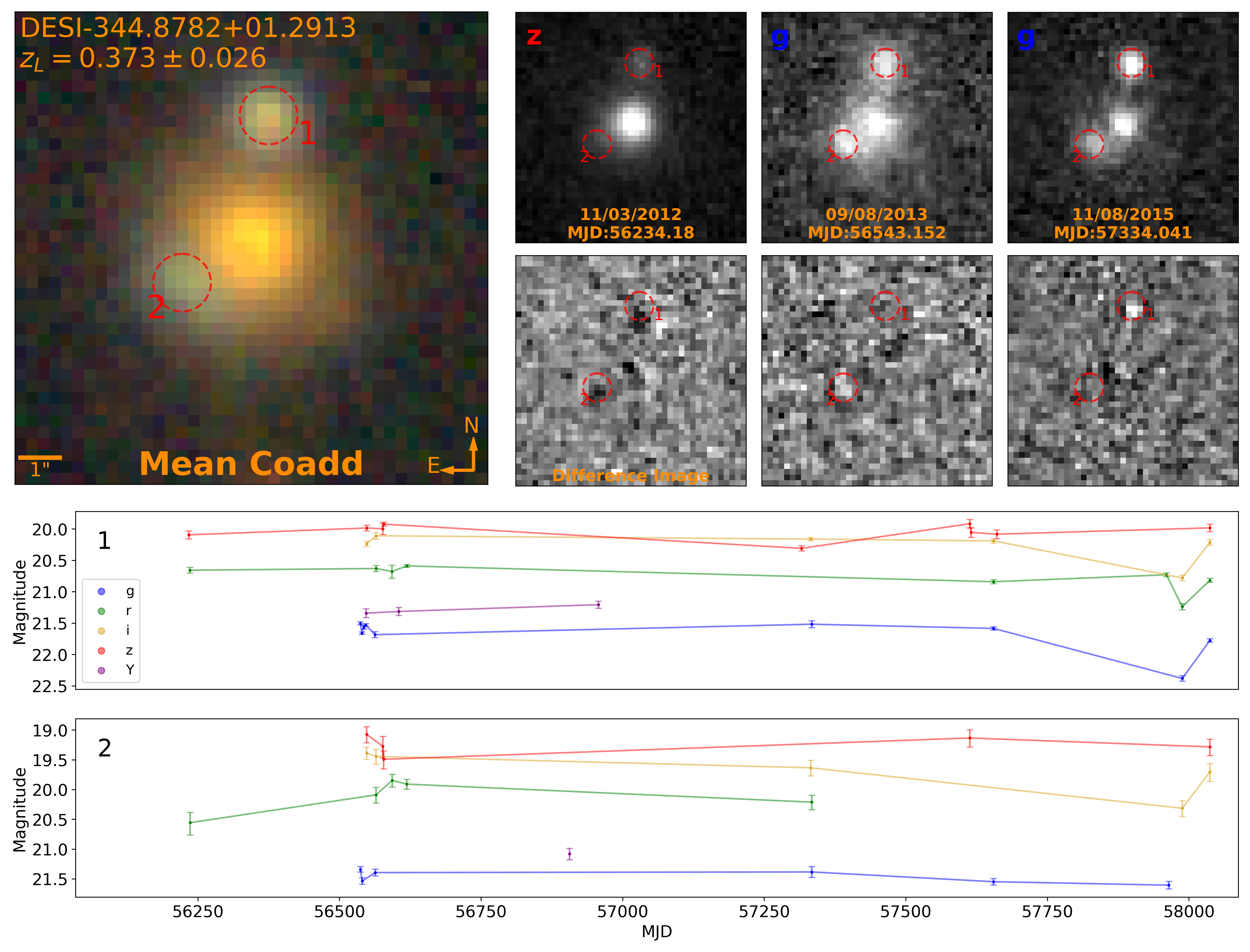}
\caption{Lensed quasar DESI-344.8782+01.2913 (see caption of Figure~\ref{454_quasars} for the full description of each subplot).}\label{437_quasars}
\end{center}
\end{figure}

\paragraph{DESI-344.8782+01.2913}\label{n3}
This system is a quadruply-lensed quasar candidate (Figure~\ref{437_quasars}), initially identified as a grade C strong lensing candidate in SLS~III.  It is \txo{possible} that the arc on the Southeast of the lens galaxy is comprised of two lensed quasar images \txr{merging}.  The brightest posited image lies directly North of the lens.  \txr{There is also a possible fourth image to the Southwest of the lens.}

\pagebreak

\begin{figure}[H]
\figurenum{A2.3}
\begin{center}
\includegraphics[width=161mm]{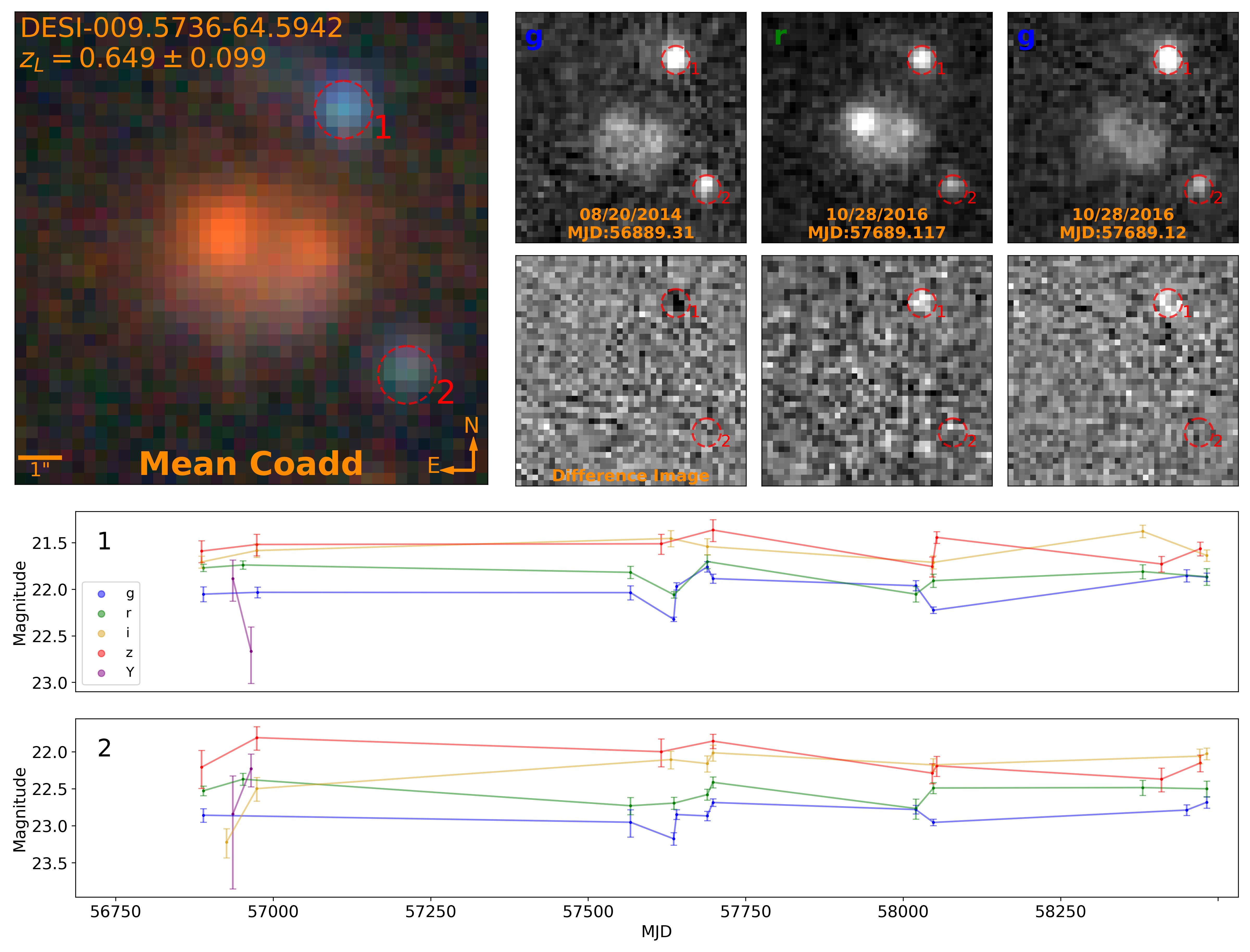}
\caption{Lensed quasar DESI-009.5736-64.5942 (see caption of Figure~\ref{454_quasars} for the full description of each subplot).}\label{436_quasars}
\end{center}
\end{figure}

\paragraph{DESI-009.5736-64.5942}\label{n16}
This system is a doubly-lensed quasar candidate (Figure~\ref{436_quasars}), initially identified as a grade C strong lensing candidate in SLS~III.  The posited lens consists of two red galaxies, with two lensed images North and West of the it.  

\pagebreak

\begin{figure}[H]
\figurenum{A2.4}
\begin{center}
\includegraphics[width=161mm]{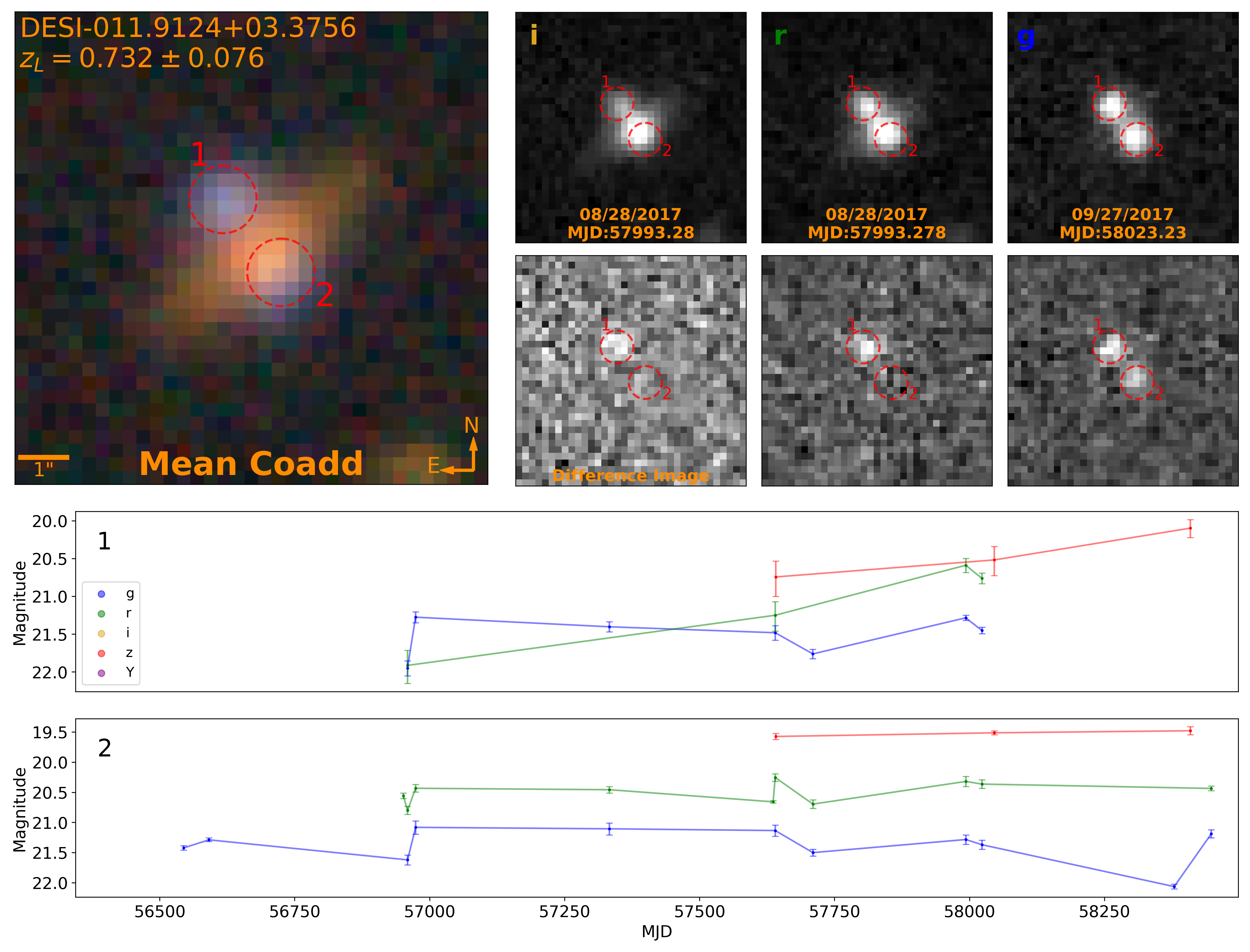}
\caption{Lensed quasar DESI-011.9124+03.3756 (see caption of Figure~\ref{454_quasars} for the full description of each subplot).}\label{453_quasars}
\end{center}
\end{figure}

\paragraph{DESI-011.9124+03.3756}\label{n6}
This system is a doubly-lensed quasar candidate (Figure~\ref{453_quasars}), initially identified as a grade A strong lensing candidate in SLS~III.  This system also exhibits \txo{a} double structure \txo{similar to} J0011-0845 (see Figure~\ref{457_quasars}).

\pagebreak

\begin{figure}[H]
\figurenum{A2.5}
\begin{center}
\includegraphics[width=161mm]{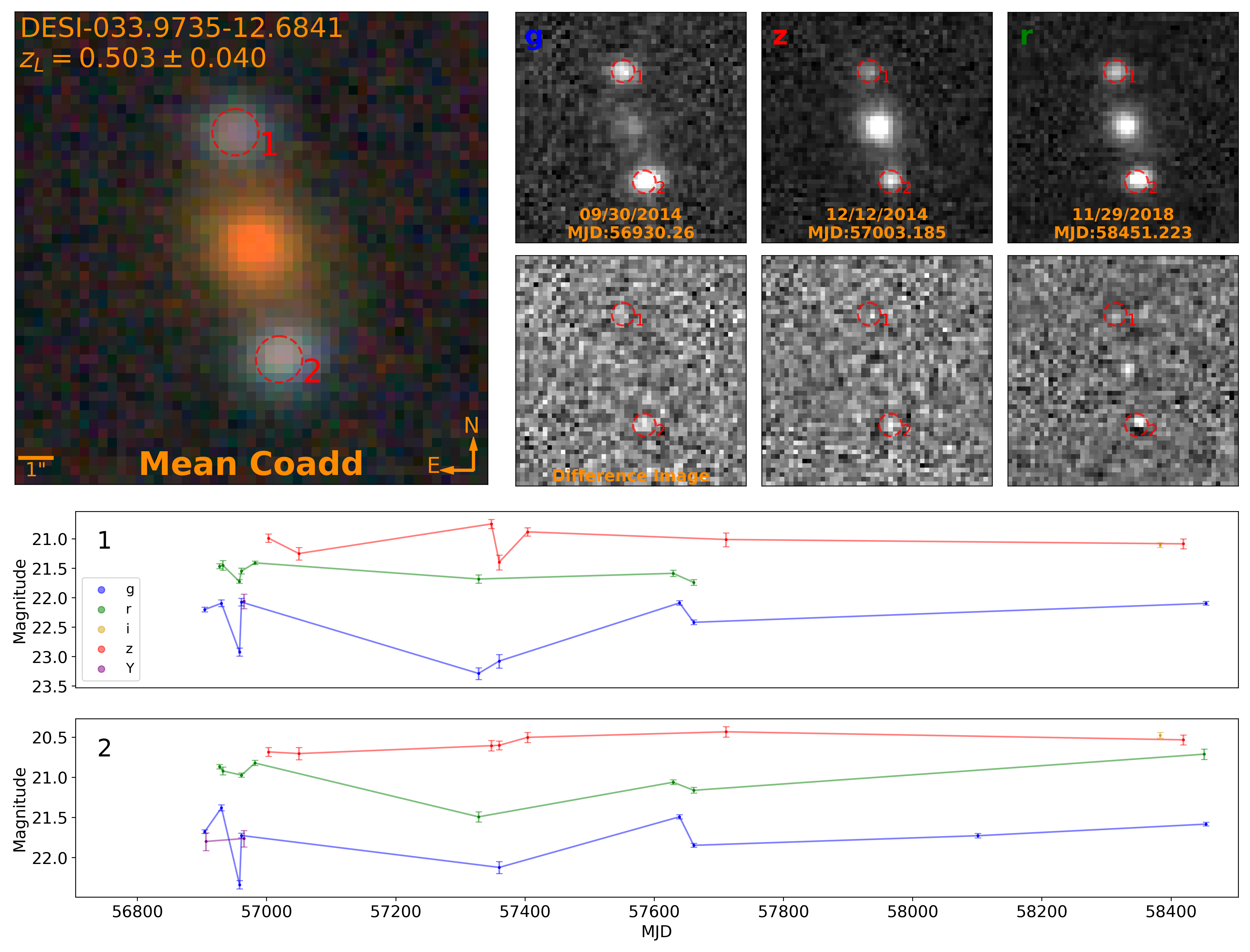}
\caption{Lensed quasar DESI-033.9735-12.6841 (see caption of Figure~\ref{454_quasars} for the full description of each subplot).}\label{458_quasars}
\end{center}
\end{figure}

\paragraph{DESI-033.9735-12.6841}\label{n9}
This system is a doubly-lensed quasar candidate (Figure~\ref{458_quasars}), initially identified as a grade C strong lensing candidate in SLS~I.  While this system also shares a typical double structure as the previous six candidates, DESI-033.9735-12.6841 has a significantly larger Einstein radius ($\sim 2''$).

\pagebreak

\begin{figure}[H]
\figurenum{A2.6}
\begin{center}
\includegraphics[width=161mm]{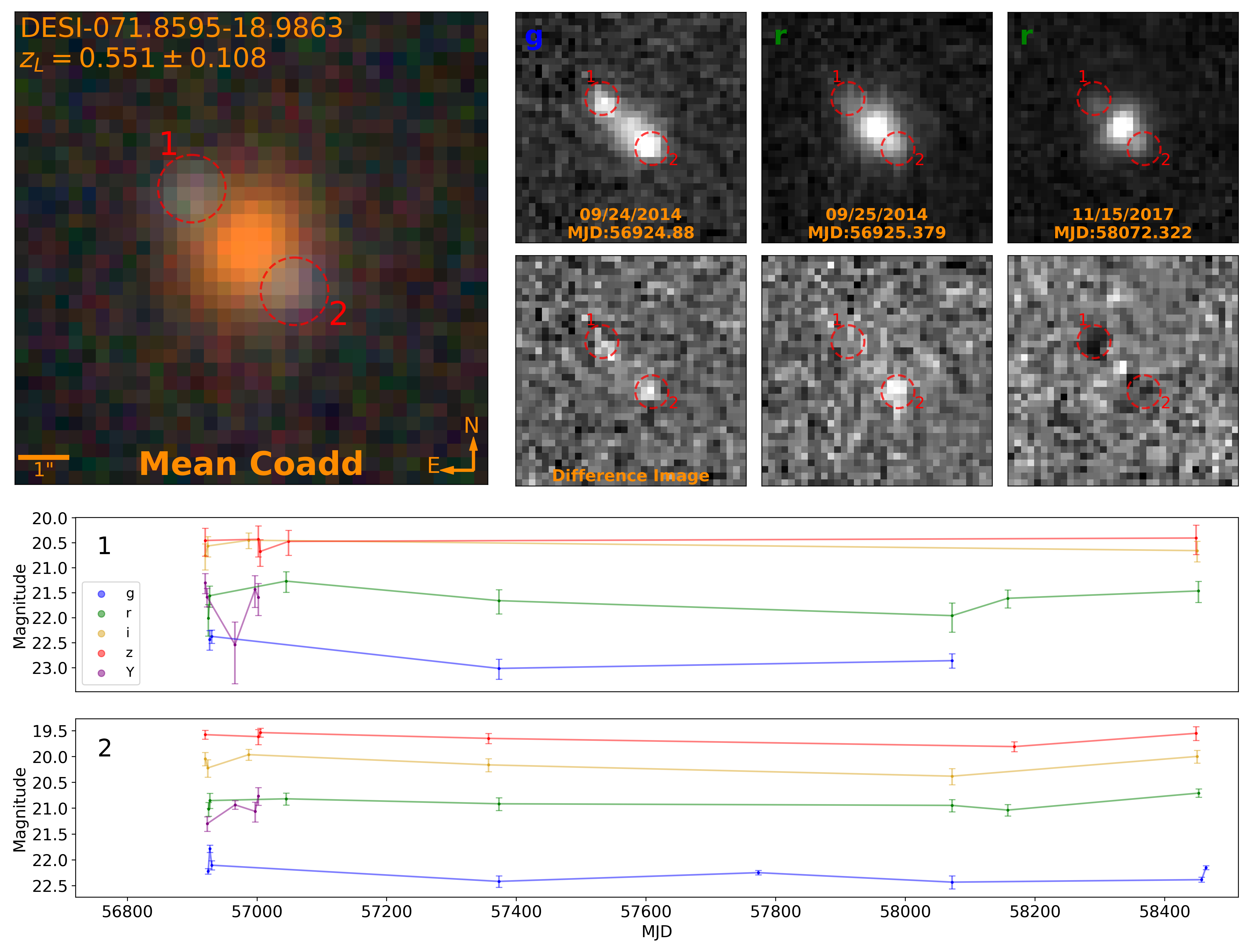}
\caption{Lensed quasar DESI-071.8595-18.9863 (see caption of Figure~\ref{454_quasars} for the full description of each subplot).}\label{439_quasars}
\end{center}
\end{figure}

\paragraph{DESI-071.8595-18.9863}\label{n5}
This system is a doubly-lensed quasar candidate (Figure~\ref{439_quasars}), initially identified as a grade B strong lensing candidate in SLS~III.  This system also exhibits \txo{a} double structure \txo{similar to} J0011-0845 (see Figure~\ref{457_quasars}). 

\pagebreak

\begin{figure}[H]
\figurenum{A2.7}
\begin{center}
\includegraphics[width=161mm]{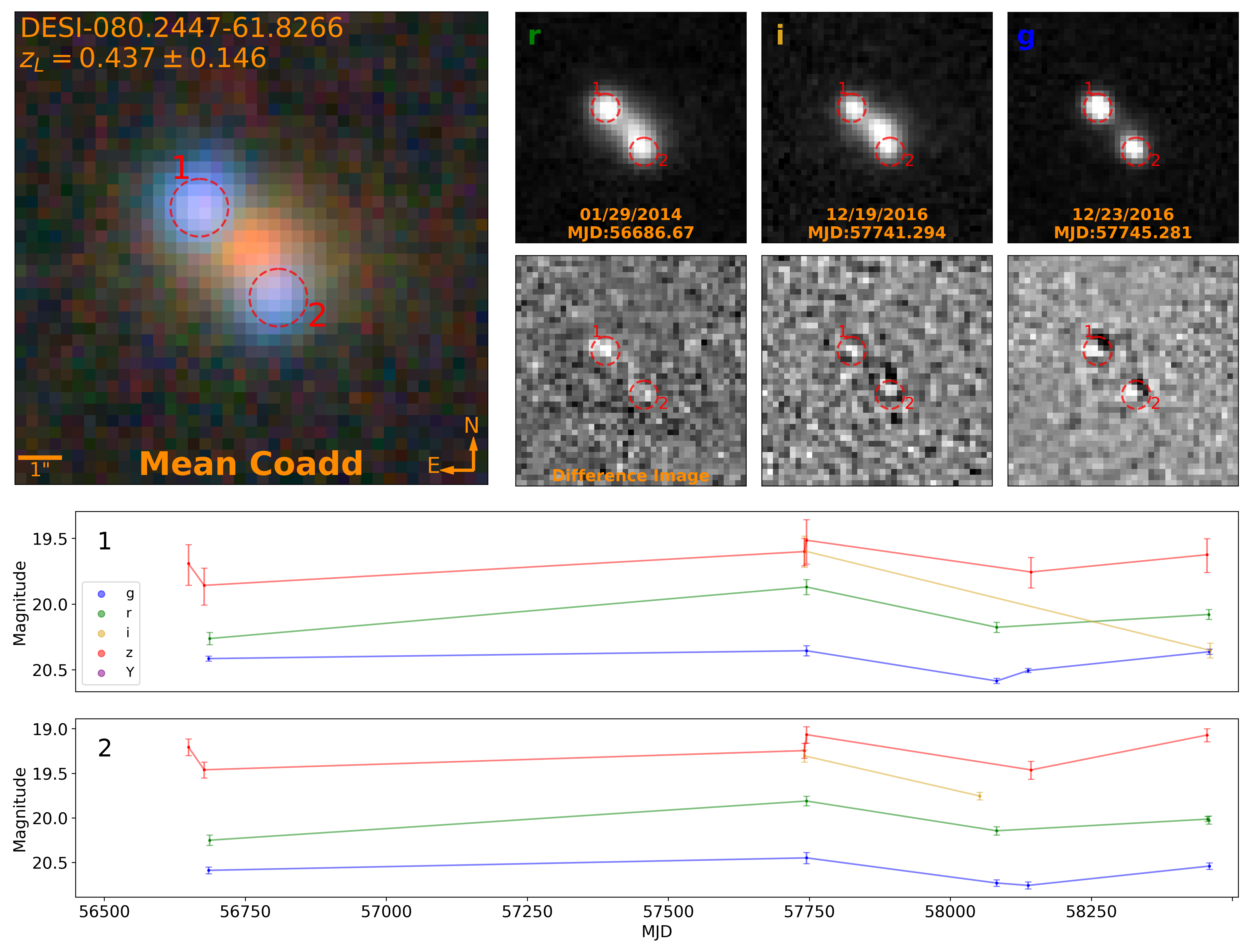}
\caption{Lensed quasar DESI-080.2447-61.8266 (see caption of Figure~\ref{454_quasars} for the full description of each subplot).}\label{447_quasars}
\end{center}
\end{figure}

\paragraph{DESI-080.2447-61.8266}\label{n10}
This system is a doubly-lensed quasar candidate (Figure~\ref{447_quasars}), initially identified as a grade C strong lensing candidate in SLS~III.  This system also exhibits \txo{a} double structure \txo{similar to} J0011-0845 (see Figure~\ref{457_quasars}).

\pagebreak

\begin{figure}[H]
\figurenum{A2.8}
\begin{center}
\includegraphics[width=161mm]{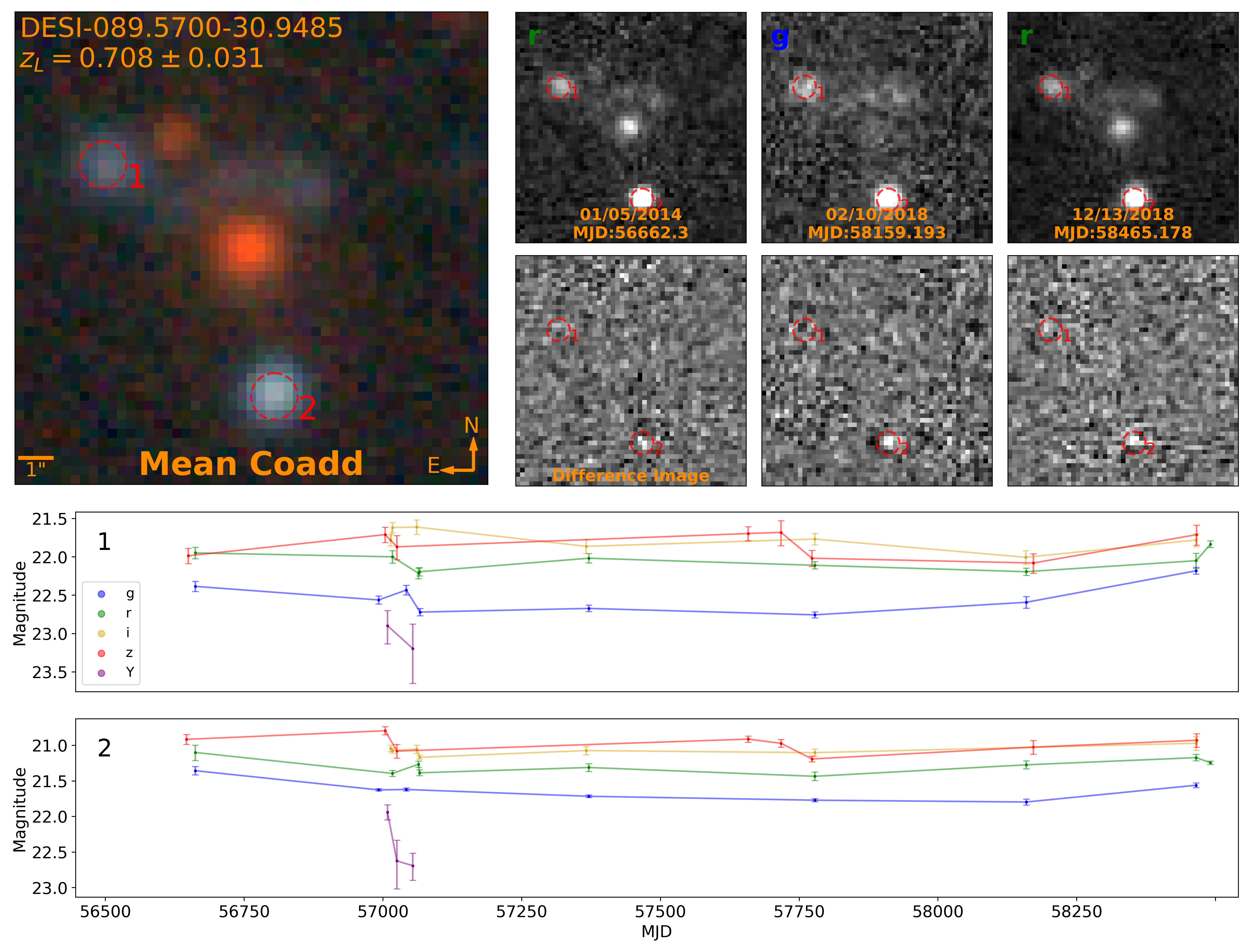}
\caption{Lensed quasar DESI-089.5700-30.9485 (see caption of Figure~\ref{454_quasars} for the full description of each subplot).}\label{443_quasars}
\end{center}
\end{figure}

\paragraph{DESI-089.5700-30.9485}\label{n15}
This system is a doubly-lensed quasar candidate (Figure~\ref{443_quasars}), initially identified as a grade B strong lensing candidate in SLS~II.  While there is a clear arc North of the lens galaxy, the posited lensed quasar images are located South and Northeast of the lens galaxy.  Hence, we posit that the lensed quasar resides in a different source galaxy as the arc, as it would be improbable for the quasar to be hosted by the arc given their separation.  The posited quasar images seem to be point sources, and exhibit variability between exposures.  The presence of an additional source can provide degeneracy-breaking modeling constraints, allowing for tighter measurements of $H_0$ \citep[e.g., ][]{shajib2020}.  

\pagebreak

\begin{figure}[H]
\figurenum{A2.9}
\begin{center}
\includegraphics[width=161mm]{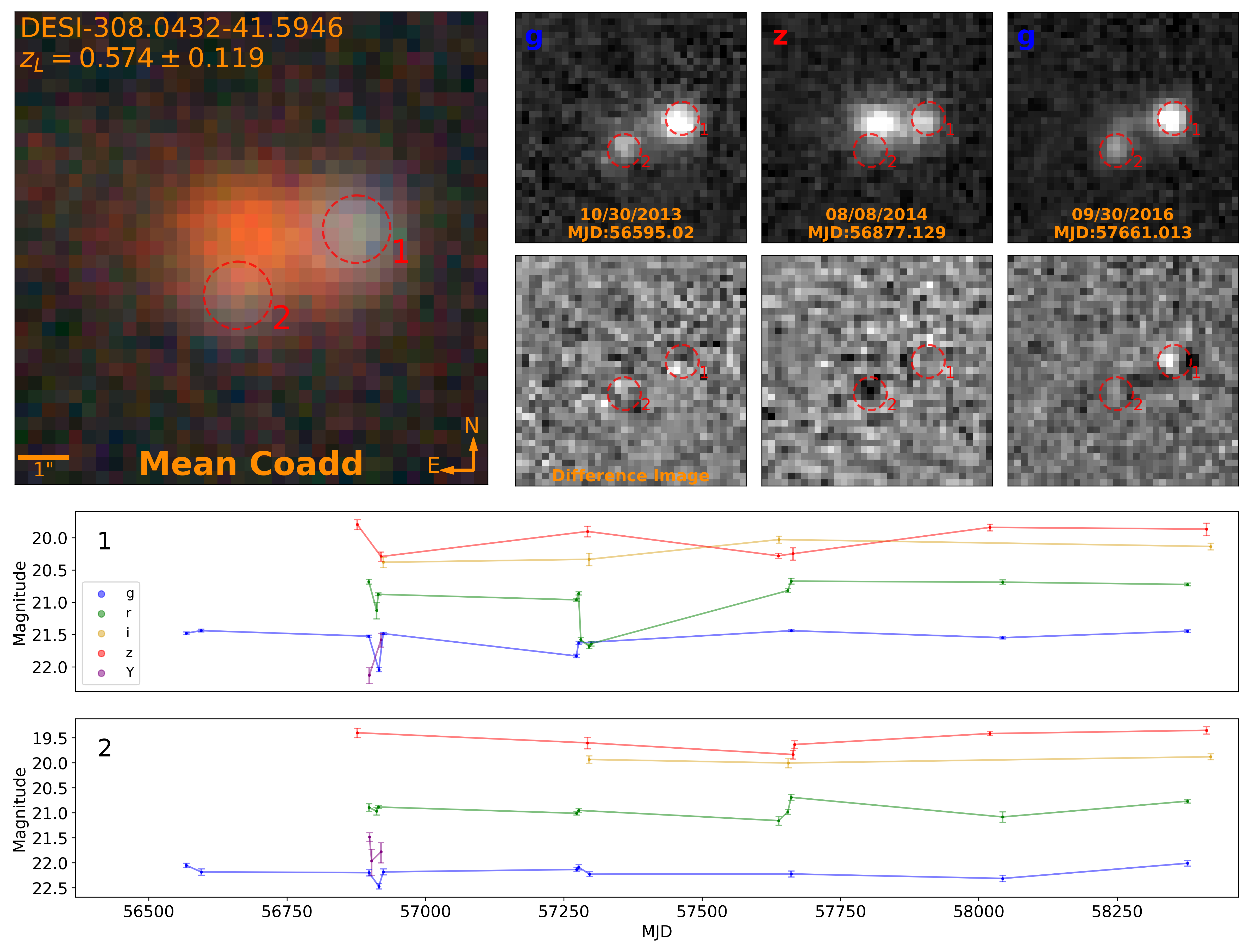}
\caption{Lensed quasar DESI-308.0432-41.5946 (see caption of Figure~\ref{454_quasars} for the full description of each subplot).}\label{438_quasars}
\end{center}
\end{figure}

\paragraph{DESI-308.0432-41.5946}\label{n12}
This system is a doubly-lensed quasar candidate (Figure~\ref{438_quasars}), initially identified as a grade C strong lensing candidate in SLS~III.  DESI-308.0432-41.5946 differs \txo{somewhat} from the typical double morphology, as the two posited lensed quasar images and the lens galaxy are far from collinear.  Since the posited images are still PSF-shaped and variable, we \txo{consider} this to be a strong lensed quasar candidate.  However, the noncollinearity could possibly indicate the presence of faint and unresolved (\txo{in} DECam observations) lensed quasar images \txo{(see an example in Figure~\ref{example_hst})}.

\pagebreak

\begin{figure}[H]
\figurenum{A2.10}
\begin{center}
\includegraphics[width=161mm]{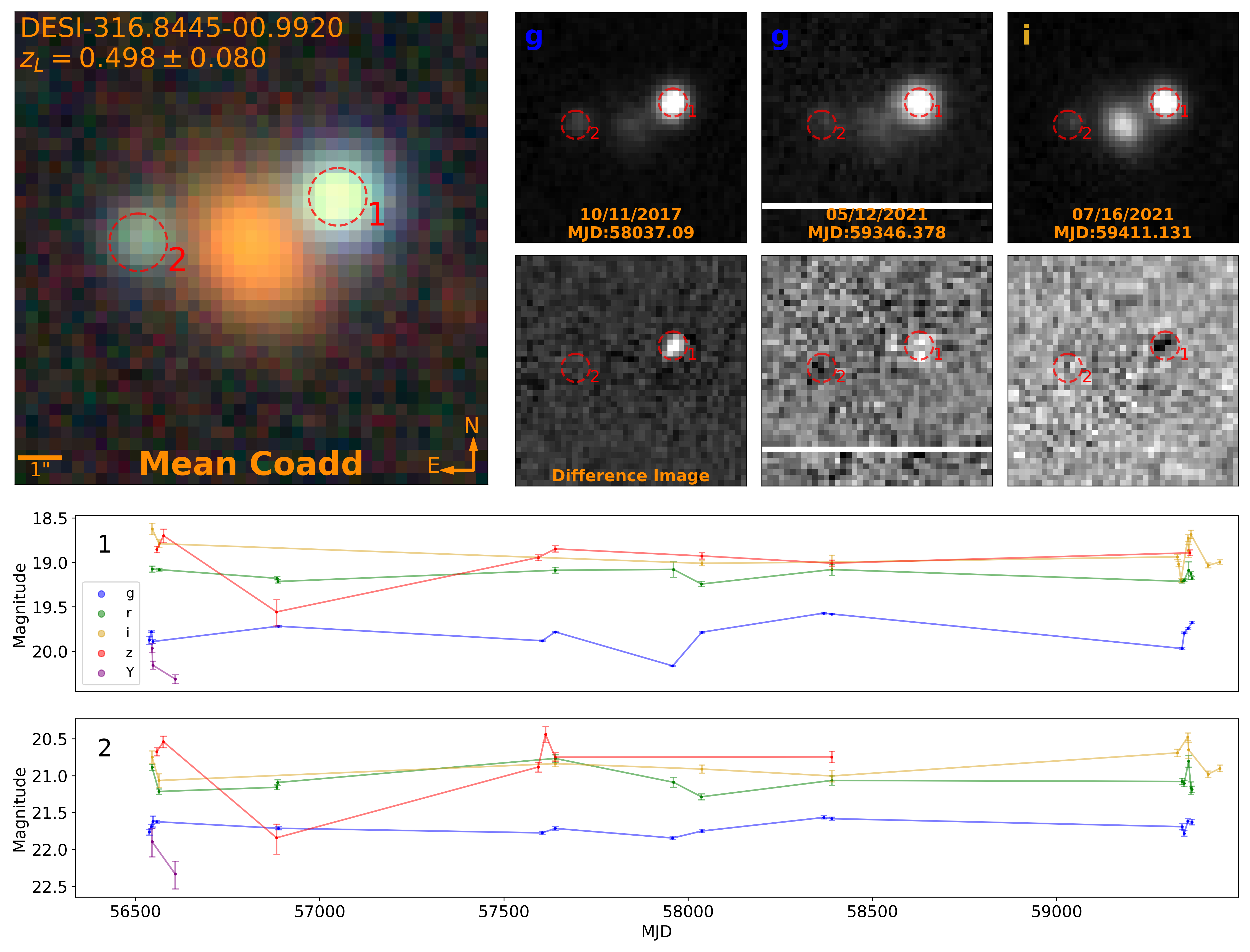}
\caption{Lensed quasar DESI-316.8445-00.9920 (see caption of Figure~\ref{454_quasars} for the full description of each subplot).}\label{444_quasars}
\end{center}
\end{figure}

\paragraph{DESI-316.8445-00.9920}\label{n11}
This system is a doubly-lensed quasar candidate (Figure~\ref{444_quasars}), initially identified as a grade B strong lensing candidate in SLS~II.  This system also exhibits \txo{a} double structure \txo{similar to} J0011-0845 (see Figure~\ref{457_quasars}).

\pagebreak

\begin{figure}[H]
\figurenum{A2.11}
\begin{center}
\includegraphics[width=161mm]{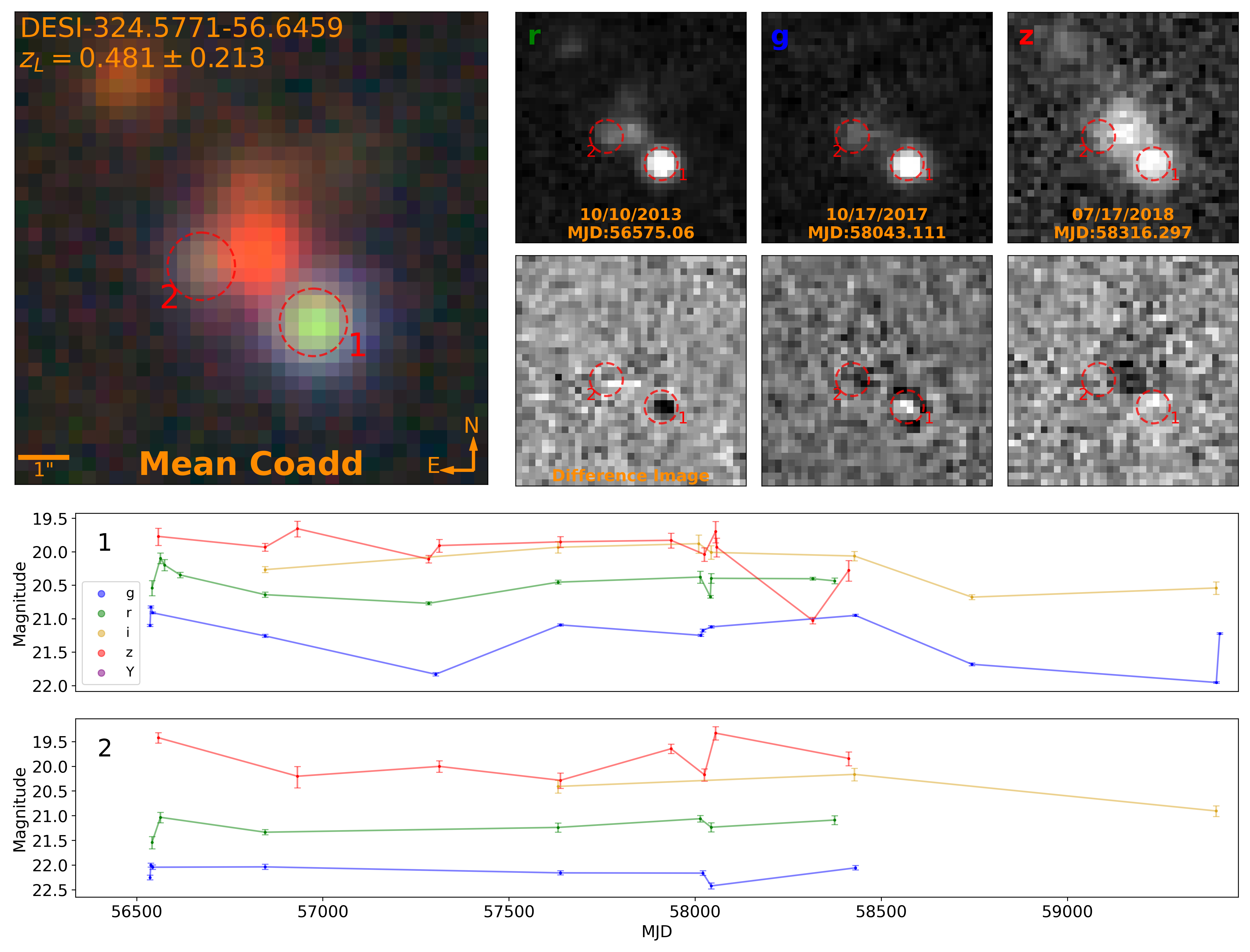}
\caption{Lensed quasar DESI-324.5771-56.6459 (see caption of Figure~\ref{454_quasars} for the full description of each subplot).}\label{452_quasars}
\end{center}
\end{figure}

\paragraph{DESI-324.5771-56.6459}\label{n13}
This system is a doubly-lensed quasar candidate (Figure~\ref{452_quasars}), initially identified as a grade C strong lensing candidate in SLS~III.  \txb{As} with the previous system, DESI-308.0432-41.5946 \txo{appears to be a} double.  \txo{But given the notable noncollinearity,  there may be other lensed images too faint and/or too close to the lens.}

\pagebreak

\begin{figure}[H]
\figurenum{A2.12}
\begin{center}
\includegraphics[width=161mm]{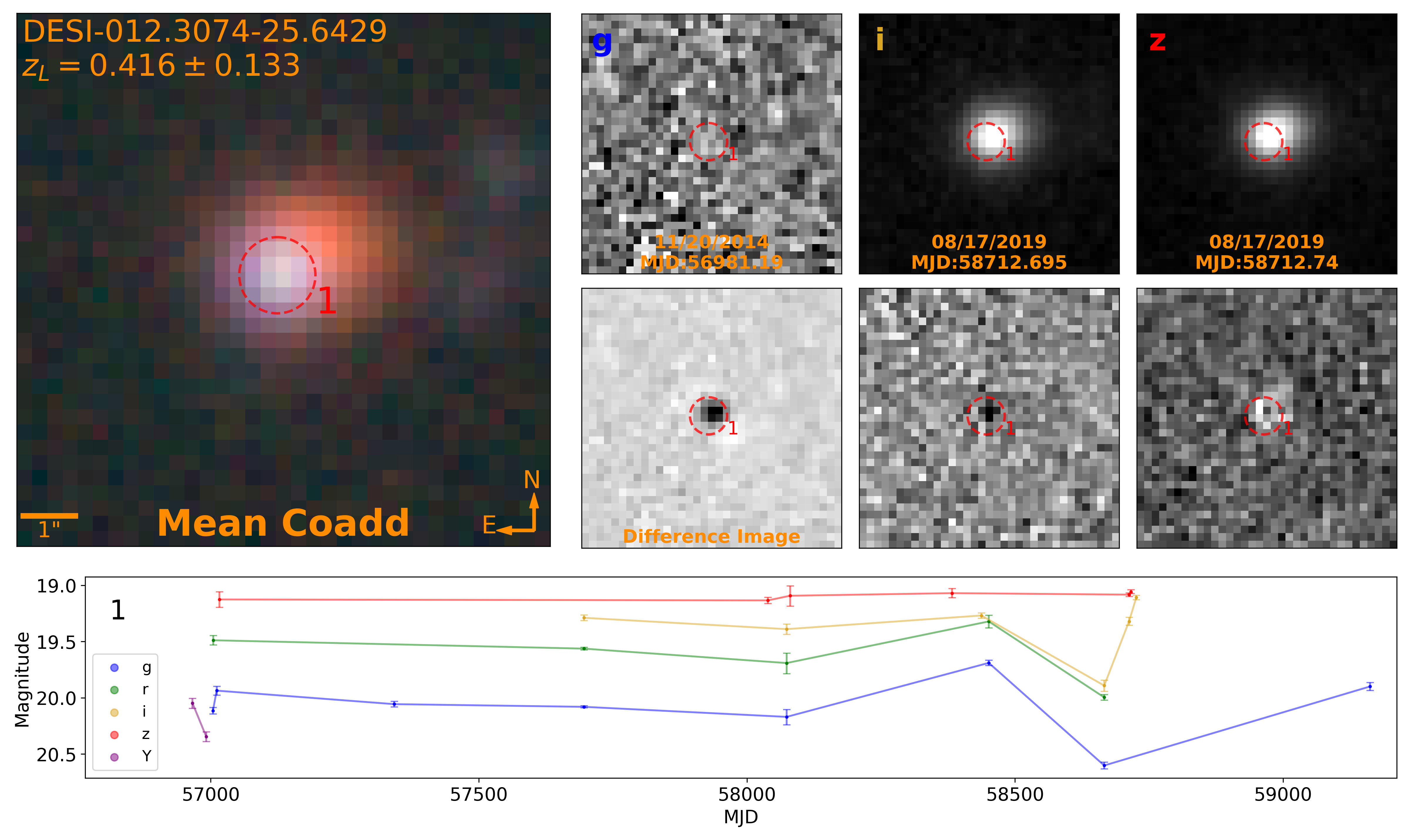}
\caption{Lensed quasar DESI-012.3074-25.6429 (see caption of Figure~\ref{454_quasars} for the full description of each subplot).}\label{446_quasars}
\end{center}
\end{figure}

\paragraph{DESI-012.3074-25.6429}\label{n14}
This system is a doubly-lensed quasar candidate (Figure~\ref{446_quasars}), initially identified as a grade C strong lensing candidate in SLS~III.  While only one variable image (directly East of the lens) is visible, \txo{given the observed configuration, this candidate system may have a counterimage on the opposite side of the lens.  If so, such an image would likely be} demagnified and would be overwhelmed by the lens galaxy light.

\pagebreak

\restartappendixnumbering
\section{Variability Confirmation of Previously Discovered Lensed Quasar Candidates} \label{append_conf}

Here, we present the systems identified by our second targeted search on strong lensed quasar candidates, shown in Figure~\ref{cd_montage} (with the exception of J0343-2828, which is presented in \S\ref{conflq}).

\begin{figure}[H]
\figurenum{B1.1}
\begin{center}
\includegraphics[width=161mm]{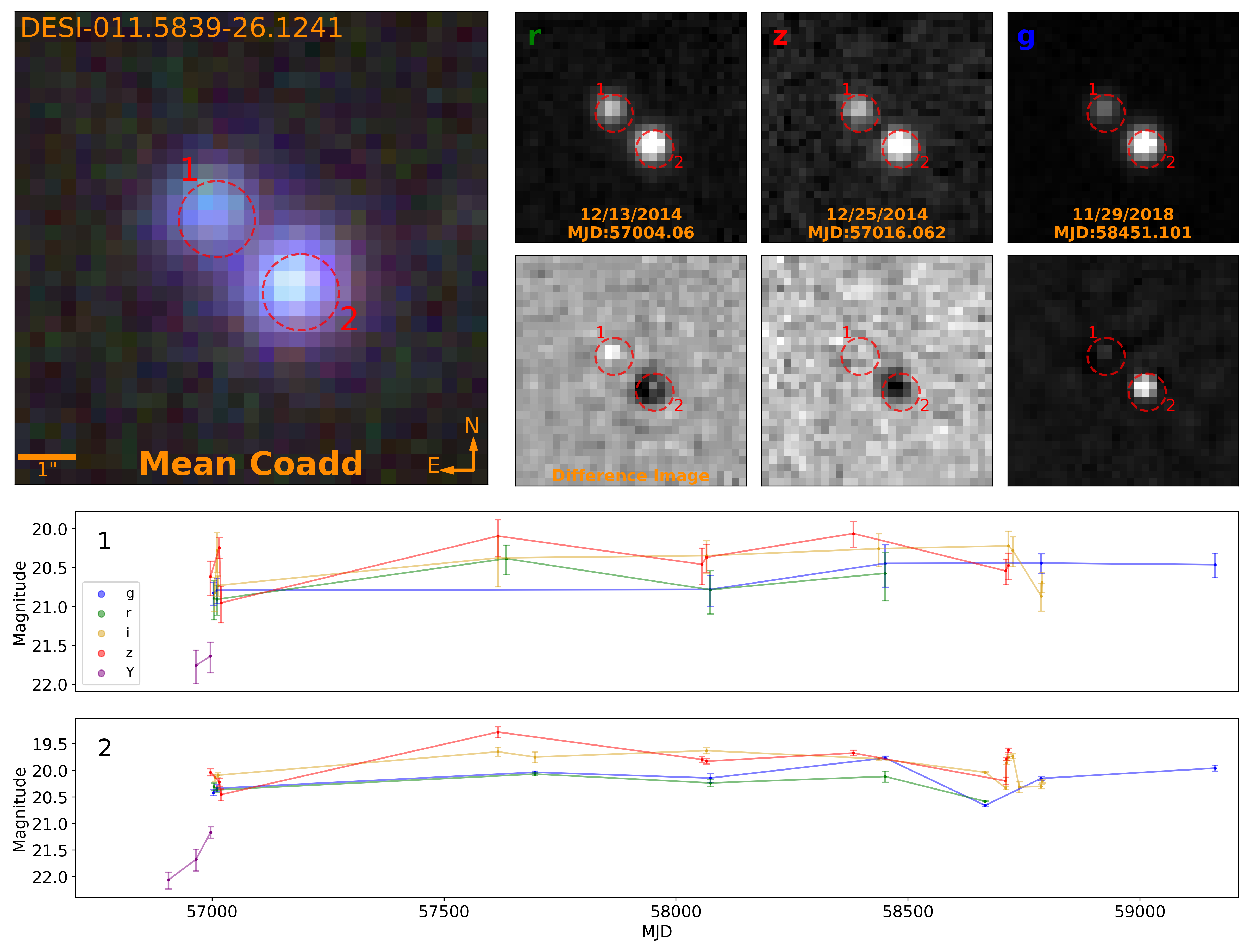}
\caption{Lensed quasar DESI-011.5839-26.1241 (see caption of Figure~\ref{454_quasars} for the full description of each subplot).}\label{108_quasars}
\end{center}
\end{figure}

\paragraph{DESI-011.5839-26.1241}\label{c6}
This system is a doubly-lensed quasar candidate (Figure~\ref{108_quasars}), initially classified as a grade B in both D22 and H23.  This system exhibits a typical double lensed quasar structure, with the lens galaxy \txo{likely} overwhelmed by the lensed quasar light.

\pagebreak

\begin{figure}[H]
\figurenum{B1.2}
\begin{center}
\includegraphics[width=161mm]{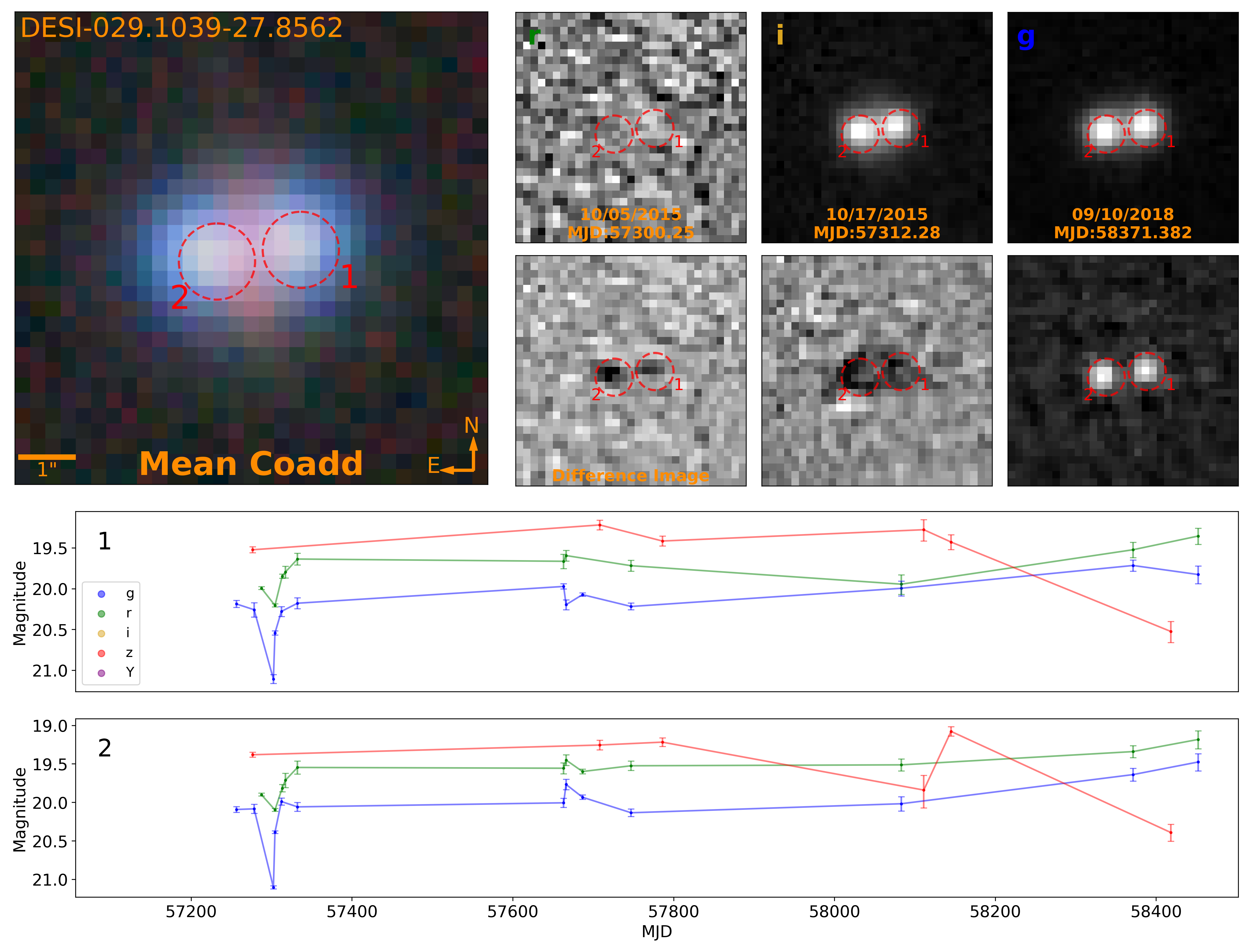}
\caption{Lensed quasar DESI-029.1039-27.8562 (see caption of Figure~\ref{454_quasars} for the full description of each subplot).}\label{10_quasars}
\end{center}
\end{figure}

\paragraph{DESI-029.1039-27.8562}\label{c1}
This system is a doubly-lensed quasar candidate (Figure~\ref{10_quasars}), initially classified as a grade A in both D22 and H23.  This system, like all the following systems, exhibits the characteristic double lensed quasar structure \txr{(see Figure~\ref{449_quasars})}.  In DESI-029.1039-27.8562, we can identify the lens galaxy (the red galaxy between the two lensed images), but this is not always possible as the lensed quasar image can sometimes overwhelm the lens galaxy.

\pagebreak

\begin{figure}[H]
\figurenum{B1.3}
\begin{center}
\includegraphics[width=161mm]{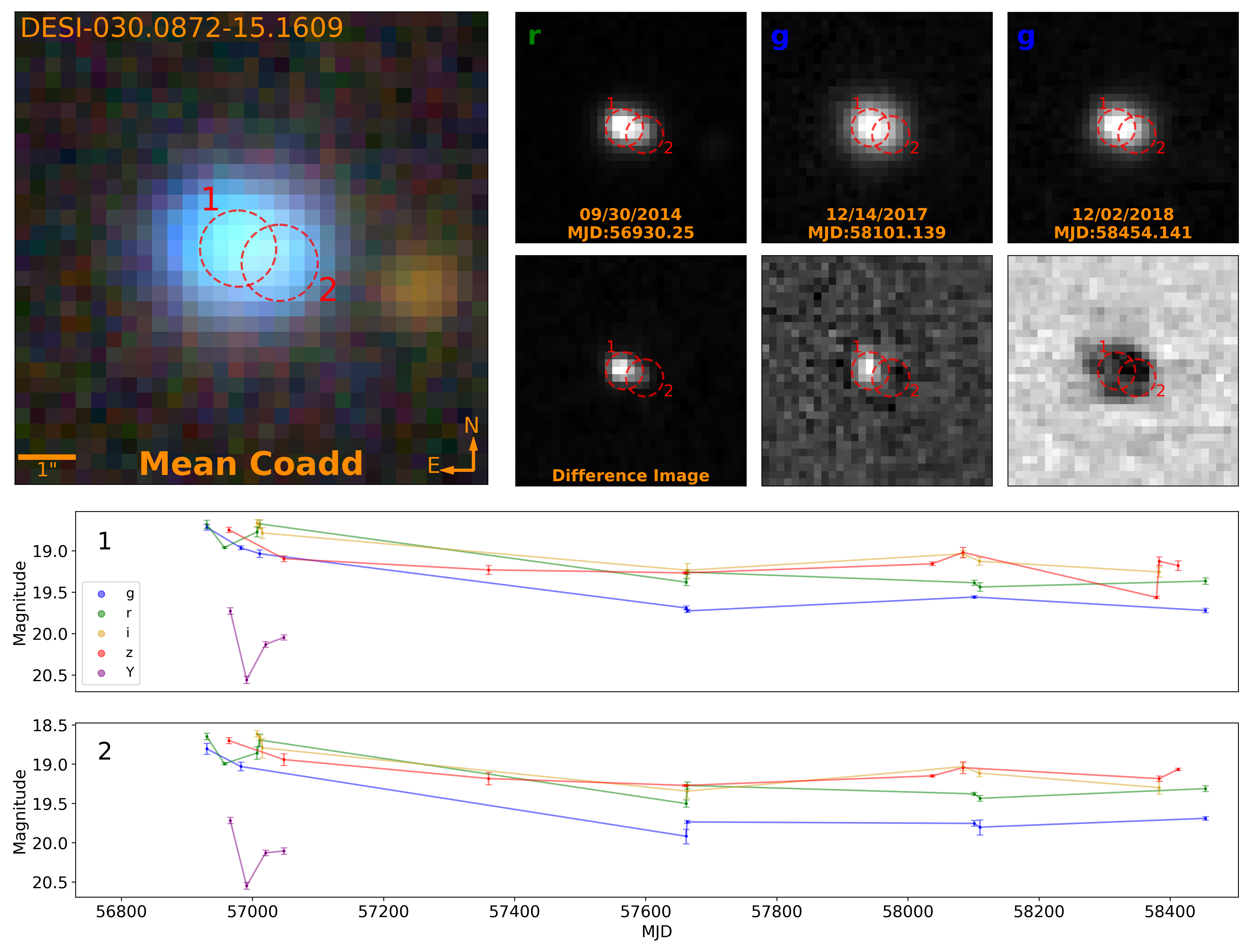}
\caption{Lensed quasar DESI-030.0872-15.1609 (see caption of Figure~\ref{454_quasars} for the full description of each subplot).}\label{11_quasars}
\end{center}
\end{figure}

\paragraph{DESI-030.0872-15.1609}\label{c2}
This system is a doubly-lensed quasar candidate (Figure~\ref{11_quasars}), initially classified as a grade A in D22.  This system also exhibits the characteristic double lensed quasar structure, but the lens galaxy cannot be discerned; it is likely that the quasar light \txr{overwhelms} the lens galaxy light.

\pagebreak

\begin{figure}[H]
\figurenum{B1.4}
\begin{center}
\includegraphics[width=161mm]{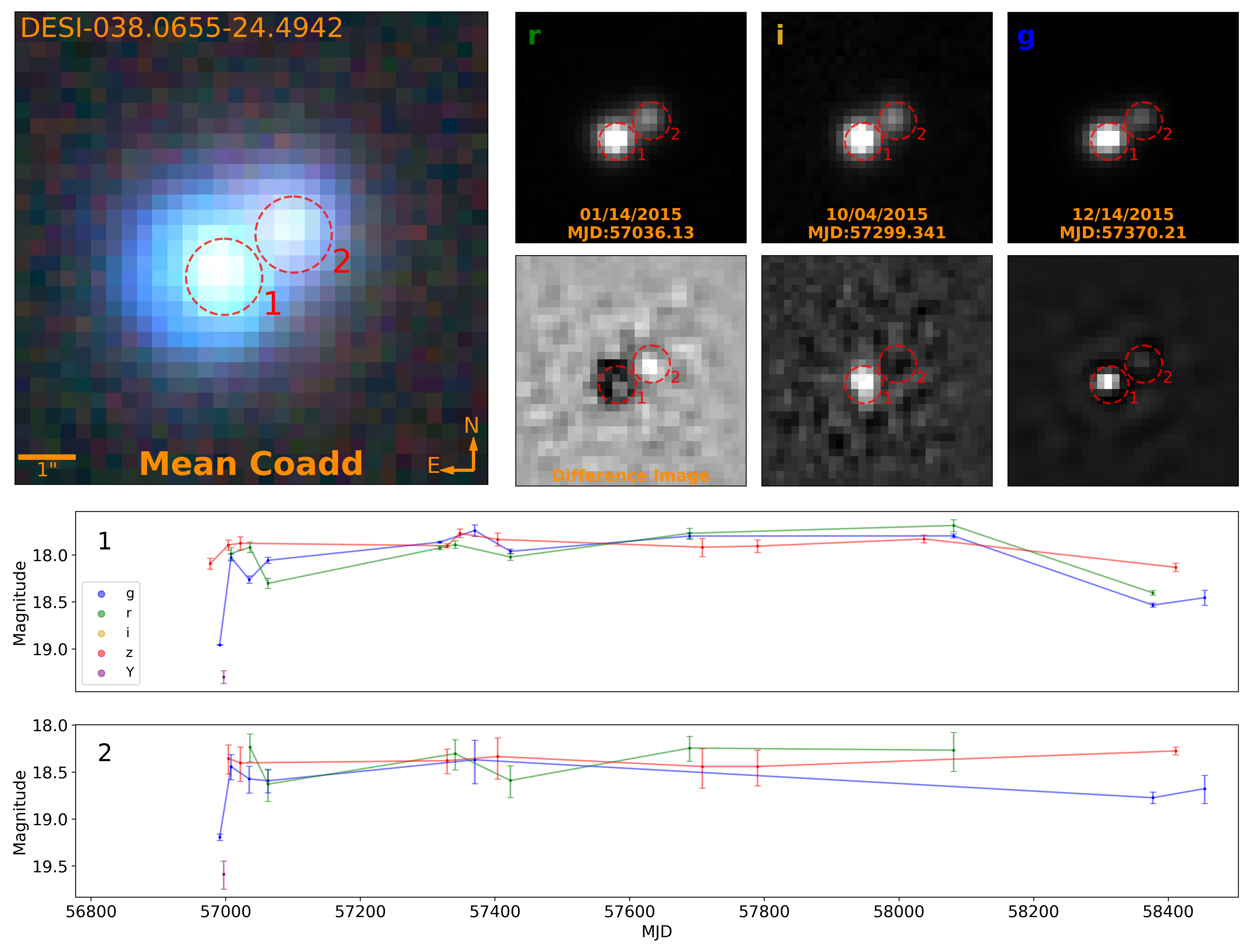}
\caption{Lensed quasar DESI-038.0655-24.4942 (see caption of Figure~\ref{454_quasars} for the full description of each subplot).}\label{18_quasars}
\end{center}
\end{figure}

\paragraph{DESI-038.0655-24.4942}\label{c3}
This system is a doubly-lensed quasar candidate (Figure~\ref{18_quasars}), initially classified as a grade A in both D22 and H23.  \txr{This system exhibits a typical double lensed quasar structure, with the lens galaxy \txo{likely} overwhelmed by the lensed quasar light.  The two images are deblended by the Tractor, despite being very close to one another.}  

\pagebreak

\begin{figure}[H]
\figurenum{B1.5}
\begin{center}
\includegraphics[width=161mm]{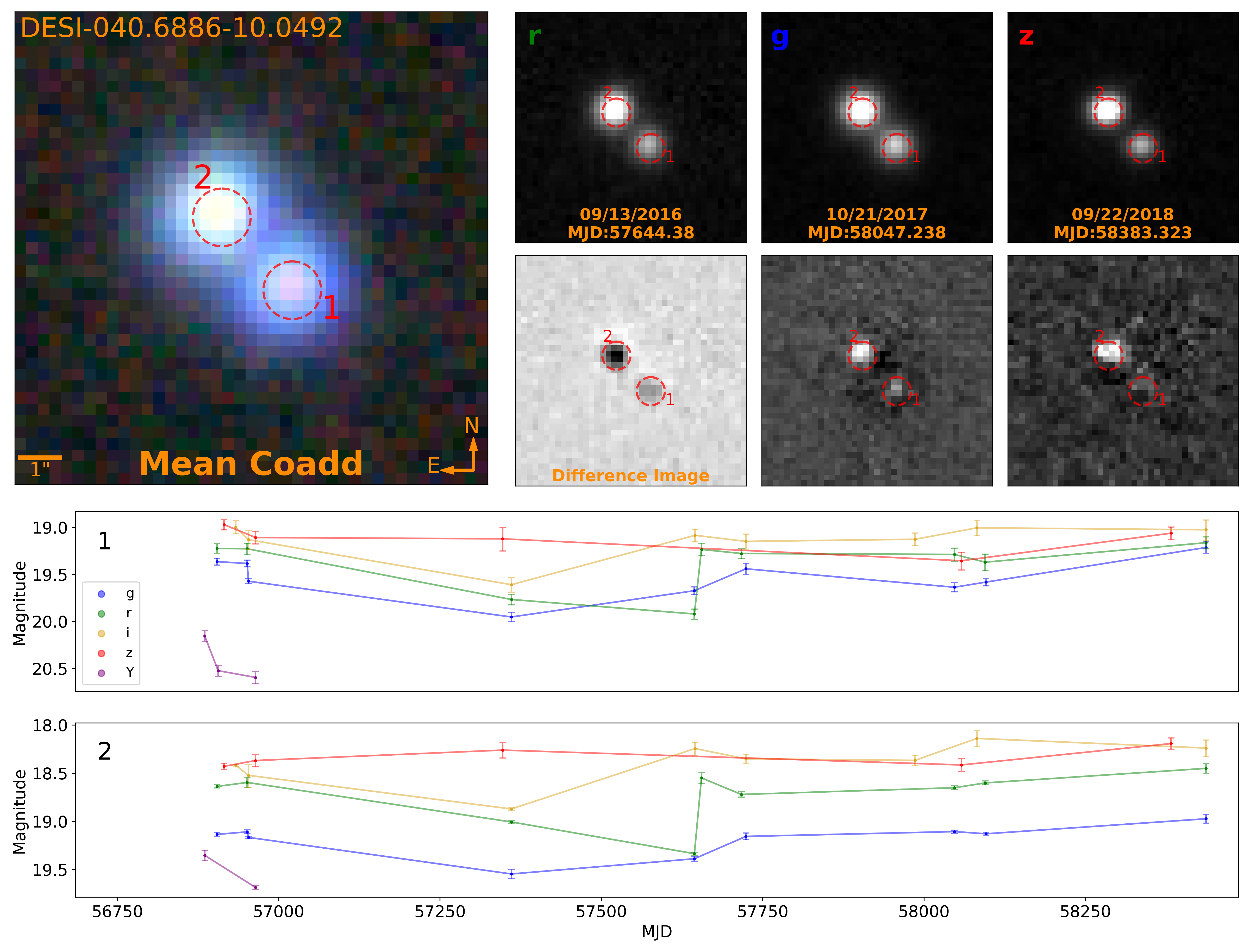}
\caption{Lensed quasar DESI-040.6886-10.0492 (see caption of Figure~\ref{454_quasars} for the full description of each subplot).}\label{254_quasars}
\end{center}
\end{figure}

\paragraph{DESI-040.6886-10.0492}\label{c10}
This system is a doubly-lensed quasar candidate (Figure~\ref{254_quasars}), initially classified as a grade C in D22 and grade A in H23.  This system exhibits a typical double lensed quasar structure, with the lens galaxy \txo{likely} overwhelmed by the lensed quasar light.

\pagebreak

\begin{figure}[H]
\figurenum{B1.6}
\begin{center}
\includegraphics[width=161mm]{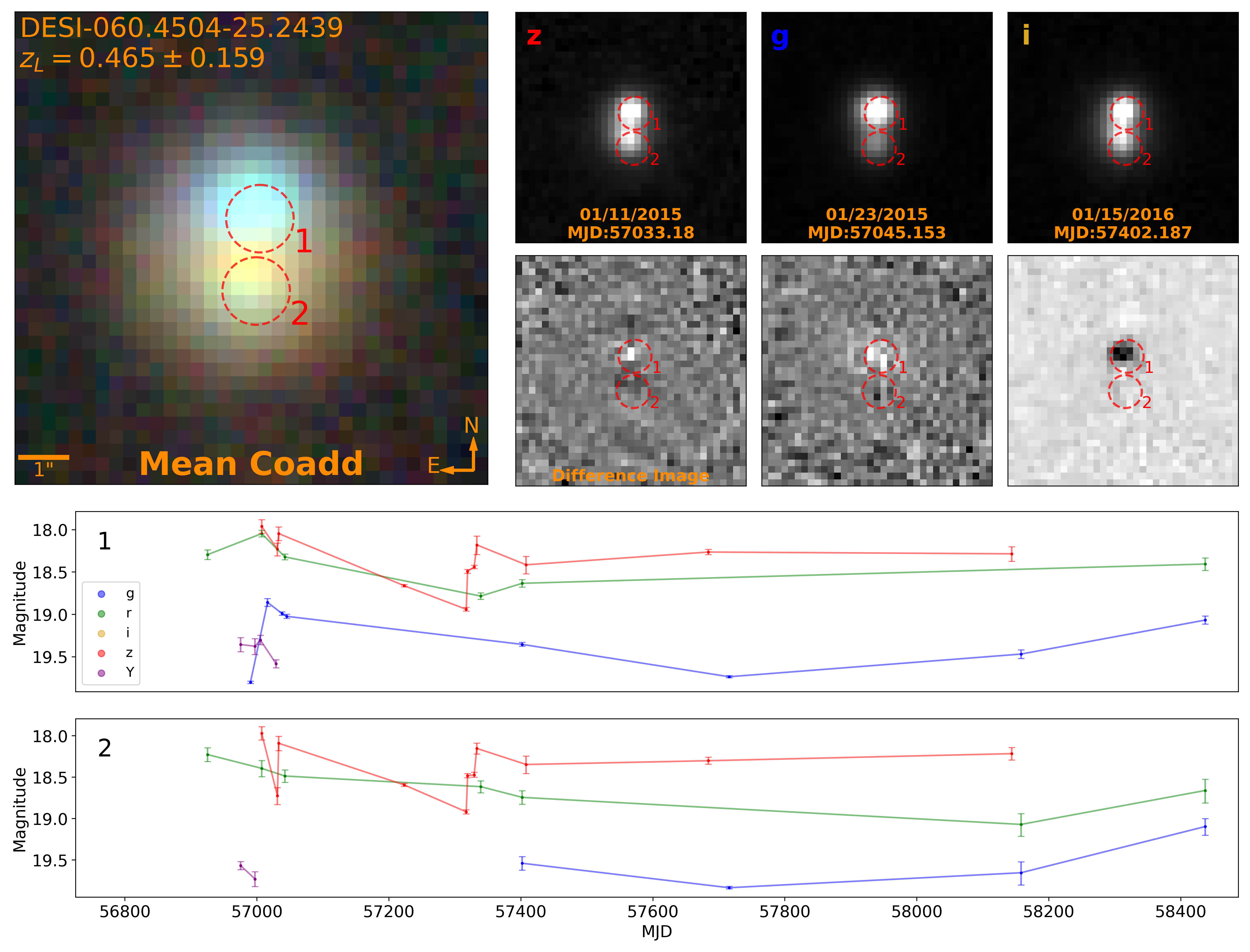}
\caption{Lensed quasar DESI-060.4504-25.2439 (see caption of Figure~\ref{454_quasars} for the full description of each subplot).}\label{25_quasars}
\end{center}
\end{figure}

\paragraph{DESI-060.4504-25.2439}\label{c4}
This system is a doubly-lensed quasar candidate (Figure~\ref{25_quasars}), initially classified as a grade A in D22.  The lens galaxy is visible as a bright red galaxy between two lensed images in this double structure.

\pagebreak

\begin{figure}[H]
\figurenum{B1.7}
\begin{center}
\includegraphics[width=161mm]{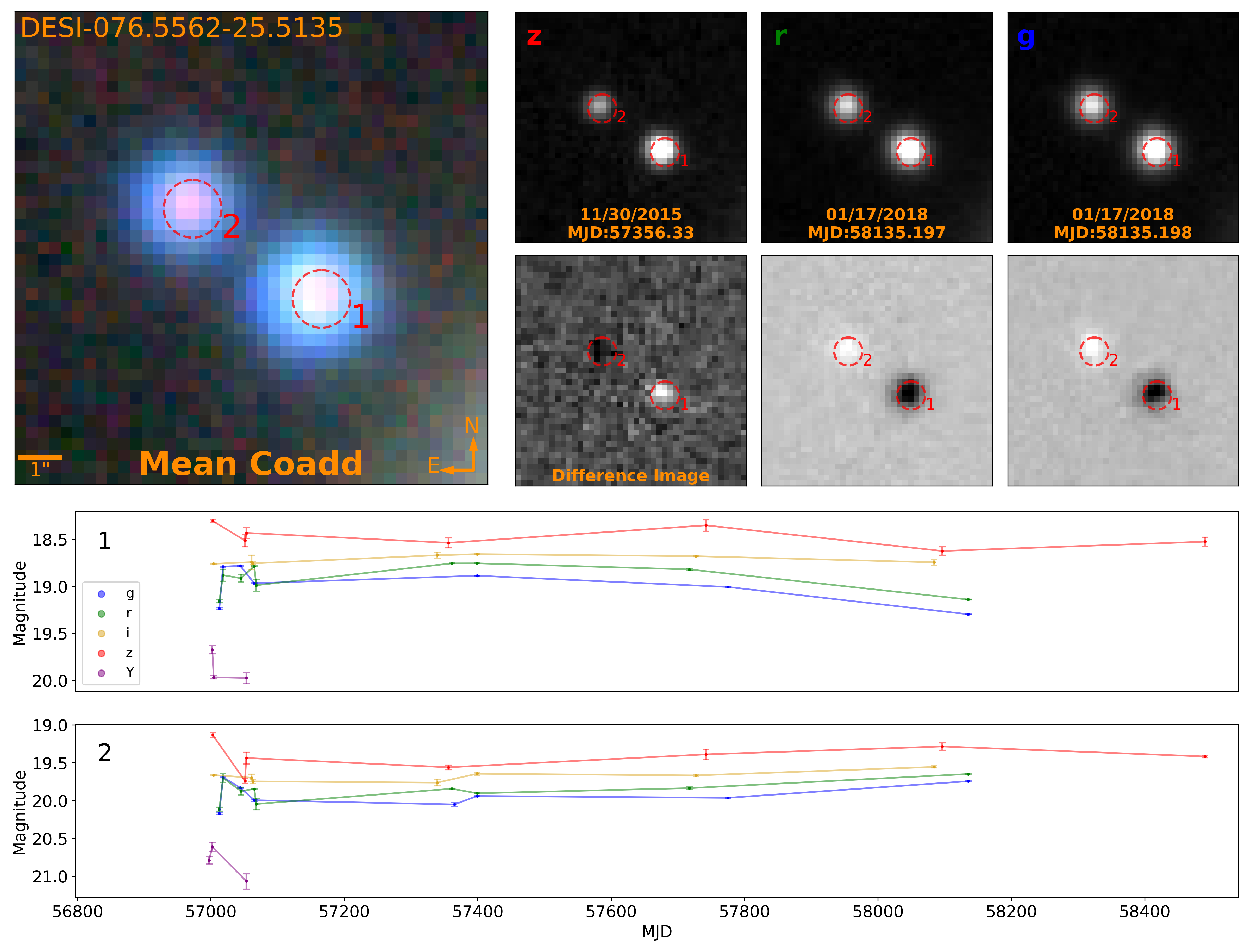}
\caption{Lensed quasar DESI-076.5562-25.5135 (see caption of Figure~\ref{454_quasars} for the full description of each subplot).}\label{144_quasars}
\end{center}
\end{figure}

\paragraph{DESI-076.5562-25.5135}\label{c7}
This system is a doubly-lensed quasar candidate (Figure~\ref{144_quasars}), initially classified as a grade B in both D22 and H23.  This system exhibits a typical double lensed quasar structure, with the lens galaxy \txo{likely} overwhelmed by the lensed quasar light.

\pagebreak

\begin{figure}[H]
\figurenum{B1.8}
\begin{center}
\includegraphics[width=161mm]{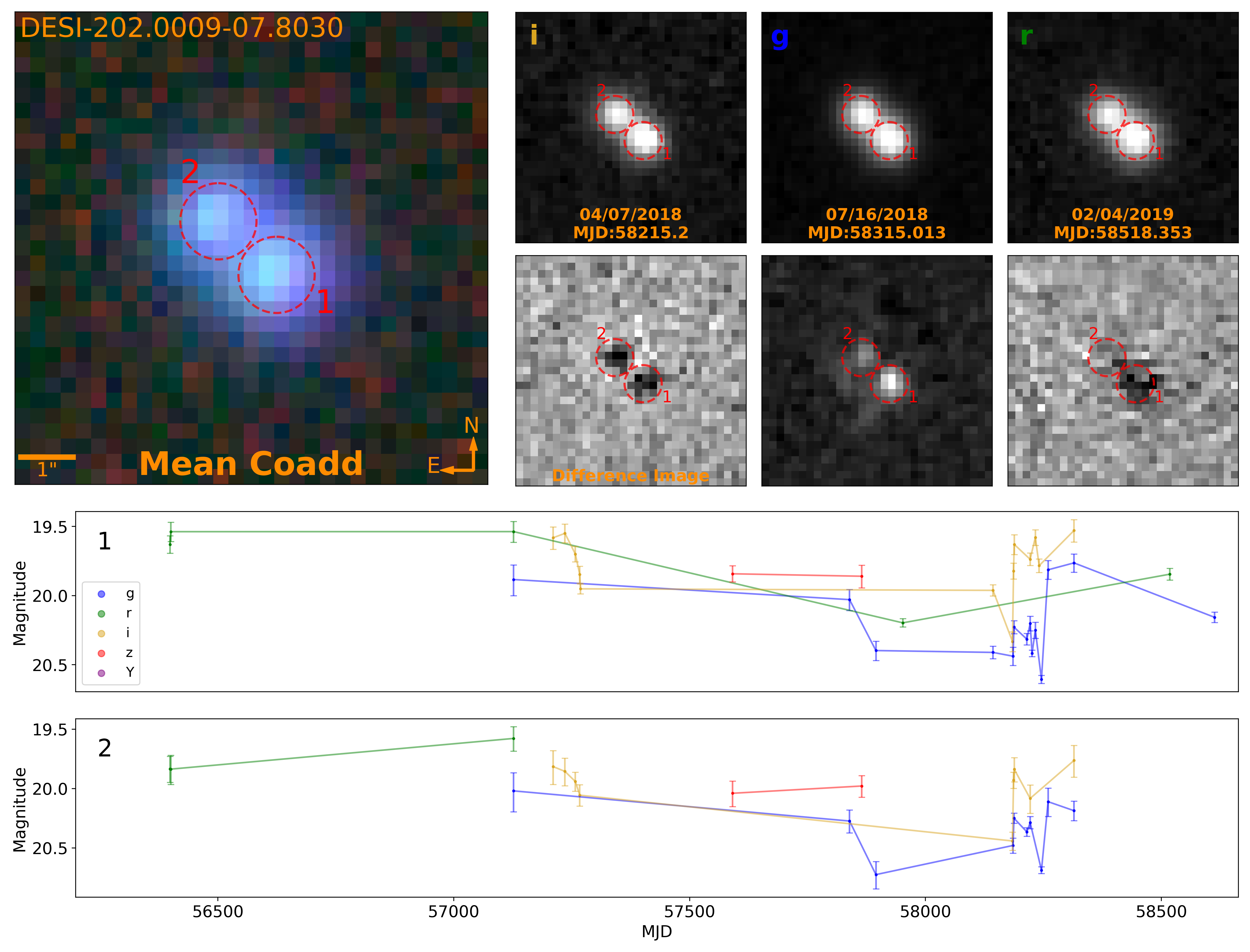}
\caption{Lensed quasar DESI-202.0009-07.8030 (see caption of Figure~\ref{454_quasars} for the full description of each subplot).}\label{71_quasars}
\end{center}
\end{figure}

\paragraph{DESI-202.0009-07.8030}\label{c5}
This system is a doubly-lensed quasar candidate (Figure~\ref{71_quasars}), initially classified as a grade A in both D22 and H23.  This system exhibits a typical double lensed quasar structure, with the lens galaxy \txo{likely} overwhelmed by the lensed quasar light.

\pagebreak

\begin{figure}[H]
\figurenum{B1.9}
\begin{center}
\includegraphics[width=161mm]{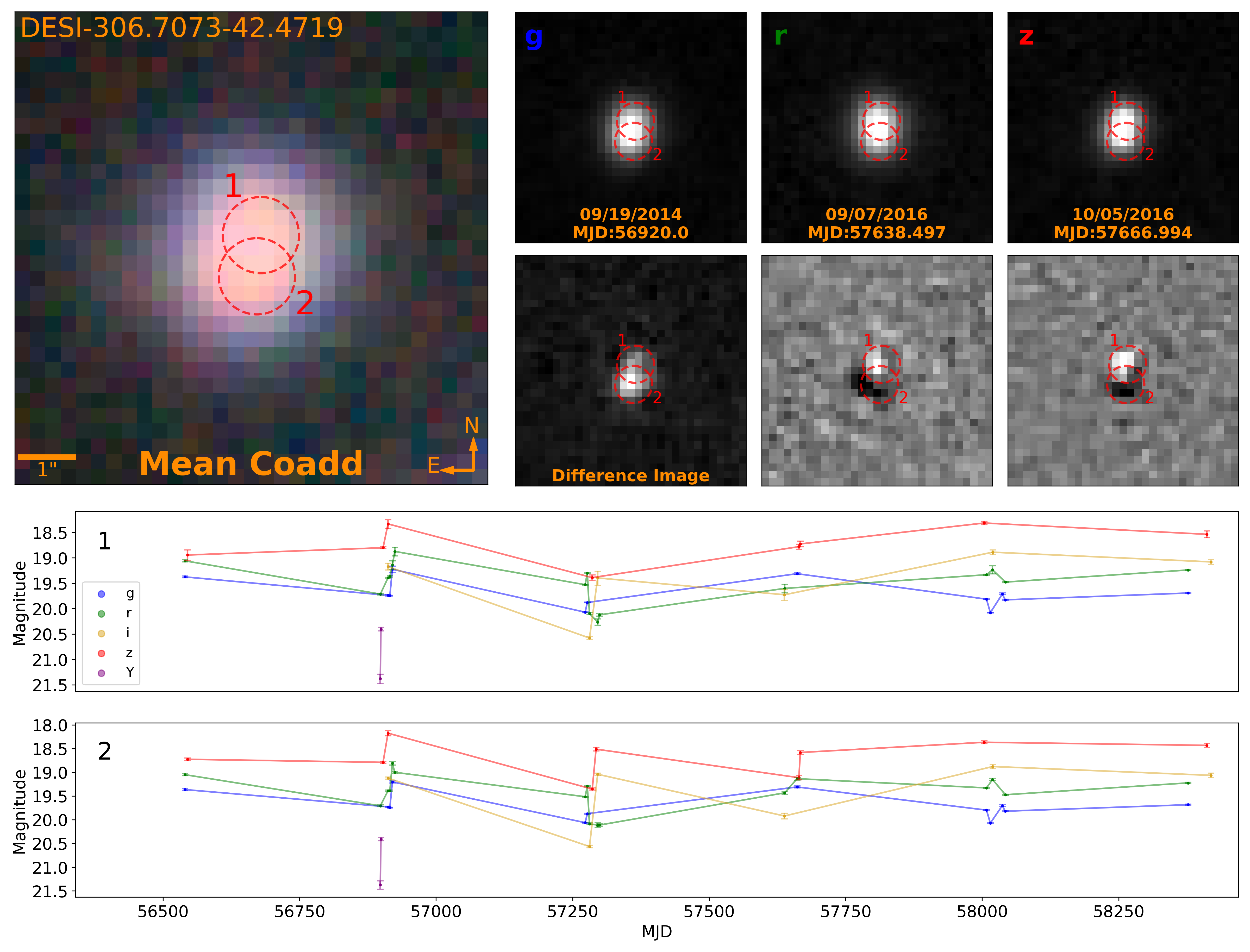}
\caption{Lensed quasar DESI-306.7073-42.4719 (see caption of Figure~\ref{454_quasars} for the full description of each subplot).}\label{205_quasars}
\end{center}
\end{figure}

\paragraph{DESI-306.7073-42.4719}\label{c8}
This system is a doubly-lensed quasar candidate (Figure~\ref{205_quasars}), initially classified as a grade B in D22.  \txr{This system exhibits a typical double lensed quasar structure, with the lens galaxy \txo{likely} overwhelmed by the lensed quasar light.  The two images are deblended by the Tractor, despite being very close to one another.}

\pagebreak

\begin{figure}[H]
\figurenum{B1.10}
\begin{center}
\includegraphics[width=161mm]{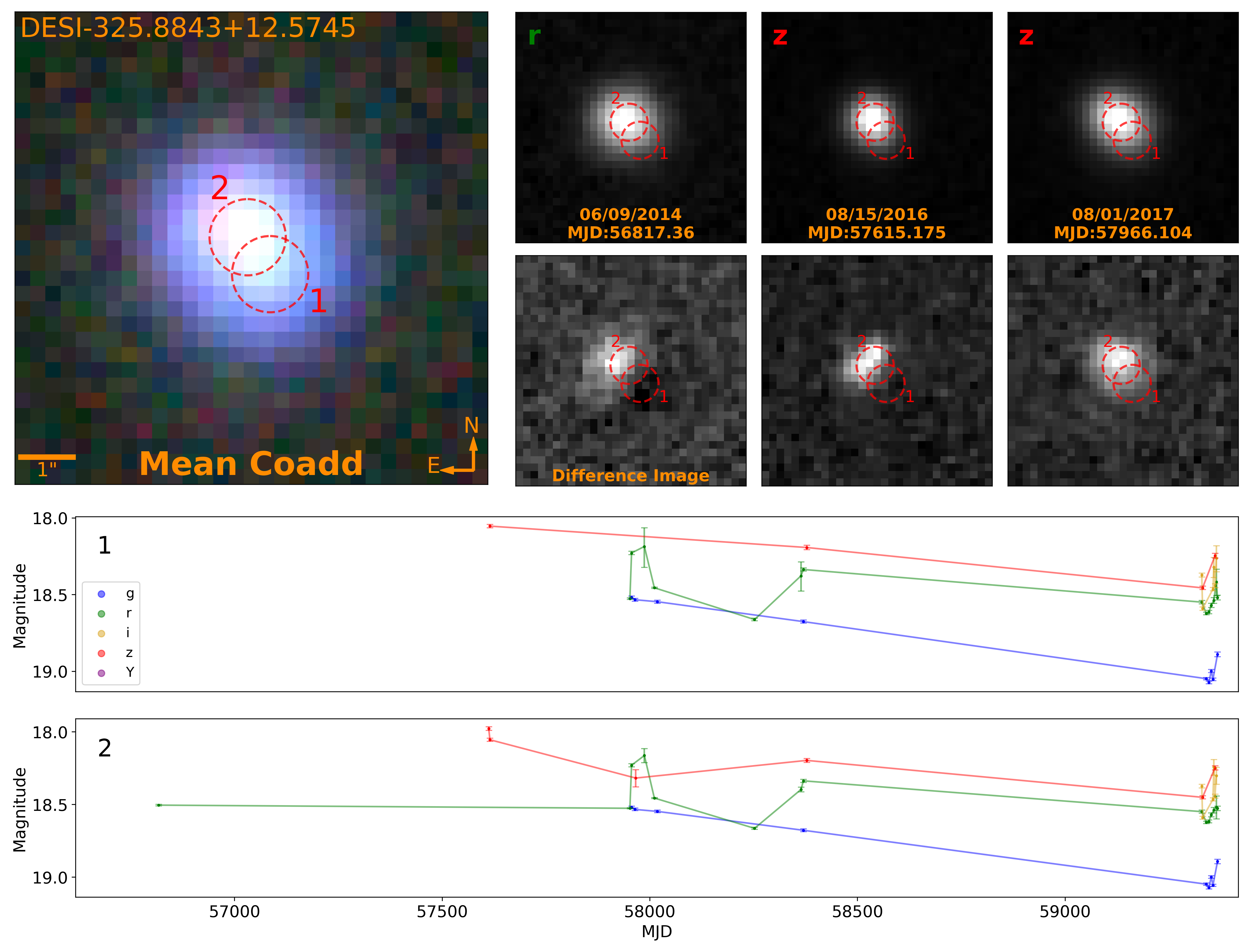}
\caption{Lensed quasar DESI-325.8843+12.5745 (see caption of Figure~\ref{454_quasars} for the full description of each subplot).}\label{406_quasars}
\end{center}
\end{figure}

\paragraph{DESI-325.8843+12.5745}\label{c11}
This system is a doubly-lensed quasar candidate (Figure~\ref{406_quasars}), initially classified as a grade C in D22 and grade A in H23.  \txr{This system exhibits a typical double lensed quasar structure, with the lens galaxy \txo{likely} overwhelmed by the lensed quasar light.  The two images are deblended by the Tractor, despite being very close to one another.}

\pagebreak

\begin{figure}[H]
\figurenum{B1.11}
\begin{center}
\includegraphics[width=161mm]{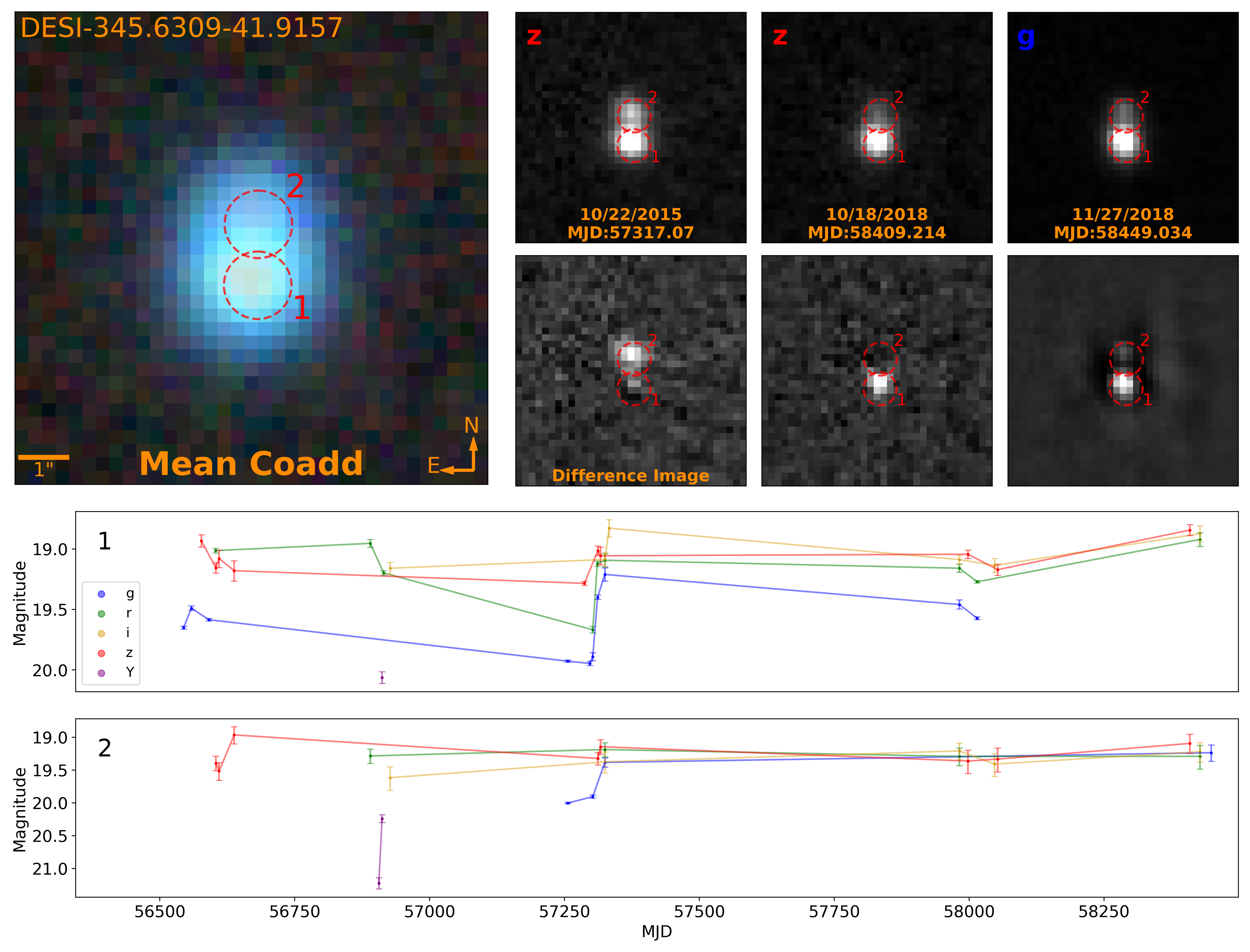}
\caption{Lensed quasar DESI-345.6309-41.9157 (see caption of Figure~\ref{454_quasars} for the full description of each subplot).}\label{425_quasars}
\end{center}
\end{figure}

\paragraph{DESI-345.6309-41.9157}\label{c12}
This system is a doubly-lensed quasar candidate (Figure~\ref{425_quasars}), initially classified as a grade C in D22 and grade A in H23.  This system exhibits a typical double lensed quasar structure, with the lens galaxy \txo{likely} overwhelmed by the lensed quasar light.

\pagebreak

\begin{figure}[H]
\figurenum{B1.12}
\begin{center}
\includegraphics[width=161mm]{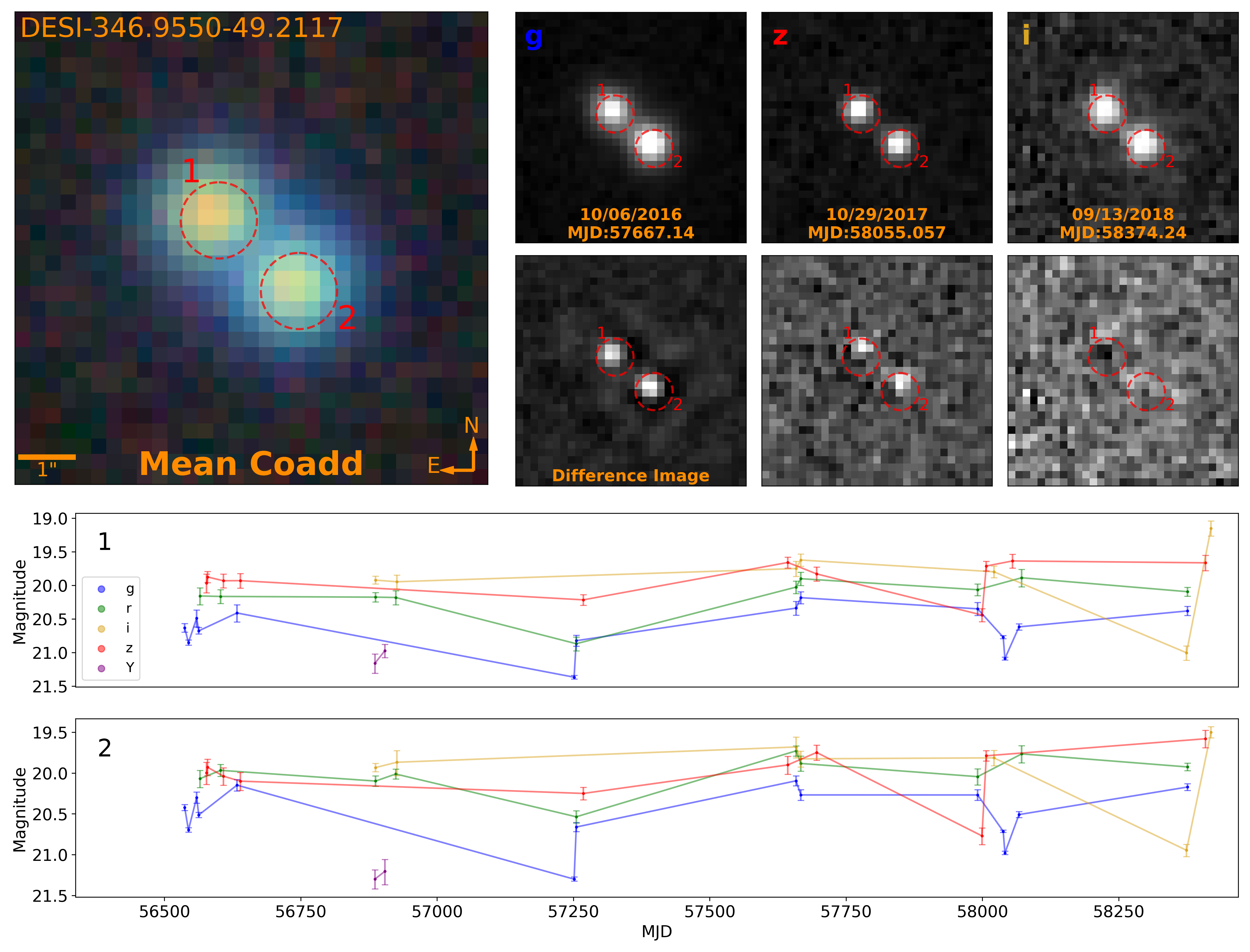}
\caption{Lensed quasar DESI-346.9550-49.2117 (see caption of Figure~\ref{454_quasars} for the full description of each subplot).}\label{216_quasars}
\end{center}
\end{figure}

\paragraph{DESI-346.9550-49.2117}\label{c9}
This system is a doubly-lensed quasar candidate (Figure~\ref{216_quasars}), initially classified as a grade B in D22.  This system exhibits a typical double lensed quasar structure, with the lens galaxy \txo{likely} overwhelmed by the lensed quasar light.

\bibliography{sample631}{}

\begin{thebibliography}{}
\expandafter\ifx\csname natexlab\endcsname\relax\def\natexlab#1{#1}\fi
\providecommand{\url}[1]{\href{#1}{#1}}
\providecommand{\dodoi}[1]{doi:~\href{http://doi.org/#1}{\nolinkurl{#1}}}
\providecommand{\doeprint}[1]{\href{http://ascl.net/#1}{\nolinkurl{http://ascl.net/#1}}}
\providecommand{\doarXiv}[1]{\href{https://arxiv.org/abs/#1}{\nolinkurl{https://arxiv.org/abs/#1}}}

\bibitem[{Agnello(2018)}]{agnello2018_single}
Agnello, A. 2018, WG021416.37-210535.3, a quadruply lensed quasar in three public surveys.
\newblock \doarXiv{1805.10296}

\bibitem[{Agnello {et~al.}(2018)Agnello, Lin, Kuropatkin, Buckley-Geer, Anguita, Schechter, Morishita, Motta, Rojas, Treu, Amara, Auger, Courbin, Fassnacht, Frieman, More, Marshall, McMahon, Meylan, Suyu, Glazebrook, Morgan, Nord, Abbott, Abdalla, Annis, Bechtol, Benoit-Lévy, Bertin, Bernstein, Brooks, Burke, Rosell, Carretero, Cunha, D’Andrea, da Costa, Desai, Drlica-Wagner, Eifler, Flaugher, García-Bellido, Gaztanaga, Gerdes, Gruen, Gruendl, Gschwend, Gutierrez, Honscheid, James, Kuehn, Lahav, Lima, Maia, March, Menanteau, Miquel, Ogando, Plazas, Sanchez, Scarpine, Schindler, Schubnell, Sevilla-Noarbe, Smith, Soares-Santos, Sobreira, Suchyta, Swanson, Tarle, Tucker, \& Wechsler}]{agnello2018}
Agnello, A., Lin, H., Kuropatkin, N., {et~al.} 2018, Monthly Notices of the Royal Astronomical Society, 479, 4345, \dodoi{10.1093/mnras/sty1419}

\bibitem[{{Astropy Collaboration} {et~al.}(2013){Astropy Collaboration}, {Robitaille}, {Tollerud}, {Greenfield}, {Droettboom}, {Bray}, {Aldcroft}, {Davis}, {Ginsburg}, {Price-Whelan}, {Kerzendorf}, {Conley}, {Crighton}, {Barbary}, {Muna}, {Ferguson}, {Grollier}, {Parikh}, {Nair}, {Unther}, {Deil}, {Woillez}, {Conseil}, {Kramer}, {Turner}, {Singer}, {Fox}, {Weaver}, {Zabalza}, {Edwards}, {Azalee Bostroem}, {Burke}, {Casey}, {Crawford}, {Dencheva}, {Ely}, {Jenness}, {Labrie}, {Lim}, {Pierfederici}, {Pontzen}, {Ptak}, {Refsdal}, {Servillat}, \& {Streicher}}]{astropy2013}
{Astropy Collaboration}, {Robitaille}, T.~P., {Tollerud}, E.~J., {et~al.} 2013, \aap, 558, A33, \dodoi{10.1051/0004-6361/201322068}

\bibitem[{{Astropy Collaboration} {et~al.}(2018){Astropy Collaboration}, {Price-Whelan}, {Sip{\H{o}}cz}, {G{\"u}nther}, {Lim}, {Crawford}, {Conseil}, {Shupe}, {Craig}, {Dencheva}, {Ginsburg}, {VanderPlas}, {Bradley}, {P{\'e}rez-Su{\'a}rez}, {de Val-Borro}, {Aldcroft}, {Cruz}, {Robitaille}, {Tollerud}, {Ardelean}, {Babej}, {Bach}, {Bachetti}, {Bakanov}, {Bamford}, {Barentsen}, {Barmby}, {Baumbach}, {Berry}, {Biscani}, {Boquien}, {Bostroem}, {Bouma}, {Brammer}, {Bray}, {Breytenbach}, {Buddelmeijer}, {Burke}, {Calderone}, {Cano Rodr{\'\i}guez}, {Cara}, {Cardoso}, {Cheedella}, {Copin}, {Corrales}, {Crichton}, {D'Avella}, {Deil}, {Depagne}, {Dietrich}, {Donath}, {Droettboom}, {Earl}, {Erben}, {Fabbro}, {Ferreira}, {Finethy}, {Fox}, {Garrison}, {Gibbons}, {Goldstein}, {Gommers}, {Greco}, {Greenfield}, {Groener}, {Grollier}, {Hagen}, {Hirst}, {Homeier}, {Horton}, {Hosseinzadeh}, {Hu}, {Hunkeler}, {Ivezi{\'c}}, {Jain}, {Jenness}, {Kanarek}, {Kendrew}, {Kern}, {Kerzendorf}, {Khvalko}, {King}, {Kirkby}, {Kulkarni},
  {Kumar}, {Lee}, {Lenz}, {Littlefair}, {Ma}, {Macleod}, {Mastropietro}, {McCully}, {Montagnac}, {Morris}, {Mueller}, {Mumford}, {Muna}, {Murphy}, {Nelson}, {Nguyen}, {Ninan}, {N{\"o}the}, {Ogaz}, {Oh}, {Parejko}, {Parley}, {Pascual}, {Patil}, {Patil}, {Plunkett}, {Prochaska}, {Rastogi}, {Reddy Janga}, {Sabater}, {Sakurikar}, {Seifert}, {Sherbert}, {Sherwood-Taylor}, {Shih}, {Sick}, {Silbiger}, {Singanamalla}, {Singer}, {Sladen}, {Sooley}, {Sornarajah}, {Streicher}, {Teuben}, {Thomas}, {Tremblay}, {Turner}, {Terr{\'o}n}, {van Kerkwijk}, {de la Vega}, {Watkins}, {Weaver}, {Whitmore}, {Woillez}, {Zabalza}, \& {Astropy Contributors}}]{astropy2018}
{Astropy Collaboration}, {Price-Whelan}, A.~M., {Sip{\H{o}}cz}, B.~M., {et~al.} 2018, \aj, 156, 123, \dodoi{10.3847/1538-3881/aabc4f}

\bibitem[{Barbary(2014)}]{sncosmo}
Barbary, K. 2014, sncosmo v0.4.2,  Zenodo, \dodoi{10.5281/zenodo.11938}

\bibitem[{{Barbary}(2018)}]{sep}
{Barbary}, K. 2018, {SEP: Source Extraction and Photometry}, Astrophysics Source Code Library, record ascl:1811.004.
\newblock \doeprint{1811.004}

\bibitem[{{Bertin} \& {Arnouts}(1996)}]{bertin1996}
{Bertin}, E., \& {Arnouts}, S. 1996, \aaps, 117, 393, \dodoi{10.1051/aas:1996164}

\bibitem[{Bramich(2008)}]{bramich2008}
Bramich, D.~M. 2008, Monthly Notices of the Royal Astronomical Society: Letters, 386, L77, \dodoi{10.1111/j.1745-3933.2008.00464.x}

\bibitem[{{Carrasco} {et~al.}(2017){Carrasco}, {Barrientos}, {Anguita}, {Garcia-Vergara}, {Bayliss}, {Gladders}, {Gilbank}, {Yee}, \& {West}}]{carrasco2017}
{Carrasco}, M., {Barrientos}, L.~F., {Anguita}, T., {et~al.} 2017, VizieR Online Data Catalog, J/ApJ/834/210

\bibitem[{Chen {et~al.}(2022)Chen, Kelly, Oguri, Broadhurst, Diego, Emami, Filippenko, Treu, \& Zitrin}]{kellysn2}
Chen, W., Kelly, P.~L., Oguri, M., {et~al.} 2022, Nature, 611, 256—259, \dodoi{10.1038/s41586-022-05252-5}

\bibitem[{Dahle {et~al.}(2015)Dahle, Gladders, Sharon, Bayliss, \& Rigby}]{dahle2015}
Dahle, H., Gladders, M.~D., Sharon, K., Bayliss, M.~B., \& Rigby, J.~R. 2015, The Astrophysical Journal, 813, 67, \dodoi{10.1088/0004-637x/813/1/67}

\bibitem[{{Dahle} {et~al.}(2013){Dahle}, {Gladders}, {Sharon}, {Bayliss}, {Wuyts}, {Abramson}, {Koester}, {Groeneboom}, {Brinckmann}, {Kristensen}, {Lindholmer}, {Nielsen}, {Krogager}, \& {Fynbo}}]{dahle2013}
{Dahle}, H., {Gladders}, M.~D., {Sharon}, K., {et~al.} 2013, \apj, 773, 146, \dodoi{10.1088/0004-637X/773/2/146}

\bibitem[{{Dawes} {et~al.}(2023){Dawes}, {Storfer}, {Huang}, {Aldering}, {Cikota}, {Dey}, \& {Schlegel}}]{dawes2022}
{Dawes}, C., {Storfer}, C., {Huang}, X., {et~al.} 2023, \apjs, 269, 61, \dodoi{10.3847/1538-4365/ad015a}

\bibitem[{{DESI Collaboration} {et~al.}(2016){DESI Collaboration}, Aghamousa, Aguilar, Ahlen, Alam, Allen, Prieto, Annis, Bailey, Balland, Ballester, Baltay, Beaufore, Bebek, Beers, Bell, Bernal, Besuner, Beutler, Blake, Bleuler, Blomqvist, Blum, Bolton, Briceno, Brooks, Brownstein, Buckley-Geer, Burden, Burtin, Busca, Cahn, Cai, Cardiel-Sas, Carlberg, Carton, Casas, Castander, Cervantes-Cota, Claybaugh, Close, Coker, Cole, Comparat, Cooper, Cousinou, Crocce, Cuby, Cunningham, Davis, Dawson, de~la Macorra, Vicente, Delubac, Derwent, Dey, Dhungana, Ding, Doel, Duan, Ealet, Edelstein, Eftekharzadeh, Eisenstein, Elliott, Escoffier, Evatt, Fagrelius, Fan, Fanning, Farahi, Farihi, Favole, Feng, Fernandez, Findlay, Finkbeiner, Fitzpatrick, Flaugher, Flender, Font-Ribera, Forero-Romero, Fosalba, Frenk, Fumagalli, Gaensicke, Gallo, Garcia-Bellido, Gaztanaga, Fusillo, Gerard, Gershkovich, Giannantonio, Gillet, de~Rivera, Gonzalez-Perez, Gott, Graur, Gutierrez, Guy, Habib, Heetderks, Heetderks, Heitmann, Hellwing,
  Herrera, Ho, Holland, Honscheid, Huff, Hutchinson, Huterer, Hwang, Laguna, Ishikawa, Jacobs, Jeffrey, Jelinsky, Jennings, Jiang, Jimenez, Johnson, Joyce, Jullo, Juneau, Kama, Karcher, Karkar, Kehoe, Kennamer, Kent, Kilbinger, Kim, Kirkby, Kisner, Kitanidis, Kneib, Koposov, Kovacs, Koyama, Kremin, Kron, Kronig, Kueter-Young, Lacey, Lafever, Lahav, Lambert, Lampton, Landriau, Lang, Lauer, Goff, Guillou, Suu, Lee, Lee, Leitner, Lesser, Levi, L'Huillier, Li, Liang, Lin, Linder, Loebman, Lukić, Ma, MacCrann, Magneville, Makarem, Manera, Manser, Marshall, Martini, Massey, Matheson, McCauley, McDonald, McGreer, Meisner, Metcalfe, Miller, Miquel, Moustakas, Myers, Naik, Newman, Nichol, Nicola, da~Costa, Nie, Niz, Norberg, Nord, Norman, Nugent, O'Brien, Oh, Olsen, Padilla, Padmanabhan, Padmanabhan, Palanque-Delabrouille, Palmese, Pappalardo, Pâris, Park, Patej, Peacock, Peiris, Peng, Percival, Perruchot, Pieri, Pogge, Pollack, Poppett, Prada, Prakash, Probst, Rabinowitz, Raichoor, Ree, Refregier, Regal, Reid,
  Reil, Rezaie, Rockosi, Roe, Ronayette, Roodman, Ross, Ross, Rossi, Rozo, Ruhlmann-Kleider, Rykoff, Sabiu, Samushia, Sanchez, Sanchez, Schlegel, Schneider, Schubnell, Secroun, Seljak, Seo, Serrano, Shafieloo, Shan, Sharples, Sholl, Shourt, Silber, Silva, Sirk, Slosar, Smith, Smoot, Som, Song, Sprayberry, Staten, Stefanik, Tarle, Tie, Tinker, Tojeiro, Valdes, Valenzuela, Valluri, Vargas-Magana, Verde, Walker, Wang, Wang, Weaver, Weaverdyck, Wechsler, Weinberg, White, Yang, Yeche, Zhang, Zhao, Zheng, Zhou, Zhou, Zhu, Zou, \& Zu}]{desi2016}
{DESI Collaboration}, Aghamousa, A., Aguilar, J., {et~al.} 2016, The DESI Experiment Part I: Science,Targeting, and Survey Design.
\newblock \doarXiv{1611.00036}

\bibitem[{Dey {et~al.}(2019)Dey, Schlegel, Lang, Blum, Burleigh, Fan, Findlay, Finkbeiner, Herrera, Juneau, Landriau, Levi, McGreer, Meisner, Myers, Moustakas, Nugent, Patej, Schlafly, Walker, Valdes, Weaver, Y{\`{e}}che, Zou, Zhou, Abareshi, Abbott, Abolfathi, Aguilera, Alam, Allen, Alvarez, Annis, Ansarinejad, Aubert, Beechert, Bell, BenZvi, Beutler, Bielby, Bolton, Brice{\~{n}}o, Buckley-Geer, Butler, Calamida, Carlberg, Carter, Casas, Castander, Choi, Comparat, Cukanovaite, Delubac, DeVries, Dey, Dhungana, Dickinson, Ding, Donaldson, Duan, Duckworth, Eftekharzadeh, Eisenstein, Etourneau, Fagrelius, Farihi, Fitzpatrick, Font-Ribera, Fulmer, Gänsicke, Gaztanaga, George, Gerdes, Gontcho, Gorgoni, Green, Guy, Harmer, Hernandez, Honscheid, Huang, James, Jannuzi, Jiang, Joyce, Karcher, Karkar, Kehoe, Jean-Paul, Kueter-Young, Lan, Lauer, Guillou, Suu, Lee, Lesser, Levasseur, Li, Mann, Marshall, Mart{\'{\i}}nez-V{\'{a}}zquez, Martini, du~Mas~des Bourboux, McManus, Meier, M{\'{e}}nard, Metcalfe,
  Mu{\~{n}}oz-Guti{\'{e}}rrez, Najita, Napier, Narayan, Newman, Nie, Nord, Norman, Olsen, Paat, Palanque-Delabrouille, Peng, Poppett, Poremba, Prakash, Rabinowitz, Raichoor, Rezaie, Robertson, Roe, Ross, Ross, Rudnick, Safonova, Saha, S{\'{a}}nchez, Savary, Schweiker, Scott, Seo, Shan, Silva, Slepian, Soto, Sprayberry, Staten, Stillman, Stupak, Summers, Tie, Tirado, Vargas-Maga{\~{n}}a, Vivas, Wechsler, Williams, Yang, Yang, Yapici, Zaritsky, Zenteno, Zhang, Zhang, Zhou, \& Zhou}]{dr9}
Dey, A., Schlegel, D.~J., Lang, D., {et~al.} 2019, The Astronomical Journal, 157, 168, \dodoi{10.3847/1538-3881/ab089d}

\bibitem[{{Diehl} {et~al.}(2017){Diehl}, {Buckley-Geer}, {Lindgren}, {Nord}, {Gaitsch}, {Gaitsch}, {Lin}, {Allam}, {Collett}, {Furlanetto}, {Gill}, {More}, {Nightingale}, {Odden}, {Pellico}, {Tucker}, {da Costa}, {Fausti Neto}, {Kuropatkin}, {Soares-Santos}, {Welch}, {Zhang}, {Frieman}, {Abdalla}, {Annis}, {Benoit-L{\'e}vy}, {Bertin}, {Brooks}, {Burke}, {Carnero Rosell}, {Carrasco Kind}, {Carretero}, {Cunha}, {D'Andrea}, {Desai}, {Dietrich}, {Drlica-Wagner}, {Evrard}, {Finley}, {Flaugher}, {Garc{\'\i}a-Bellido}, {Gerdes}, {Goldstein}, {Gruen}, {Gruendl}, {Gschwend}, {Gutierrez}, {James}, {Kuehn}, {Kuhlmann}, {Lahav}, {Li}, {Lima}, {Maia}, {Marshall}, {Menanteau}, {Miquel}, {Nichol}, {Nugent}, {Ogando}, {Plazas}, {Reil}, {Romer}, {Sako}, {Sanchez}, {Santiago}, {Scarpine}, {Schindler}, {Schubnell}, {Sevilla-Noarbe}, {Sheldon}, {Smith}, {Sobreira}, {Suchyta}, {Swanson}, {Tarle}, {Thomas}, {Walker}, \& {DES Collaboration}}]{diehl2017}
{Diehl}, H.~T., {Buckley-Geer}, E.~J., {Lindgren}, K.~A., {et~al.} 2017, The Astrophysical Journal Supplement, 232, 15, \dodoi{10.3847/1538-4365/aa8667}

\bibitem[{Dux {et~al.}(2023)Dux, Lemon, Courbin, Neira, Anguita, Galan, Kim, Hempel, Hempel, \& Lachaume}]{dux2023}
Dux, F., Lemon, C., Courbin, F., {et~al.} 2023, Nine lensed quasars and quasar pairs discovered through spatially-extended variability in Pan-STARRS.
\newblock \doarXiv{2307.13729}

\bibitem[{Fawcett {et~al.}(2020)Fawcett, Alexander, Rosario, Klindt, Fotopoulou, Lusso, Morabito, \& Calistro Rivera}]{fawcett2020}
Fawcett, V.~A., Alexander, D.~M., Rosario, D.~J., {et~al.} 2020, Monthly Notices of the Royal Astronomical Society, 494, 4802, \dodoi{10.1093/mnras/staa954}

\bibitem[{{Flaugher} {et~al.}(2015){Flaugher}, {Diehl}, {Honscheid}, {Abbott}, {Alvarez}, {Angstadt}, {Annis}, {Antonik}, {Ballester}, {Beaufore}, {Bernstein}, {Bernstein}, {Bigelow}, {Bonati}, {Boprie}, {Brooks}, {Buckley-Geer}, {Campa}, {Cardiel-Sas}, {Castander}, {Castilla}, {Cease}, {Cela-Ruiz}, {Chappa}, {Chi}, {Cooper}, {da Costa}, {Dede}, {Derylo}, {DePoy}, {de Vicente}, {Doel}, {Drlica-Wagner}, {Eiting}, {Elliott}, {Emes}, {Estrada}, {Fausti Neto}, {Finley}, {Flores}, {Frieman}, {Gerdes}, {Gladders}, {Gregory}, {Gutierrez}, {Hao}, {Holland}, {Holm}, {Huffman}, {Jackson}, {James}, {Jonas}, {Karcher}, {Karliner}, {Kent}, {Kessler}, {Kozlovsky}, {Kron}, {Kubik}, {Kuehn}, {Kuhlmann}, {Kuk}, {Lahav}, {Lathrop}, {Lee}, {Levi}, {Lewis}, {Li}, {Mandrichenko}, {Marshall}, {Martinez}, {Merritt}, {Miquel}, {Mu{\~n}oz}, {Neilsen}, {Nichol}, {Nord}, {Ogando}, {Olsen}, {Palaio}, {Patton}, {Peoples}, {Plazas}, {Rauch}, {Reil}, {Rheault}, {Roe}, {Rogers}, {Roodman}, {Sanchez}, {Scarpine}, {Schindler}, {Schmidt},
  {Schmitt}, {Schubnell}, {Schultz}, {Schurter}, {Scott}, {Serrano}, {Shaw}, {Smith}, {Soares-Santos}, {Stefanik}, {Stuermer}, {Suchyta}, {Sypniewski}, {Tarle}, {Thaler}, {Tighe}, {Tran}, {Tucker}, {Walker}, {Wang}, {Watson}, {Weaverdyck}, {Wester}, {Woods}, {Yanny}, \& {DES Collaboration}}]{flaugher2015}
{Flaugher}, B., {Diehl}, H.~T., {Honscheid}, K., {et~al.} 2015, \aj, 150, 150, \dodoi{10.1088/0004-6256/150/5/150}

\bibitem[{Freedman(2021)}]{freedman2021}
Freedman, W.~L. 2021, The Astrophysical Journal, 919, 16, \dodoi{10.3847/1538-4357/ac0e95}

\bibitem[{{Frye} {et~al.}(2023){Frye}, {Pascale}, {Cohen}, {Summers}, {Foo}, {Kamieneski}, {Carleton}, {Jansen}, {Pierel}, {Engesser}, {Chen}, {Austin}, {Marshall}, {Trussler}, {Meena}, {Leimbach}, {Garuda}, {Honor}, {Furtak}, {Strolger}, {Windhorst}, {Koekemoer}, {Zitrin}, {Diego}, {Kelly}, {Coe}, {Conselice}, {Dai}, {D{\^a}Silva}, {Dole}, {Driver}, {Grogin}, {Nonino}, {Pirzkal}, {Polletta}, {Robotham}, {Rutkowski}, {Ryan}, {Tompkins}, {Willmer}, {Willner}, {Yan}, \& {Yun}}]{snh0pe}
{Frye}, B., {Pascale}, M., {Cohen}, S., {et~al.} 2023, Transient Name Server AstroNote, 96, 1

\bibitem[{Goobar {et~al.}(2017)Goobar, Amanullah, Kulkarni, Nugent, Johansson, Steidel, Law, Mörtsell, Quimby, Blagorodnova, \& et~al.}]{goobar}
Goobar, A., Amanullah, R., Kulkarni, S.~R., {et~al.} 2017, Science, 356, 291–295, \dodoi{10.1126/science.aal2729}

\bibitem[{{Goobar} {et~al.}(2022){Goobar}, {Johansson}, {Dhawan}, {Schulze}, {Arendse}, {Carracedo}, {Joseph}, {Nordin}, \& {Townsend}}]{astronotegoobar}
{Goobar}, A., {Johansson}, J., {Dhawan}, S., {et~al.} 2022, Transient Name Server AstroNote, 180, 1

\bibitem[{Green {et~al.}(2022)Green, Pulgarin-Duque, Anderson, MacLeod, Eracleous, Ruan, Runnoe, Graham, Roulston, Schneider, Ahlf, Bizyaev, Brownstein, del Casal, Dodd, Hoover, Matt, Merloni, Pan, Ramirez, Ridder, \& Moseley}]{green2022}
Green, P.~J., Pulgarin-Duque, L., Anderson, S.~F., {et~al.} 2022, The Astrophysical Journal, 933, 180, \dodoi{10.3847/1538-4357/ac743f}

\bibitem[{Harris {et~al.}(2020)Harris, Millman, van~der Walt, Gommers, Virtanen, Cournapeau, Wieser, Taylor, Berg, Smith, Kern, Picus, Hoyer, van Kerkwijk, Brett, Haldane, del R{\'{\i}}o, Wiebe, Peterson, G{\'{e}}rard-Marchant, Sheppard, Reddy, Weckesser, Abbasi, Gohlke, \& Oliphant}]{numpy}
Harris, C.~R., Millman, K.~J., van~der Walt, S.~J., {et~al.} 2020, Nature, 585, 357, \dodoi{10.1038/s41586-020-2649-2}

\bibitem[{He {et~al.}(2023)He, Li, Cao, Li, Zou, \& Dye}]{he2023}
He, Z., Li, N., Cao, X., {et~al.} 2023, Discovering strongly lensed quasar candidates with catalogue-based methods from DESI Legacy Surveys.
\newblock \doarXiv{2301.11080}

\bibitem[{Hu {et~al.}(2022)Hu, Wang, Chen, \& Yang}]{hu2021}
Hu, L., Wang, L., Chen, X., \& Yang, J. 2022, The Astrophysical Journal, 936, 157, \dodoi{10.3847/1538-4357/ac7394}

\bibitem[{{Huang} {et~al.}(2020){Huang}, {Storfer}, {Ravi}, {Pilon}, {Domingo}, {Schlegel}, {Bailey}, {Dey}, {Gupta}, {Herrera}, {Juneau}, {Landriau}, {Lang}, {Meisner}, {Moustakas}, {Myers}, {Schlafly}, {Valdes}, {Weaver}, {Yang}, \& {Y{\`e}che}}]{huang2020}
{Huang}, X., {Storfer}, C., {Ravi}, V., {et~al.} 2020, The Astrophysical Journal, 894, 78, \dodoi{10.3847/1538-4357/ab7ffb}

\bibitem[{{Huang} {et~al.}(2021){Huang}, {Storfer}, {Gu}, {Ravi}, {Pilon}, {Sheu}, {Venguswamy}, {Banka}, {Dey}, {Landriau}, {Lang}, {Meisner}, {Moustakas}, {Myers}, {Sajith}, {Schlafly}, \& {Schlegel}}]{huang2021}
{Huang}, X., {Storfer}, C., {Gu}, A., {et~al.} 2021, The Astrophysical Journal, 909, 27, \dodoi{10.3847/1538-4357/abd62b}

\bibitem[{Hunter(2007)}]{matplotlib}
Hunter, J.~D. 2007, Computing in Science \& Engineering, 9, 90, \dodoi{10.1109/MCSE.2007.55}

\bibitem[{Inada {et~al.}(2012)Inada, Oguri, Shin, Kayo, Strauss, Morokuma, Rusu, Fukugita, Kochanek, Richards, Schneider, York, Bahcall, Frieman, Hall, \& White}]{inada2012}
Inada, N., Oguri, M., Shin, M.-S., {et~al.} 2012, The Astronomical Journal, 143, 119, \dodoi{10.1088/0004-6256/143/5/119}

\bibitem[{Jacob {et~al.}(2010)Jacob, Katz, Berriman, Good, Laity, Deelman, Kesselman, Singh, Su, Prince, \& Williams}]{montage}
Jacob, J.~C., Katz, D.~S., Berriman, G.~B., {et~al.} 2010, arXiv e-prints, \dodoi{10.48550/ARXIV.1005.4454}

\bibitem[{{Jacobs} {et~al.}(2017){Jacobs}, {Glazebrook}, {Collett}, {More}, \& {McCarthy}}]{jacobs2017}
{Jacobs}, C., {Glazebrook}, K., {Collett}, T., {More}, A., \& {McCarthy}, C. 2017, \mnras, 471, 167, \dodoi{10.1093/mnras/stx1492}

\bibitem[{{Jacobs} {et~al.}(2019){Jacobs}, {Collett}, {Glazebrook}, {Buckley-Geer}, {Diehl}, {Lin}, {McCarthy}, {Qin}, {Odden}, {Caso Escudero}, {Dial}, {Yung}, {Gaitsch}, {Pellico}, {Lindgren}, {Abbott}, {Annis}, {Avila}, {Brooks}, {Burke}, {Carnero Rosell}, {Carrasco Kind}, {Carretero}, {da Costa}, {De Vicente}, {Fosalba}, {Frieman}, {Garc{\'\i}a-Bellido}, {Gaztanaga}, {Goldstein}, {Gruen}, {Gruendl}, {Gschwend}, {Hollowood}, {Honscheid}, {Hoyle}, {James}, {Krause}, {Kuropatkin}, {Lahav}, {Lima}, {Maia}, {Marshall}, {Miquel}, {Plazas}, {Roodman}, {Sanchez}, {Scarpine}, {Serrano}, {Sevilla-Noarbe}, {Smith}, {Sobreira}, {Suchyta}, {Swanson}, {Tarle}, {Vikram}, {Walker}, {Zhang}, \& {DES Collaboration}}]{jacobs2019}
{Jacobs}, C., {Collett}, T., {Glazebrook}, K., {et~al.} 2019, The Astrophysical Journal Supplement, 243, 17, \dodoi{10.3847/1538-4365/ab26b6}

\bibitem[{Jaelani {et~al.}(2021)Jaelani, Rusu, Kayo, More, Sonnenfeld, Silverman, Schramm, Anguita, Inada, Kondo, Schechter, Lee, Oguri, Chan, Wong, \& Inoue}]{jaelani2021}
Jaelani, A.~T., Rusu, C.~E., Kayo, I., {et~al.} 2021, Monthly Notices of the Royal Astronomical Society, 502, 1487, \dodoi{10.1093/mnras/stab145}

\bibitem[{Kelly {et~al.}(2013)Kelly, Treu, Malkan, Pancoast, \& Woo}]{kelly2013}
Kelly, B.~C., Treu, T., Malkan, M., Pancoast, A., \& Woo, J.-H. 2013, The Astrophysical Journal, 779, 187, \dodoi{10.1088/0004-637x/779/2/187}

\bibitem[{{Kelly} {et~al.}(2022){Kelly}, {Zitrin}, {Oguri}, {Diego}, {Williams}, {Broadhurst}, {Chen}, {Koekemoer}, {Pierel}, {Strolger}, \& {Treu}}]{astronotekelly}
{Kelly}, P., {Zitrin}, A., {Oguri}, M., {et~al.} 2022, Transient Name Server AstroNote, 169, 1

\bibitem[{Kelly {et~al.}(2015)Kelly, Rodney, Treu, Foley, Brammer, Schmidt, Zitrin, Sonnenfeld, Strolger, Graur, Filippenko, Jha, Riess, Bradac, Weiner, Scolnic, Malkan, von~der Linden, Trenti, Hjorth, Gavazzi, Fontana, Merten, McCully, Jones, Postman, Dressler, Patel, Cenko, Graham, \& Tucker}]{kelly2015}
Kelly, P.~L., Rodney, S.~A., Treu, T., {et~al.} 2015, Science, 347, 1123, \dodoi{10.1126/science.aaa3350}

\bibitem[{Kelly {et~al.}(2023)Kelly, Rodney, Treu, Oguri, Chen, Zitrin, Birrer, Bonvin, Dessart, Diego, Filippenko, Foley, Gilman, Hjorth, Jauzac, Mandel, Millon, Pierel, Sharon, Thorp, Williams, Broadhurst, Dressler, Graur, Jha, McCully, Postman, Schmidt, Tucker, \& von~der Linden}]{kelly2023}
Kelly, P.~L., Rodney, S., Treu, T., {et~al.} 2023, Science, 380, \dodoi{10.1126/science.abh1322}

\bibitem[{Khramtsov {et~al.}(2019)Khramtsov, Sergeyev, Spiniello, Tortora, Napolitano, Agnello, Getman, de~Jong, Kuijken, Radovich, Shan, \& Shulga}]{khramtsov2019}
Khramtsov, V., Sergeyev, A., Spiniello, C., {et~al.} 2019, Astronomy \& Astrophysics, 632, A56, \dodoi{10.1051/0004-6361/201936006}

\bibitem[{{Kochanek} {et~al.}(2006){Kochanek}, {Mochejska}, {Morgan}, \& {Stanek}}]{kochanek2006}
{Kochanek}, C.~S., {Mochejska}, B., {Morgan}, N.~D., \& {Stanek}, K.~Z. 2006, \apjl, 637, L73, \dodoi{10.1086/500559}

\bibitem[{Kostrzewa-Rutkowska {et~al.}(2018)Kostrzewa-Rutkowska, Kozłowski, Lemon, Anguita, Greiner, Auger, Wyrzykowski, Apostolovski, Bolmer, Udalski, Szymański, Soszyński, Poleski, Pietrukowicz, Skowron, Mróz, Ulaczyk, \& Pawlak}]{kostrzewa2018}
Kostrzewa-Rutkowska, Z., Kozłowski, S., Lemon, C., {et~al.} 2018, Monthly Notices of the Royal Astronomical Society, 476, 663–672, \dodoi{10.1093/mnras/sty259}

\bibitem[{{Lacki} {et~al.}(2009){Lacki}, {Kochanek}, {Stanek}, {Inada}, \& {Oguri}}]{lacki2009}
{Lacki}, B.~C., {Kochanek}, C.~S., {Stanek}, K.~Z., {Inada}, N., \& {Oguri}, M. 2009, \apj, 698, 428, \dodoi{10.1088/0004-637X/698/1/428}

\bibitem[{{Lang} {et~al.}(2016){Lang}, {Hogg}, \& {Mykytyn}}]{lang2016}
{Lang}, D., {Hogg}, D.~W., \& {Mykytyn}, D. 2016, {The Tractor: Probabilistic astronomical source detection and measurement}, Astrophysics Source Code Library, record ascl:1604.008.
\newblock \doeprint{1604.008}

\bibitem[{{Leighly} {et~al.}(2018){Leighly}, {Terndrup}, {Gallagher}, {Richards}, \& {Dietrich}}]{leighly2018}
{Leighly}, K.~M., {Terndrup}, D.~M., {Gallagher}, S.~C., {Richards}, G.~T., \& {Dietrich}, M. 2018, \apj, 866, 7, \dodoi{10.3847/1538-4357/aadee6}

\bibitem[{{Leighly} {et~al.}(2019){Leighly}, {Terndrup}, {Lucy}, {Choi}, {Gallagher}, {Richards}, {Dietrich}, \& {Raney}}]{leighly2019}
{Leighly}, K.~M., {Terndrup}, D.~M., {Lucy}, A.~B., {et~al.} 2019, \apj, 879, 27, \dodoi{10.3847/1538-4357/ab212a}

\bibitem[{Lemon {et~al.}(2020)Lemon, Auger, McMahon, Anguita, Apostolovski, Chen, Fassnacht, Melo, Motta, Shajib, Treu, Agnello, Buckley-Geer, Schechter, Birrer, Collett, Courbin, Rusu, Abbott, Allam, Annis, Avila, Bertin, Brooks, Burke, Carnero Rosell, Carrasco Kind, Carretero, Costanzi, da Costa, De Vicente, Desai, Eifler, Flaugher, Frieman, García-Bellido, Gaztanaga, Gerdes, Gruen, Gruendl, Gschwend, Gutierrez, Honscheid, James, Kim, Krause, Kuehn, Kuropatkin, Lahav, Lima, Lin, Maia, March, Marshall, Menanteau, Miquel, Palmese, Paz-Chinchón, Plazas, Roodman, Sanchez, Schubnell, Serrano, Smith, Soares-Santos, Suchyta, Tarle, \& Walker}]{lemon2020}
Lemon, C., Auger, M.~W., McMahon, R., {et~al.} 2020, Monthly Notices of the Royal Astronomical Society, 494, 3491–3511, \dodoi{10.1093/mnras/staa652}

\bibitem[{{Lemon} {et~al.}(2023){Lemon}, {Anguita}, {Auger-Williams}, {Courbin}, {Galan}, {McMahon}, {Neira}, {Oguri}, {Schechter}, {Shajib}, {Treu}, {Agnello}, \& {Spiniello}}]{lemon2023}
{Lemon}, C., {Anguita}, T., {Auger-Williams}, M.~W., {et~al.} 2023, \mnras, 520, 3305, \dodoi{10.1093/mnras/stac3721}

\bibitem[{Lemon {et~al.}(2018{\natexlab{a}})Lemon, Auger, \& McMahon}]{lemon2019}
Lemon, C.~A., Auger, M.~W., \& McMahon, R.~G. 2018{\natexlab{a}}, Monthly Notices of the Royal Astronomical Society, 483, 4242, \dodoi{10.1093/mnras/sty3366}

\bibitem[{Lemon {et~al.}(2018{\natexlab{b}})Lemon, Auger, McMahon, \& Ostrovski}]{lemon2018}
Lemon, C.~A., Auger, M.~W., McMahon, R.~G., \& Ostrovski, F. 2018{\natexlab{b}}, Monthly Notices of the Royal Astronomical Society, 479, 5060, \dodoi{10.1093/mnras/sty911}

\bibitem[{Liao(2019)}]{liao2019}
Liao, K. 2019, The Astrophysical Journal, 871, 113, \dodoi{10.3847/1538-4357/aaf733}

\bibitem[{MacLeod {et~al.}(2019)MacLeod, Green, Anderson, Bruce, Eracleous, Graham, Homan, Lawrence, LeBleu, Ross, Ruan, Runnoe, Stern, Burgett, Chambers, Kaiser, Magnier, \& Metcalfe}]{macLeod2019}
MacLeod, C.~L., Green, P.~J., Anderson, S.~F., {et~al.} 2019, The Astrophysical Journal, 874, 8, \dodoi{10.3847/1538-4357/ab05e2}

\bibitem[{More {et~al.}(2015)More, Oguri, Kayo, Zinn, Strauss, Santiago, Mosquera, Inada, Kochanek, Rusu, Brownstein, da~Costa, Kneib, Maia, Quimby, Schneider, Streblyanska, \& York}]{more2016}
More, A., Oguri, M., Kayo, I., {et~al.} 2015, Monthly Notices of the Royal Astronomical Society, 456, 1595, \dodoi{10.1093/mnras/stv2813}

\bibitem[{{Moustakas}(2012)}]{moustakas2012}
{Moustakas}, L. 2012, in 10th Hellenic Astronomical Conference, ed. I.~{Papadakis} \& A.~{Anastasiadis}, 14--14

\bibitem[{{Planck Collaboration} {et~al.}(2020){Planck Collaboration}, {Aghanim}, {Akrami}, {Ashdown}, {Aumont}, {Baccigalupi}, {Ballardini}, {Banday}, {Barreiro}, {Bartolo}, {Basak}, {Battye}, {Benabed}, {Bernard}, {Bersanelli}, {Bielewicz}, {Bock}, {Bond}, {Borrill}, {Bouchet}, {Boulanger}, {Bucher}, {Burigana}, {Butler}, {Calabrese}, {Cardoso}, {Carron}, {Challinor}, {Chiang}, {Chluba}, {Colombo}, {Combet}, {Contreras}, {Crill}, {Cuttaia}, {de Bernardis}, {de Zotti}, {Delabrouille}, {Delouis}, {Di Valentino}, {Diego}, {Dor{\'e}}, {Douspis}, {Ducout}, {Dupac}, {Dusini}, {Efstathiou}, {Elsner}, {En{\ss}lin}, {Eriksen}, {Fantaye}, {Farhang}, {Fergusson}, {Fernandez-Cobos}, {Finelli}, {Forastieri}, {Frailis}, {Fraisse}, {Franceschi}, {Frolov}, {Galeotta}, {Galli}, {Ganga}, {G{\'e}nova-Santos}, {Gerbino}, {Ghosh}, {Gonz{\'a}lez-Nuevo}, {G{\'o}rski}, {Gratton}, {Gruppuso}, {Gudmundsson}, {Hamann}, {Handley}, {Hansen}, {Herranz}, {Hildebrandt}, {Hivon}, {Huang}, {Jaffe}, {Jones}, {Karakci}, {Keih{\"a}nen},
  {Keskitalo}, {Kiiveri}, {Kim}, {Kisner}, {Knox}, {Krachmalnicoff}, {Kunz}, {Kurki-Suonio}, {Lagache}, {Lamarre}, {Lasenby}, {Lattanzi}, {Lawrence}, {Le Jeune}, {Lemos}, {Lesgourgues}, {Levrier}, {Lewis}, {Liguori}, {Lilje}, {Lilley}, {Lindholm}, {L{\'o}pez-Caniego}, {Lubin}, {Ma}, {Mac{\'\i}as-P{\'e}rez}, {Maggio}, {Maino}, {Mandolesi}, {Mangilli}, {Marcos-Caballero}, {Maris}, {Martin}, {Martinelli}, {Mart{\'\i}nez-Gonz{\'a}lez}, {Matarrese}, {Mauri}, {McEwen}, {Meinhold}, {Melchiorri}, {Mennella}, {Migliaccio}, {Millea}, {Mitra}, {Miville-Desch{\^e}nes}, {Molinari}, {Montier}, {Morgante}, {Moss}, {Natoli}, {N{\o}rgaard-Nielsen}, {Pagano}, {Paoletti}, {Partridge}, {Patanchon}, {Peiris}, {Perrotta}, {Pettorino}, {Piacentini}, {Polastri}, {Polenta}, {Puget}, {Rachen}, {Reinecke}, {Remazeilles}, {Renzi}, {Rocha}, {Rosset}, {Roudier}, {Rubi{\~n}o-Mart{\'\i}n}, {Ruiz-Granados}, {Salvati}, {Sandri}, {Savelainen}, {Scott}, {Shellard}, {Sirignano}, {Sirri}, {Spencer}, {Sunyaev}, {Suur-Uski}, {Tauber}, {Tavagnacco},
  {Tenti}, {Toffolatti}, {Tomasi}, {Trombetti}, {Valenziano}, {Valiviita}, {Van Tent}, {Vibert}, {Vielva}, {Villa}, {Vittorio}, {Wandelt}, {Wehus}, {White}, {White}, {Zacchei}, \& {Zonca}}]{planck2018}
{Planck Collaboration}, {Aghanim}, N., {Akrami}, Y., {et~al.} 2020, \aap, 641, A6, \dodoi{10.1051/0004-6361/201833910}

\bibitem[{Potts \& Villforth(2021)}]{potts2021}
Potts, B., \& Villforth, C. 2021, Astronomy \& Astrophysics, 650, A33, \dodoi{10.1051/0004-6361/202140597}

\bibitem[{{Pourrahmani} {et~al.}(2018){Pourrahmani}, {Nayyeri}, \& {Cooray}}]{pourrahmani2018}
{Pourrahmani}, M., {Nayyeri}, H., \& {Cooray}, A. 2018, The Astrophysical Journal, 856, 68, \dodoi{10.3847/1538-4357/aaae6a}

\bibitem[{Quimby {et~al.}(2014)Quimby, Oguri, More, More, Moriya, Werner, Tanaka, Folatelli, Bersten, Maeda, \& Nomoto}]{quimby2014}
Quimby, R.~M., Oguri, M., More, A., {et~al.} 2014, Science, 344, 396, \dodoi{10.1126/science.1250903}

\bibitem[{Riess {et~al.}(2022)Riess, Yuan, Macri, Scolnic, Brout, Casertano, Jones, Murakami, Anand, Breuval, Brink, Filippenko, Hoffmann, Jha, Kenworthy, Mackenty, Stahl, \& Zheng}]{reiss2021}
Riess, A.~G., Yuan, W., Macri, L.~M., {et~al.} 2022, The Astrophysical Journal Letters, 934, L7, \dodoi{10.3847/2041-8213/ac5c5b}

\bibitem[{{Rodney} {et~al.}(2021){Rodney}, {Brammer}, {Pierel}, {Richard}, {Toft}, {O'Connor}, {Akhshik}, \& {Whitaker}}]{rodney2021}
{Rodney}, S.~A., {Brammer}, G.~B., {Pierel}, J. D.~R., {et~al.} 2021, Nature Astronomy, 5, 1118, \dodoi{10.1038/s41550-021-01450-9}

\bibitem[{{Schmidt} {et~al.}(2023){Schmidt}, {Treu}, {Birrer}, {Shajib}, {Lemon}, {Millon}, {Sluse}, {Agnello}, {Anguita}, {Auger-Williams}, {McMahon}, {Motta}, {Schechter}, {Spiniello}, {Kayo}, {Courbin}, {Ertl}, {Fassnacht}, {Frieman}, {More}, {Schuldt}, {Suyu}, {Aguena}, {Andrade-Oliveira}, {Annis}, {Bacon}, {Bertin}, {Brooks}, {Burke}, {Carnero Rosell}, {Carrasco Kind}, {Carretero}, {Conselice}, {Costanzi}, {da Costa}, {Pereira}, {De Vicente}, {Desai}, {Doel}, {Everett}, {Ferrero}, {Friedel}, {Garc{\'\i}a-Bellido}, {Gaztanaga}, {Gruen}, {Gruendl}, {Gschwend}, {Gutierrez}, {Hinton}, {Hollowood}, {Honscheid}, {James}, {Kuehn}, {Lahav}, {Menanteau}, {Miquel}, {Palmese}, {Paz-Chinch{\'o}n}, {Pieres}, {Plazas Malag{\'o}n}, {Prat}, {Rodriguez-Monroy}, {Romer}, {Sanchez}, {Scarpine}, {Sevilla-Noarbe}, {Smith}, {Suchyta}, {Tarle}, {To}, {Varga}, \& {DES Collaboration}}]{schmidt2023}
{Schmidt}, T., {Treu}, T., {Birrer}, S., {et~al.} 2023, \mnras, 518, 1260, \dodoi{10.1093/mnras/stac2235}

\bibitem[{Schulze {et~al.}(2017)Schulze, Schramm, Zuo, Wu, Urrutia, Kotilainen, Reynolds, Terao, Nagao, \& Izumiura}]{schulze2017}
Schulze, A., Schramm, M., Zuo, W., {et~al.} 2017, The Astrophysical Journal, 848, 104, \dodoi{10.3847/1538-4357/aa8e4c}

\bibitem[{Shajib {et~al.}(2020)Shajib, Birrer, Treu, Agnello, Buckley-Geer, Chan, Christensen, Lemon, Lin, Millon, Poh, Rusu, Sluse, Spiniello, Chen, Collett, Courbin, Fassnacht, Frieman, Galan, Gilman, More, Anguita, Auger, Bonvin, McMahon, Meylan, Wong, Abbott, Annis, Avila, Bechtol, Brooks, Brout, Burke, Rosell, Kind, Carretero, Castander, Costanzi, da~Costa, Vicente, Desai, Dietrich, Doel, Drlica-Wagner, Evrard, Finley, Flaugher, Fosalba, Garc{\'{\i} }a-Bellido, Gerdes, Gruen, Gruendl, Gschwend, Gutierrez, Hollowood, Honscheid, Huterer, James, Jeltema, Krause, Kuropatkin, Li, Lima, MacCrann, Maia, Marshall, Melchior, Miquel, Ogando, Palmese, Paz-Chinch{\'{o}}n, Plazas, Romer, Roodman, Sako, Sanchez, Santiago, Scarpine, Schubnell, Scolnic, Serrano, Sevilla-Noarbe, Smith, Soares-Santos, Suchyta, Tarle, Thomas, Walker, \& Zhang}]{shajib2020}
Shajib, A.~J., Birrer, S., Treu, T., {et~al.} 2020, Monthly Notices of the Royal Astronomical Society, 494, 6072, \dodoi{10.1093/mnras/staa828}

\bibitem[{Shajib {et~al.}(2023)Shajib, Mozumdar, Chen, Treu, Cappellari, Knabel, Suyu, Bennert, Frieman, Sluse, Birrer, Courbin, Fassnacht, Villafa{\~{n} }a, \& Williams}]{shajib2023}
Shajib, A.~J., Mozumdar, P., Chen, G. C.-F., {et~al.} 2023, Astronomy {\&} Astrophysics, 673, A9, \dodoi{10.1051/0004-6361/202345878}

\bibitem[{{Sheu} {et~al.}(2023){Sheu}, {Huang}, {Cikota}, {Suzuki}, {Schlegel}, \& {Storfer}}]{sheu2023}
{Sheu}, W., {Huang}, X., {Cikota}, A., {et~al.} 2023, \apj, 952, 10, \dodoi{10.3847/1538-4357/acd1e4}

\bibitem[{{Sonnenfeld} \& {Leauthaud}(2018)}]{sonnenfeld2018}
{Sonnenfeld}, A., \& {Leauthaud}, A. 2018, \mnras, 477, 5460, \dodoi{10.1093/mnras/sty935}

\bibitem[{Sonnenfeld {et~al.}(2020)Sonnenfeld, Verma, More, Baeten, Macmillan, Wong, Chan, Jaelani, Lee, Oguri, Rusu, Veldthuis, Trouille, Marshall, Hutchings, Allen, Donnell, Cornen, Davis, McMaster, Lintott, \& Miller}]{sonnenfeld2020}
Sonnenfeld, A., Verma, A., More, A., {et~al.} 2020, \aap, 642, A148, \dodoi{10.1051/0004-6361/202038067}

\bibitem[{Storfer {et~al.}(2022)Storfer, Huang, Gu, Sheu, Banka, Dey, Jain, Kwon, Lang, Lee, Meisner, Moustakas, Myers, Tabares-Tarquinio, Schlafly, \& Schlegel}]{storfer2022}
Storfer, C., Huang, X., Gu, A., {et~al.} 2022, arXiv e-prints, \dodoi{10.48550/ARXIV.2206.02764}

\bibitem[{Suyu {et~al.}(2020)Suyu, Huber, Ca{\~{n} }ameras, Kromer, Schuldt, Taubenberger, Y{\i}ld{\i}r{\i}m, Bonvin, Chan, Courbin, Nöbauer, Sim, \& Sluse}]{suyu2020}
Suyu, S.~H., Huber, S., Ca{\~{n} }ameras, R., {et~al.} 2020, Astronomy {\&} Astrophysics, 644, A162, \dodoi{10.1051/0004-6361/202037757}

\bibitem[{{Walsh} {et~al.}(1979){Walsh}, {Carswell}, \& {Weymann}}]{walsh1979}
{Walsh}, D., {Carswell}, R.~F., \& {Weymann}, R.~J. 1979, \nat, 279, 381, \dodoi{10.1038/279381a0}

\bibitem[{Wethers {et~al.}(2020)Wethers, Kotilainen, Schramm, \& Schulze}]{wethers2020}
Wethers, C.~F., Kotilainen, J., Schramm, M., \& Schulze, A. 2020, Monthly Notices of the Royal Astronomical Society, 498, 1469–1479, \dodoi{10.1093/mnras/staa2017}

\bibitem[{{Weymann} {et~al.}(1979){Weymann}, {Chaffee}, {Davis}, {Carleton}, {Walsh}, \& {Carswell}}]{weymann1979}
{Weymann}, R.~J., {Chaffee}, F.~H., J., {Davis}, M., {et~al.} 1979, \apjl, 233, L43, \dodoi{10.1086/183073}

\bibitem[{{Weymann} {et~al.}(1980){Weymann}, {Latham}, {Angel}, {Green}, {Liebert}, {Turnshek}, {Turnshek}, \& {Tyson}}]{weymann1980}
{Weymann}, R.~J., {Latham}, D., {Angel}, J. R.~P., {et~al.} 1980, \nat, 285, 641, \dodoi{10.1038/285641a0}

\bibitem[{Wong {et~al.}(2018)Wong, Sonnenfeld, Chan, Rusu, Tanaka, Jaelani, Lee, More, Oguri, Suyu, \& Komiyama}]{wong2018}
Wong, K.~C., Sonnenfeld, A., Chan, J. H.~H., {et~al.} 2018, The Astrophysical Journal, 867, 107, \dodoi{10.3847/1538-4357/aae381}

\bibitem[{Wong {et~al.}(2019)Wong, Suyu, Chen, Rusu, Millon, Sluse, Bonvin, Fassnacht, Taubenberger, Auger, Birrer, Chan, Courbin, Hilbert, Tihhonova, Treu, Agnello, Ding, Jee, Komatsu, Shajib, Sonnenfeld, Blandford, Koopmans, Marshall, \& Meylan}]{wong2019}
Wong, K.~C., Suyu, S.~H., Chen, G. C.-F., {et~al.} 2019, Monthly Notices of the Royal Astronomical Society, 498, 1420, \dodoi{10.1093/mnras/stz3094}

\bibitem[{Zhou {et~al.}(2020)Zhou, Newman, Mao, Meisner, Moustakas, Myers, Prakash, Zentner, Brooks, Duan, Landriau, Levi, Prada, \& Tarle}]{zhou2020}
Zhou, R., Newman, J.~A., Mao, Y.-Y., {et~al.} 2020, Monthly Notices of the Royal Astronomical Society, 501, 3309, \dodoi{10.1093/mnras/staa3764}

\end{thebibliography}
\bibliographystyle{aasjournal}
\end{document}